\documentclass{article}
\usepackage{fancybox,amsfonts} 
\usepackage{graphicx}
\usepackage{color}
\usepackage{amssymb}
\usepackage{latexsym}
\RequirePackage{ifpdf}
\makeatletter

\@addtoreset{equation}{section} \makeatother
\newtheorem{theorem}{Theorem}[section]
\newtheorem{remark}[theorem]{Remark}
\newtheorem{lemma}[theorem]{Lemma}
\newtheorem{proposition}[theorem]{Proposition}
\newtheorem{corollary}[theorem]{Corollary}
\newtheorem{definition}[theorem]{Definition}
\newtheorem{example}[theorem]{Example}
\newtheorem{convention}[theorem]{Convention}

\newcommand{\cvd}{\ \rule{0.5em}{0.5em}}
\newcommand{\be}{\begin{equation}}
\newcommand{\ee}{\end{equation}}

\newcommand{\N}{{\mathbb N}}

\newcommand{\R}{{\mathbb R}}
\newcommand{\LL}{{\mathbb L}}
\newcommand{\SSS}{{\mathbb S}}

\newcommand{\ben}{\begin{enumerate}}
\newcommand{\een}{\end{enumerate}}
\newcommand{\bit}{\begin{itemize}}
\newcommand{\eit}{\end{itemize}}
\newcommand{\edoc}{\end{document}}

\newcommand{\bdefi}{\begin{definition}}
\newcommand{\btheo}{\begin{theorem}}
\newcommand{\bprop}{\begin{proposition}}
\newcommand{\brema}{\begin{remark}}
\newcommand{\bcoro}{\begin{corollary}}
\newcommand{\blemm}{\begin{lemma}}
\newcommand{\bexam}{\begin{example}}

\newcommand{\edefi}{\end{definition}}
\newcommand{\etheo}{\end{theorem}}
\newcommand{\eprop}{\end{proposition}}
\newcommand{\erema}{\end{remark}}
\newcommand{\ecoro}{\end{corollary}}
\newcommand{\elemm}{\end{lemma}}
\newcommand{\eexam}{\end{example}}

\newcommand{\deform}{timelike deformable}
\newcommand{\tamech}{chronologically tame}
\newcommand{\tameca}{causally tame}
\begin{document}
\textwidth=140mm \textheight=200mm
\parindent=5mm
\date{}

\medskip


\title{{\bf\LARGE On the final definition of the causal
boundary and its relation with the conformal boundary}}

\author{{\bf\large J.L. Flores$^*$,
J. Herrera$^*$,
M. S\'anchez$^\dagger$}\\
{\it\small $^*$Departamento de \'Algebra, Geometr\'{\i}a y Topolog\'{\i}a,}\\
{\it \small Facultad de Ciencias, Universidad de M\'alaga,}\\
{\it\small Campus Teatinos, 29071 M\'alaga, Spain}\\
{\it\small $^\dagger$Departamento de Geometr\'{\i}a y Topolog\'{\i}a,}\\
{\it\small Facultad de Ciencias, Universidad de Granada,}\\
{\it\small Avenida Fuentenueva s/n, 18071 Granada, Spain}}

\maketitle

\begin{abstract}

The notion of {\em causal boundary} $\partial M$ for a strongly
causal spacetime $M$ has been a controversial topic along last
decades: on one hand, some attempted definitions were not fully
consistent, on the other, there were simple examples where an open
conformal embedding $i:M\hookrightarrow M_0$  could be defined,
but the corresponding conformal boundary $\partial_i M$ disagreed
drastically with the causal one. Nevertheless, the recent progress
in this topic suggests that a  final option for $\partial M$ is
available in most cases. Our study has two parts:

(I) To give general arguments on a boundary in order to ensure
that it is {\em admissible} as a causal boundary at the three
natural levels, i.e., as a point set, as a chronological space and
as a topological space. Then, the essential uniqueness of our
choice is stressed, and the relatively few admissible alternatives
are discussed.

(II) To analyze the role of the conformal boundary $\partial_iM$.
We show that, in general, $\partial_iM$ may present a very
undesirable structure. Nevertheless,  it is  well-behaved under
certain general assumptions, and its accessible part
$\partial_i^*M$ agrees with the causal boundary.

This study justifies both boundaries. On one hand, the conformal
boundary $\partial_i^*M$, which cannot be defined for a general
spacetime but is easily computed in particular examples, appears
now as a special case of the causal boundary. On the other, the
new redefinition of the causal boundary not only is free of
inconsistencies and applicable to any strongly causal spacetime,
but also recovers the expected structure in the  cases where a
natural simple conformal boundary is available. The cases of
globally hyperbolic spacetimes and asymptotically conformally flat
ends are especially studied.
\end{abstract}
\begin{quote}
\end{quote}
\begin{quote}
\end{quote}

\newpage

\tableofcontents

\section{Presentation of the results}

One of the major issues in Lorentzian Geometry is to find a
natural boundary for any spacetime, which would encode relevant
information on it. In fact, a good number of boundaries have been
defined, among them Geroch's geodesic boundary \cite{GeJMP},
Schmidt bundle boundary \cite{Sc, Sc2} and Scott and Szekeres
abstract boundary \cite{SS}. Two boundaries have a specially
important role in General Relativity, the conformal and the causal
ones. The conformal boundary is clearly the most applied in
Mathematical Relativity. Notions such as asymptotic flatness or
tools as Penrose-Carter diagrams rely on the conformal boundary.
Moreover, this boundary is also important in the framework of  the
AdS-CFT correspondence, as here one typically assumes that the
field theory lives on the conformal boundary of the ambient
spacetime. However, the conformal boundary has important
limitations, as it is an {\em ad hoc} construction: no general
formalism determines when the boundary of a reasonably general
spacetime is definable, intrinsic, unique  and containing useful
information on the spacetime. This was also stressed by Marolf and
Ross \cite{MR1} in the framework of the AdS-CFT correspondence: it
is trivial to consider a conformal boundary for anti-de Sitter
spacetime but, for  other backgrounds such as plane waves
\cite{BMN, BN}, this boundary may no exist (or yield suspicious
properties) 
so that the causal boundary must be used \cite{MR1, FS}.

One of the main motivations of the causal boundary (c-boundary,
for short) is to fill this gap of the conformal one --as well as
gaps of the other type of boundaries, see \cite{GLW, S}. The
c-boundary is intrinsic to the spacetime and conformally
invariant, according to the seminal construction by Geroch,
Kronheimer and Penrose \cite{GKP}. As a difference with the
conformal one, the c-boundary takes into account only timelike
curves and directions. Thus, the naive expectation  is that the
part of the conformal boundary which is accessible by means of
timelike directions of the original spacetime, must agree with the
c-boundary.  In this case, the (accessible) conformal boundary
would be endowed with a natural intrinsic meaning. There are other
motivations for the study of the c-boundary. As suggested in
\cite{S}, this boundary yields an appealing extended causal
picture of the spacetime. Indeed, the possible lack of global
hyperbolicity of the spacetime is associated to the existence of a
timelike point of the c-boundary (naked singularity); the
non-timelike points generate naturally a future and a past
infinity. From the viewpoint of the hyperbolic equations and the
initial value problem, the  solution on any neighborhood of the
timelike points and the past (or future) infinity, would determine
the solution. In simple cases,  the boundary conditions for the
equations might be imposed on the timelike points, and the initial
conditions might be imposed in the past infinity, or regarded as a
sort of limit on it.

Nevertheless, there has been a serious obstacle for the c-boundary
until now. This is the so-called {\em identification problem}
between future and past preboundary points, a problem of
consistency which also affects the choice of a natural topology
for the c-completion. Many authors have tried to solve or
circumvent it \cite{GKP, BS, Ra, Sz, Sz2, H1, H2}  (see
\cite{GpScqg05,S} for detailed reviews). However, among other
problems, the choices of identifications and topologies have been
affected by a major drawback: the so defined causal boundaries did
not agree with the conformal one, even in some simple open subsets
of Lorentz Minkowski --where a choice of the conformal boundary
seemed obvious \cite{KLLPRD, KLJMP, KLPRD, Ru, S}. The
unsatisfactory behavior appeared clearly on the side of the
c-boundary, especially in its topology\footnote{Kuang and Liang,
who found quite a few of these examples, commented in the last
one: ``We are inclined to believe that the whole project of
constructing a singular boundary has to be given up''
\cite{KLPRD}.}.

However, in the last years, some new ideas have been introduced on
these problems. First, Harris introduced the notion of
chronological set \cite{H1}, crucial to justify the universal
character of the causal boundary approach, and suggested a limit
operator on it \cite{H2}. Independently, Marolf and Ross
\cite{MR}, introduced a different viewpoint for the identification
problem: no future and past preboundary point must be identified,
but any c-boundary point must be regarded as a pair of two subsets
of the spacetime (a future and a past one), which satisfy a
certain binary relation (the {\em Szabados relation}, or simply,
{\em $S$-relation}). Then, these authors introduced a reasonable
c-boundary as a point set, endowed with a natural extended
chronological relation. They also suggested a pair of possible
topologies, with the hope that some topology between them were
good enough. Taken into account this viewpoint of pairs, the
first-named author introduced in \cite{F} a very general notion of
conformally invariant completion for spacetimes, or, in general,
for Harris' chronological sets. In spite of its generality, a
natural assumption on minimality for the completions made the
S-relation to appear, providing in particular a good support for
Marolf-Ross construction. Moreover, this author: (1) stretched to
the limit the assumption on minimality, arriving to the {\em
chronological completions}, which may differ from Marolf-Ross one,
and  (2) introduced a different topology, the {\em chronological
(chr.) topology}, inspired in some ideas by Harris. These two
fresh ideas, however, introduced more doubts on the existence of a
unique satisfactory definition of c-boundary. Nevertheless, the
critical review of the c-boundary by the third-named author
\cite{S}, announced that, with all the previous ingredients at
hand, there is a  natural choice for the c-boundary in most cases.
Moreover, this choice will be consistent with the conformal
boundary in the natural cases, giving a good support to both
boundaries. Summing up, the moral is the following. Even though
our choice of the c-boundary may be revised and modified, if some
mild hypotheses are satisfied, then it is both,  unique and
endowed with a good number of satisfactory properties. As these
characteristics should be reproduced by any further redefinition,
the c-boundary can be used safely in most cases, with independence
of future redefinitions.

Our purpose here is to develop this idea in detail. The plan of
work is the following. After some preliminaries in Section
\ref{s2}, the paper is divided into two parts. The first
one is devoted to 
the c-boundary (Section \ref{s3}), and the second to the conformal
one (Section \ref{s4}).

In the first part, our main aim is to  find a definition of the
c-boundary for any strongly causal spacetime which is supported by
simple and general requirements. These requirements are widely
discussed and compared with possible alternatives. Once the
c-boundary is defined, its properties are analyzed. More
precisely, in Section \ref{s3.1} we isolate which minimal
conditions must be fulfilled, in order to have a satisfactory
completion. So, starting at the general framework derived from
Geroch et al. \cite{GKP}, Marolf and Ross \cite{MR} and Flores
\cite{F},  we introduce some conditions on admissibility for the
c-boundary at the three levels --point set, chronologically and
topologically. At each level, we prove that there are few possible
alternatives and, essentially, only one reasonable option.
Concretely, we show that, as a point set, the admissible
boundaries may lie between Flores' chronological completions
(minimal options) and Marolf-Ross one (maximum one). All of them
will coincide in the relevant cases. Our canonical choice here
will be the univocally determined Marolf-Ross' one
---even though the framework of chronological ones may be required
further.  As a chronological space, different arguments will
confirm that the extended chronological relation, redefined by
Marolf and Ross from the work by Szabados \cite{Sz}, is the
undisputable admissible chronology. However, the question is
subtler as a topological space. Here, the essential conditions of
admissibility are  two: (A1) chronological futures and pasts must
be open, and (A2) the limits of the converging sequences in the
completions and the point-set structure of the futures and pasts
of the terms of the sequences, must satisfy a minimum
compatibility. As a third condition (A3), we also impose to have
the coarsest topology which fulfills the other requirements. This
last condition not only avoids undesirable topologies, such as the
discrete one, but it is also necessary to fix one topology. In
fact, we choose Flores' chr. topology and show that, when it is of
first order (Defn. \ref{overline}), then it is selected univocally
---it is the unique sequential admissible topology, Th. \ref{spc}.
As such a hypothesis is very mild (see Appendix \ref{sA1}, in
particular Proposition \ref{p6.2}), any other admissible topology
would differ only in the rather pathological cases where it is not
satisfied. So, our choices for the c-boundary $\partial M$ and
c-completion $\overline{M}$ are concluded here. In Section
\ref{sMR} we discuss Marolf-Ross topology in \cite{MR}. In fact,
they introduced two topologies,
 the first one did not satisfy
 the condition of admissibility
(A2), and the second one did not satisfy (A1). As (A2) is somewhat
subtler,  we analyze further both, (A2) and the first Marolf-Ross
topology.  Of course, this topology will also agree with the chr.
topology in most cases. However, some properties which can be
derived by using (A2) (as the $T_1$ character of the topology)
allows to fix a single one and, so, more non-$T_1$ topologies
similar to Marolf-Ross one should be conceivable.
 In Section \ref{s3.6} we show
that our choice of c-boundary fulfills satisfactory intrinsic
properties, (Th. \ref{rev}) --and they also hold for any
admissible completion as a point set. In Section \ref{s3.7} we
focus on the c-boundary of globally hyperbolic spacetimes, and
provide a natural definition of (asymptotically) conformally flat
end. Finally, in Section \ref{scausalcompletion} we discuss the
following question. The c-completion ensures that any timelike
curve will have an endpoint either in the spacetime or in the
boundary, at what extent does this remain true for causal curves?
We show that, even though this property may not hold in some cases
(and, eventually, the c-boundary might be redefined to ensure it),
it holds for some relevant and general families of spacetimes (Th.
\ref{ge}, Remark \ref{rge}). Finally, in the Appendix to this
first part,  we explain some subtleties about the topologies
determined by a limit operator, such as the chr and Harris ones,
which turn out important to
understand their possible pathologies. 

The second part has a threefold aim: first, to stress the
difficulties for a useful definition of the conformal boundary
(Sections \ref{s4.1}, \ref{s4.2}), second to provide general
conditions so that the (accessible) conformal boundary agrees with
the causal one (Section \ref{s4.3}) and third to obtain  easily
computable conditions for the conformal boundary points which are
$C^1$ (Section \ref{s4.4}). More precisely, in Section \ref{s4.1}
we provide the general notions of conformal envelopment $i:
M\hookrightarrow M_0$, boundary $\partial_iM$ and completion
$\overline{M}_i$ (Defn. \ref{d4.1}), discuss the possible
chronological and causal relations definable in $\overline{M}_i$
(Remark \ref{rconfchr}), focus on the accessible parts
$\partial_i^*M, \overline{M}^*_i$ (Defn. \ref{dacc}), and impose
chronological completeness (Defn. \ref{dcc}). Under this
assumption, there exist well-defined projections $\hat \pi$,
$\check \pi$ from the past and future causal completions to the
conformal one (Th. \ref{tproy}). However, these projections induce
a well-defined projection of the c-completion
$\pi:\overline{M}\rightarrow \overline{M}^*_i$ only when some
compatibility between $\hat \pi$ and $\check \pi$ holds (formula
(\ref{estarc})). Of course, we would like  not only that $\pi$
were well-defined, but also that it were an isomorphism at the
three levels (point-set, chronology, topology) ---so that one can
identify $\overline{M}\equiv \overline{M}^*_i$. The difficulties
are discussed in Section \ref{s4.2}. Then, in Section \ref{s4.3}
we give the general notion of {\em regular accessibility} for
conformal boundary points (Defn. \ref{ta}), which ensures the
required identification
(see the main result  Th. \ref{t1}). This notion comprises two
properties, {\em timelike deformability} (Defn. \ref{d1}) and {\em
timelike transitivity} (Defn. \ref{d2}). Even though such
properties are not difficult to check in practice, in Section
\ref{s4.4} we focus on $C^1$ boundary points, and find natural
interpretations to ensure them. In fact, there are two natural
chronological relations $\ll_i, \ll_i^S$, and two natural causal
relations $\le_i, \le_i^S$ definable in $\overline{M}_i$. When
these relations agree (i.e, the conformal completion is {\em
chronologically tame}, $\ll_i=\ll_i^S$, and {\em causally tame},
$\le_i = \le_i^S$), then the identification of the two completions
is ensured, $\overline{M}\equiv \overline{M}^*_i$ (Th. \ref{t10}).
Moreover, in practice one has to check only  the following
property: if a point of the boundary $z\in
\partial_iM$ is the endpoint of a future-directed (resp.
past-directed) timelike curve $\gamma$  in $M$,  then $\gamma$ can
be chosen such that it is also smooth and timelike at $z$ ({\em
strong accessibility}, Defn. \ref{s-a}, see Cor. \ref{t12} and
Remark \ref{rbclm}). We end this section by studying conformal
boundaries which are $C^1$ at any accessible boundary point $z\in
\partial^*_iM$. In this case,
 $z$ is called {\em timelike} if so is its tangent hyperplane
$T_z(\partial^*_iM)$. We prove that the absence of timelike points
implies the equivalence of the boundaries $\overline{M}\equiv
\overline{M}^*_i$ (Th. \ref{ttt1}). Moreover, among other results,
we prove rigourously a frequently claimed assertion: the absence
of timelike points is equivalent to the global hyperbolicity of
the spacetime (Cor. \ref{cfinal}). So, for globally hyperbolic
spacetimes with $C^1$ boundary  the conformal and causal
completions are  fully equivalent ($\overline{M}\equiv
\overline{M}^*_i \equiv \overline{M}_i$).
 Finally, quite a few  examples about the differences between the causal and conformal
 boundaries  are collected in the last Appendix.
They are referred along  all Section \ref{s4}, but its independent
reading may help to understand the process and choices we have
carried out.

\section{Preliminaries}\label{s2}

Here, we introduce some basic concepts and terminology. However,
some background and motivation on the causal boundary is assumed
and, so,  the survey \cite{S} is recommended. For the basic
elements of causality, we follow the conventions in \cite{O, MS}.
So, a {\em spacetime} is a Lorentzian manifold $M\equiv (M,g)$ of
dimension $N>1$  where a time-orientation is implicitly assumed. A
tangent vector $v\in T_pM, p\in M$ is {\em causal} if it is either
timelike ($g(v,v)<0$) or lightlike ($g(v,v)=0, v\neq 0$).

\subsection{Background on the causal boundary}
The causal boundary of a spacetime was introduced by  Geroch,
Kronheimer and Penrose  \cite{GKP} under an  appealing philosophy:
to add {\em ideal points} to the spacetime so that any timelike
curve acquires  some endpoint in the completed space. To formalize
this GKP seminal idea, the following elements are basic.

 A subset $P\subset M$ is called a {\em past set} if it coincides
with its past; i.e. $P=I^{-}[P]:=\{p\in M: p\ll q\;\hbox{for
some}\; q\in P\}$. The {\em common past} of $S\subset M$ is
defined by $\downarrow S:=I^{-}[\{p\in M:\;\; p\ll q\;\;\forall
q\in S\}]$. A past set that cannot be written as the union of two
proper subsets, both of which are also past sets, is called {\em
indecomposable past} set, IP. An IP which does coincide with the
past of some point of the spacetime $P=I^{-}(p)$, $p\in M$ is
called {\em proper indecomposable past set}, PIP. Otherwise,
$P=I^{-}[\gamma]$ for some inextensible future-directed timelike
curve $\gamma$, and it is called {\em terminal indecomposable past
set}, TIP. The dual notions of {\em future set}, {\em common
future}, IF, TIF and PIF, are obtained just by interchanging the
roles of past and future in previous definitions.

To construct the GKP future causal completion, first identify
every event $p\in M$ with its PIP, $I^{-}(p)$ (this is consistent
for distinguishing spacetimes, see definition below). Then, the
{\em future causal boundary} $\hat{\partial}M$ of $M$ is defined
as the set of all the TIPs in $M$. Therefore, {\em the future
causal completion} $\hat{M}$ becomes the set of all the IPs:
\[
M\equiv \hbox{PIPs},\qquad \hat{\partial}M\equiv
\hbox{TIPs},\qquad\hat{M}\equiv \hbox{IPs}.
\]
Analogously, every event $p\in M$ can be identified with its PIF,
$I^{+}(p)$. Then, the {\em past causal boundary}
$\check{\partial}M$ of $M$ is defined as the set of all the TIFs
in $M$, and thus, {\em the past causal completion} $\check{M}$ is
the set of all the IFs:
\[
M\equiv \hbox{PIFs},\qquad \check{\partial}M\equiv
\hbox{TIFs},\qquad\check{M}\equiv \hbox{IFs}.
\]
In order to define the (total) causal completion, consider first
the precompletion $M^\sharp = (\hat{M} \cup \check{M})/\sim$,
equal to the space $\hat{M}\cup\check{M}$ with PIP's and PIF's
identified in an obvious way ($I^{-}(p)\sim I^{+}(p)$ on
$\hat{M}\cup\check{M}$ for all $p\in M$). However, some additional
identifications  between the preboundary points
$\hat{\partial}M\cup\check{\partial}M$ seemed necessary in order
to obtain a reasonably consistent definition. The study of this
identification problem was already initiated in \cite{GKP}, and
 developed further in \cite{BS, Ra, Sz, Sz2}, but without fully
satisfactory results. In the initial GKP idea the problem is
linked to the choice of an appropriate topology: $M^\sharp$ is
made a topological space, and $\sim$ yields the minimum
identifications so that the quotient is Hausdorff. The
so-constructed completion was expected to work for strongly causal
spacetimes, although the existence of  $\sim$ required stable
causality \cite{Sz}. But even in this case,  some examples showed
that the boundary was unsatisfactory \cite{KLLPRD, KLJMP}. Budic
and Sachs \cite{BS} introduced a direct identification between IPs
and IFs, later refined by Szabados \cite{Sz}, which is then called
the S-relation $\sim_S$ (see (\ref{eSz}), (\ref{eSz2}) below).
However, the choices of topologies by these and other authors did
not fulfill the expectations \cite{KLLPRD, KLJMP, KLPRD}.
According to the new viewpoint developed by Marolf and Ross
\cite{MR}, the identifications of preboundary points are
abandoned. Instead, one takes all the pairs $(P,F)$ such that $P$
is an IP, $F$ is an IF and they are related by the S-relation.
This pairing process is justified in \cite{F} by general arguments
involving a minimality condition.

Related important ingredients were introduced by Harris in
\cite{H1,H2}. On one hand, a limit operator, which lies in the
core of the topology defined by Flores \cite{F}. On the other,
Harris defined the following notion, in order to describe
simultaneously the properties of the spacetimes and their
completions.
\begin{definition}\label{chronrel} A {\em chronological set}
$(X,\ll)$ is a set $X$ endowed with a {\em chronology}, i.e., a
binary relation $\ll$ which is transitive, anti-reflexive and
satisfies:
\begin{itemize}
\item[(i)] it contains no isolates: each $x\in X$ satisfies $x\ll
y$ or $y\ll x$ for some $y\in X$, and \item[(ii)] it is {\em
chronologically separable}, that is, there exists a countable set
${\cal S}\subset X$ which is {\em chronologically dense}: for all
$x\ll y$ there exists $s\in {\cal S}$ such that $x\ll s\ll y$.
\end{itemize}
\end{definition}
Notice that definitions for spacetimes on causality and GKP
construction such as chronological futures and pasts $I^\pm(x)$,
IPs, TIPs, etc., can be translated immediately to chronological
sets. In particular, future-directed timelike curves are replaced
by {\em future-directed {\em (}chronological{\em )} chains}, i.e.,
sequences $\varsigma=\{x_n\}_n$ such that $x_n \ll x_{n+1}$ for
all $n$. Chronological separability ensures that IPs coincide with
pasts of future-directed chains \cite[Th. 3]{H1}. A point $x\in X$
is  {\em regular} if both, $I^-(x)$ is an IP and $I^+(x)$ an IF;
notice that this happens in spacetimes (where they are PIP and
PIF, resp.) but not in general, see Fig. \ref{f5}. The
chronological set will be {\em distinguishing} when $x\neq y
\Rightarrow I^-(x)\neq I^-(y)$ and $ I^+(x)\neq I^+(y)$, i.e.,
each point $x\in X$ is identifiable with both, $I^-(x)$ and
$I^+(x)$.

\begin{remark} {\em Chronological sets constitute the general framework for
causal boundaries and, trivially, our choice of c-boundary can be
 formulated on them. However, we will focus on
 minimally well-behaved spacetimes, being strong causality enough
 for our purposes. The reason is that we are looking
for the admissible conditions which  must be satisfied by a causal
boundary. A chronological set does not have a simple distinguished
topology --except the Alexandrov one, which is {\em not}
appropriate for the completions of spacetimes, even the strongly
causal ones. Although the chronological topology, which will be
our final choice, can be formulated in chronological sets, the
motivations for this topology become more transparent in
spacetimes.}

In what follows, all spacetimes will be strongly causal, except if
otherwise is said explicitly.

\end{remark}

\subsection{Some technicalities related to the conformal boundary}

Once the causal boundary is constructed, our aim is to test it by
means of  the conformal one.
For this boundary, one embeds conformally the original spacetime
$M$ in a new ({\em aphysical}) spacetime $M_0$ of the same
dimension, and regards the topological boundary of $M$ in $M_0$ as
the conformal boundary. We will discuss widely this idea below;
here, we just point out some technicalities.

\brema \label{r2.2o} {\em (1) In Causality Theory, the causal and
timelike curves with two fixed endpoints are typically assumed
{\em piecewise smooth}, i.e., they have a finite number of breaks.
In principle, there is no problem to smooth such curves retaining
its causal or timelike character. Nevertheless,  a difficulty
arises when the conformal boundary is studied. Here, we consider
inextensible timelike curves $\gamma$ in the original spacetime
$M$ which will have an endpoint in the aphysical spacetime $M_0$.
In principle, one can allow that these curves have an infinite
number of breaks, and all  the breaks may be smoothed. However,
the extended curve $\bar \gamma$ to $M_0$
may not be smooth at its endpoint in $M_0\setminus M$ (Fig.
\ref{fig2''}). The difficulty to smooth it satisfactorily is
inherent to the fact that the endpoint lies in the boundary  but
the curve $\gamma$ must be smoothed in $M$.

(2) The most general  space for causal curves is the one of, say,
future-directed {\em continuous causal curves}. Such a curve
$\rho$ is a continuous one which satisfies $t_{1}<t_{2}\Rightarrow
\rho(t_{1})\le\rho(t_{2})$ (usually, this condition is assumed to
hold for arbitrarily small neighborhoods, however we can drop this
requirement as our spacetimes will be always strongly causal).
Future-directed continuous causal curves can be characterized as
those locally absolutely continuous curves such that its velocity
is almost everywhere future-directed causal \cite{CFS} (in
particular, if a continuous curve lies in the conformal closure of
a spacetime, its possible causal character depends only on this
closure). Continuous causal curves  will play a role in the
interplay between the causal relations in $M$ and
$M_0$. 
 }\erema Summing up, in what follows all the causal curves in $M$ will be
assumed smooth with no loss of generality and, if a causal curve
in $M_0$ is regarded only as continuous, then it will be called
explicitly continuous causal curve. Finally, as a technical result
to be used frequently\footnote{All the results will be stated for
future-directed curves, TIPs, etc., and the dual versions for
past-directed curves, TIFs or dual elements, will be used without
further mention.}:

\bprop\label{paux} Let $M$ be a strongly causal spacetime and
$\gamma: [a,b) \rightarrow M$ a future-directed timelike curve.
The curve $\gamma$ can be continuously extended to some $p\in M$
if either

(i)  for some sequence $\{t_n\}_n\nearrow  b$ the sequence $\{
\gamma(t_n)\}_n$ converges to $p$, or

(ii) $I^-[\gamma] = I^-(p)$.

\eprop {\em Proof}. For (i), notice that, otherwise, strong
causality would be violated at $p$. For (ii), just recall that
$I^-[\gamma]$ is a PIP and, thus, $\gamma$ is continuously
extensible \cite[Prop. 6.14]{BEE}. \cvd

\section{Admissible completions: uniqueness of the c-boundary}
\label{s3}

In this part we analyze the admissible possibilities for a causal
boundary. We retain the original idea that  any inextensible
future or past-directed timelike curve must converge to a boundary
point.  Of course, one can think that non--timelike curves should
be also included in a full completion. This will be discussed for
lightlike curves in Section \ref{scausalcompletion}. However,  we
will not try to include spacelike curves, as this would require a
very different  viewpoint --causal boundary is based on Causality,
which is the global conformal invariant of a spacetime. At any
case, if a bigger conformally invariant completion were developed,
the causal boundary should be included in some way (see also
Section \ref{scausalcompletion}).

\subsection{Conditions on admissibility}\label{s3.1}

These conditions will appear at the three levels: point-set,
chronological and topological.

\subsubsection{Admissible completions as point sets}

We will identify $M$ with the subset of $\hat{M}\times \check{M}$
formed by all the pairs $(I^-(p),I^+(p))$. Denote also
$\hat{M}_{\emptyset}=\hat{M}\cup \{\emptyset\}$ (resp.
$\check{M}_{\emptyset}=\check{M}\cup \{\emptyset\}$). The
S-relation $\sim_S$ is defined in $\hat{M}_{\emptyset}\times
\check{M}_{\emptyset}$ as follows. First, for $(P,F)\in
\hat{M}\times \check{M}$:

\be \label{eSz}  P\sim_S F \Longleftrightarrow \left\{
\begin{array}{l}
F \quad
\hbox{is included and is a maximal  IF  in} \quad \uparrow P \\
P \quad \hbox{is included and is a maximal  IP  in} \quad
\downarrow F.
\end{array} \right.
\ee (Recall that, as proved by Szabados \cite{Sz}, $I^-(p) \sim_S
I^+(p)$ for all $p\in M$, and these are the unique S-relations
involving proper indecomposable sets, according to (\ref{eSz})).
For $(P,F)\in \hat{M}_{\emptyset}\times \check{M}_{\emptyset}$,
with $(P,F)\neq (\emptyset,\emptyset)$ we also put \be
\label{eSz2} P\sim_S \emptyset, \quad \quad (\hbox{resp.} \;
\emptyset \sim_S F )\ee if $P$ (resp. $F$) is a (non-empty,
necessarily terminal) indecomposable set
 and is not S-related by (\ref{eSz}) to any other
indecomposable set. Notice that $\emptyset$ is never S-related to
itself.

\begin{definition}\label{def} A subset $\overline{M}\subset \hat{M}_{\emptyset}\times
\check{M}_{\emptyset}$ is called an {\em admissible completion as
a point set} for $M$ if it contains $M$ (i.e. $\{(I^-(p),I^+(p)):
p\in M\}\subset \overline{M}$) and satisfies the following
conditions:
\begin{itemize}
\item[(1)] Completeness:  every TIP and TIF in $M$ is the
component of some pair in the {\em boundary of the completion},
i.e. in the subset $\partial M:= \overline{M}\setminus M$,
\item[(2)] S-relation: if a pair $(P,F)\in
\hat{M}_{\emptyset}\times \check{M}_{\emptyset}$ lies in
$\overline{M}$ then
$P\sim_{S} F$.
\end{itemize}
\end{definition}
\begin{remark} \label{r2.2} (Justification of the conditions on
admissibility.)

 {\em (A) To regard $\overline M$ as a subset of $\overline{M}\subset \hat{M}_{\emptyset}\times
\check{M}_{\emptyset}$ is just a general framework (introduced in
\cite{MR}).
 The axiom (1) is necessary in order
to formalize the idea that any timelike curve which is
inextensible towards the future (or past) will have an {\em
endpoint} in the boundary. This can be formalized more accurately
in the general setting of chronological spaces by requiring that
any chain has an {\em endpoint} (see \cite[Sect. 3]{F} for
definitions).

In principle, it is possible to consider even a more general
framework. The notion of {\em completion} introduced in
\cite[Defn. 3.2]{F} allows $\overline{M}$ to be included not only
in $\hat{M}_{\emptyset}\times \check{M}_{\emptyset}$ but also in
the larger set $M_P\times M_F$ which is formed by all the pairs
composed by a future and a past set (see also \cite[Defn. 5.3,
Prop. 5.2]{S}). In this case, the notion of endpoint is a bit
stronger than the one of limit point. However, the analysis of
this a priori more general completions  in \cite[Th. 7.4]{F} (see
also below), shows that the relevant completions will turn out
subsets of $\hat{M}_{\emptyset}\times \check{M}_{\emptyset}$
(i.e.,  IPs and IFs of the original GKP idea are recovered) where
axiom (2) holds. In this case, the notions of endpoint and limit
point become equivalent for chains, and our simplified framework
is sufficient.

(B) Condition (2) plus the following one,
\begin{itemize}
\item[{\it (NR)}] {\em Non-redundancy}: if $(P,F_{1}),
(P,F_{2})\in\partial M$ and $F_{1}\neq F_{2}$ (resp. $(P_{1},F),
(P_{2},F)\in\partial M$ and $P_{1}\neq P_{2}$) then $F_{i}$ (resp.
$P_{i}$), $i=1,2$, do not appear in another pair of $\partial M$,
\end{itemize}
characterize those completions which are {\em minimal} in a
natural sense. This minimality condition was introduced in the
general setting of completions in $M_P\times M_F$ for a
chronological set \cite[Defn. 7.3, 7.1]{F}. In this reference,
completions satisfying (NR) are called {\em chronological
completions}, and it is proved that, when  a chronological
completion of a strongly causal spacetime is considered, it
satisfies axiom (2) automatically. That is, the axiomatically
imposed condition (2) can be deduced from a more fundamental
viewpoint, supporting the appearance of the S-relation.
}\end{remark}

\begin{remark} \label{r2.2b} (Uniqueness of the admissible
completions as a point set.) {\em Marolf-Ross (MR) completion
\cite{MR} is the one which contains all the possible pairs
compatible with (2), that is, if $P\sim_{S}F$ then
$(P,F)\in\overline{M}$. Notice that, in this case, condition (1)
becomes superfluous, that is, MR completion is characterized by
the following strengthening of (1) and (2):
\begin{itemize}
 \item[{\it (MR)}] A pair $(P,F)\in \hat{M}_{\emptyset}\times
\check{M}_{\emptyset}$ lies in $\overline{M}$ if and only if
$P\sim_{S} F$.
\end{itemize}
So, as a point set, MR completion is characterized as the maximum
admissible completion. In particular, MR completion is unique, as
a difference with chronological completions. However, it may be
redundant, even in cases where the chronological completion is
unique, see \cite[Example 10.6]{F}, \cite[Appendix A]{MR}.

}\end{remark} Summing up:
\begin{itemize}
\item[1.] The axioms for admissible completions, including the
S-relation, are not only intuitively sound  but also appear
naturally from the general notion of chronological completions.
\item[2.] Then, MR completion (characterized by {\em (MR)} above)
is selected here as the maximum admissible completion --in
principle, the unique admissible completion canonically
determined. \item[3.] At any case, the differences between the
different admissible completions, from the minimal (chronological)
completions to the maximum (Marolf-Ross) one are essentially
irrelevant from the practical viewpoint: in the natural physical
examples all of them coincide (this includes, for example, the
subtler case of minimally well-behaved wave-type spacetimes
\cite{FS}).

So, in order to construct more than one admissible completion for
the same spacetime, one has to consider some rather sophisticated
examples as those exhibited in \cite{F}, where some regions are
artificially removed, or in  \cite{FHSst}, where  standard
stationary spacetimes with a very peculiar behavior close to its
Busemann boundary are considered.
\end{itemize}


\subsubsection{Admissible chronology}

\begin{definition}\label{def1} Let $\overline{M}$ be an admissible completion as a point set for some
strongly causal spacetime $M$. A  chronology $\tilde{\ll}$ on the
completion $\overline{M}$ (Defn. \ref{chronrel}) is called:

(i) An {\em extended chronology} if
\begin{eqnarray}\label{eea}
p \in  P & \Rightarrow & (I^-(p),I^+(p))\; \tilde \ll \;(P,F), \\
\label{eeb} q \in  F & \Rightarrow & \; (P,F) \tilde \ll \;
(I^-(q),I^+(q)).
\end{eqnarray}
for all $p,q\in M$ and $(P,F)\in \overline{ M}$.

(ii) An {\em admissible chronology} if
$I^{\pm}((P,F))\subset\overline{M}$ computed with $\tilde{\ll}$
satisfies:
\begin{equation}\label{ee0}
I^{-}((P,F))\cap M=P \quad \mbox{and} \quad I^{+}((P,F))\cap M=F
\quad  \; \forall (P,F)\in\overline{M}.
\end{equation}
\end{definition}

\begin{remark}\label{rchr}{\em (A) When restricted to points in $M$,
conditions (\ref{eea}), (\ref{eeb}) only say:
$$p\ll q \Rightarrow 
p\; \tilde \ll \; q \quad \forall p,q\in M. $$
When $(P,F)\in \partial M$, those conditions mean only that
$(P,F)$ is (future or past) chronologically related with the
points in the TIP or TIF which define the pair. That is, {\em an
extended chronology contains both, the original chronology $\ll$
and the chronological relations which define the boundary
points\footnote{In the discussion at the end of Section
\ref{s4.2}, in point 4, we will see that the conformal boundary
may lead to some dilemma in the definition of a chronological
relation for the conformal completion. In fact, from the viewpoint
of this boundary, it seems consistent  to admit $p\in P$ but
$p\tilde{\not\ll} (P,F)$ for some $p\in M$ and $(P,F)$ in the
boundary. However, such a possibility appears only because of the
non-intrinsic character of the conformal boundary, and would
generate new problems. Moreover, if such a possibility were
admitted in general, one could also speculate with very
undesirable chronological relations on $\overline M$, as the one
which maintains only the original chronological relations in
$M$.}.}

(B) Of course, an admissible chronology is an extended one. In
fact, condition (\ref{ee0}) not only implies  (\ref{eea}) and
(\ref{eeb}) but also says that no other relations are introduced
between points in the spacetime or a point in the spacetime and
one in the boundary. That is, {\em the chronology $\tilde{\ll}$ is
called  admissible when it is an extended chronology which does
not introduce new (spurious) relations between a point in $M$ and
any point in $\overline{M}$.}}
\end{remark}

Remark \ref{rchr} (B) suggests the possibility to single out an
admissible chronology just by requiring that no new relations
between two points in $\partial M$ are introduced. This will be
our choice of extended chronology, which was firstly suggested by
Szabados \cite{Sz}, refined to make it consistent by Marolf and
Ross
 \cite{MR} and
also considered in \cite{F,S}.
\begin{definition}\label{def1'} Let $\overline{M}$ be an admissible completion as a point set for some strongly causal spacetime $M$.
The {\em extended chronological relation} $\overline{\ll}$ is the
relation on $\overline{M}$ given by:
\begin{equation}\label{ee}
(P,F)\overline{\ll} (P',F')\quad\Leftrightarrow\quad F\cap
P'\neq\emptyset,\qquad \forall (P,F),(P',F')\in \overline{M}.
\end{equation}
\end{definition}
The name of this relation suggests that it will be the canonical
one on $\overline{M}$. To justify this, it is easy to check  first
that $\overline{\ll}$ is an admissible
 chronology (see, for example, \cite[Th. 4.3]{F}),
as well as the following characterization.
\begin{theorem}\label{tminimum}
Let $\overline{M}$ be an admissible completion as a point set for
some strongly causal spacetime $M$, and $\tilde \ll$  an extended
chronology on $\overline{M}$. Then, the extended chronological
relation $\overline{\ll}$ satisfies:
\begin{equation}\label{eminimum} (P,F) \overline{\ll} (P'F')
\Rightarrow (P,F) \tilde{\ll} (P'F'), \quad \quad \forall (P,F),
(P',F') \in \overline{M}. \end{equation} That is, the extended
chronological relation $\overline{\ll}$ is the minimum extended
chronology for an admissible completion as a point set
$\overline{M}$ of a strongly causal spacetime.
\end{theorem}
{\it Proof.} Assume that $(P,F)\overline{\ll}(P',F')$ for some
$(P,F), (P',F')\in \overline{M}$. From (\ref{ee}), there exists
some $p\in F\cap P'\neq\emptyset$. From (\ref{eea}), (\ref{eeb}),
$(P,F)\tilde{\ll}(I^{-}(p),I^{+}(p))\tilde{\ll} (P',F')$. Hence,
by the transitivity of $\tilde{\ll}$, we deduce
$(P,F)\tilde{\ll}(P',F')$. \cvd

\smallskip

\smallskip

\noindent Even though this result singles out the extended
chronological relation, one could still speculate with the
possibility of an admissible chronology such that additional
chronological relations are introduced between the boundary
points. However, the following result suggests that this is not
reasonable --at least if the construction is intended to has a
minimum compatibility with a reasonable topology.

\begin{theorem}\label{topen}
Let $\overline{M}$ be an admissible completion as a point set for
some strongly causal spacetime $M$, and $\tilde \ll$  an
admissible chronology on $\overline{M}$. Assume that
$\overline{M}$ is endowed with any topology such that:\footnote{We
anticipate here some properties of consistency  for the topology,
which will be developed below.}
(i) $M$ is dense into $\overline{M}$,  and
(ii) the topology is compatible with $\tilde \ll$ in the sense
that
$I^{\pm}((P,F))$ is open in $\overline{M}$ for all $(P,F)\in
\overline{M}$.

Then, $\tilde \ll$ is equal to the extended chronological relation
$\overline{\ll}$.
\end{theorem}
{\it Proof.} Taking into account Theorem \ref{tminimum}, we have
just to prove the converse of (\ref{eminimum}), under our
additional hypotheses for $\tilde{\ll}$. Assume that
$(P,F)\tilde{\ll} (P',F')$. Since $\tilde{\ll}$ is chronologically
separable  (Defn. \ref{chronrel}), there exists
$(P'',F'')\in I^{+}((P,F))\cap I^{-}((P',F'))$ 
(notice that $I^\pm$  are computed with $\tilde{\ll}$ in
$\overline{M}$). Since $I^{\pm}(\cdot)$ are open and $M$ is dense
in $\overline{M}$, we can choose $(P'',F'')$ equal to a point
$p\in M$. As $\tilde{\ll}$ is admissible, $I^{+}((P,F))\cap M=F$,
$I^{-}((P',F'))\cap M=P'$, and $p\in F\cap P'\neq \emptyset$,
i.e., $(P,F)\overline{\ll} (P',F')$, as required. \cvd

%
%

\smallskip

\noindent Summing up:

\begin{quote} The extended chronological relation $\overline{\ll}$ (Defn.
\ref{def1'}) is singled out as:
\begin{itemize}
\item[(1)] The minimum extended chronology (Th. \ref{tminimum}).
\item[(2)] The unique admissible chronology compatible with a
minimally well-behaved topology (Th. \ref{topen}).
\end{itemize}
Moreover, it is, in principle, the unique admissible chronology
canonically defined. So, even though other chronologies have been
defined on $\overline{M}$ (see \cite[Remark 5.1, Subsect.
3.3.1]{S} for a discussion), the properties above make
$\overline{\ll}$ the standard choice of chronology in
$\overline{M}$.

So, $\overline{\ll}$ will be taken in any admissible completion as
a point set $\overline{M}$ along the remainder of this paper.

\end{quote}

\begin{convention} {\em In what follows, the points of the spacetime
are denoted by lower case letters $p,q\in M$, and their
chronological future and past in $M$ by $I^{+}(p)$, $I^{-}(q)$.
Points in the completion (eventually also in $M$) are denoted as
pairs $(P,F)$, and $I^{+}((P,F)), I^{-}((P,F))\subset
\overline{M}$ denote their chronological future and past in
$\overline{M}$ for the extended chronological relation. In case of
possibility of confusion, the notation $I^{+}(p,M)$,
$I^{+}(p,\overline{M})$ is used for the chronological future of
$p\in M$ in $M$ and $\overline{M}$, resp. Consistently, a
superscript $c$ will denote the complementary of the corresponding
subset; for example, $I^{+}((P,F))^{c} =\overline{M}\setminus
I^{+}((P,F))$.}
\end{convention}

\subsubsection{Admissible topologies}\label{ss}

In the following, we are going to assume that $\overline{M}$ is an
admissible completion as a point set (Defn. \ref{def}) endowed
with the extended chronological relation $\overline{\ll}$ (Defn.
\ref{def1'}). Our next goal is to determine when a topology on
$\overline M$ can be admitted as admissible and, even more, if one
such topology can be singled out. In principle, there are two good
candidates, one of the two choices by Marolf and Ross in \cite{MR}
(see Section \ref{sMR}) and the chronological topology introduced
in \cite{F}. First, we consider the latter in order to make clear
its point-set viewpoint.

\begin{definition}\label{overline}
Let $\overline{M}$ be an admissible completion as a point set for
some strongly causal spacetime $M$. The {\em (chr) limit operator}
$L$ maps each sequence $\sigma=\{(P_{n},F_{n})\}\subset
\overline{M}$ in the subset $L(\sigma)\subset \overline M$ defined
as:
\[
(P,F)\in L(\sigma)\;\quad \Leftrightarrow\;\quad \hbox{if} \;
P\neq \emptyset, \; P\in \hat{L}(P_{n})\quad\hbox{and}\quad
\hbox{if} \; F\neq \emptyset, \; F\in \check{L}(F_{n}),
\]
where
\[
P\in \hat{L}(P_{n})\quad\Leftrightarrow\quad
\left\{\begin{array}{l} P\subset LI(P_{n}) \\
P\;\;\hbox{is a}\;\;\hbox{maximal}\;\; IP\;\;\hbox{in}\;\;
LS(P_{n})\end{array}\right.
\]
\[
F\in
\check{L}(F_{n})\quad\Leftrightarrow\quad
\left\{\begin{array}{l} F\subset LI(F_{n}) \\
F\;\;\hbox{is a}\;\;\hbox{maximal}\;\; IF\;\;\hbox{in}\;\;
LS(F_{n})\end{array}\right.
\]
and $LS$, $LI$ denote the lim-sup and lim-inf operators in set
theory.

The {\em chr. topology} is the topology in $\overline M$  such
that $C\subset \overline{M}$ is defined as closed if and only if
for any sequence $\sigma$ in $C$, necessarily $L(\sigma)\subset
C$. The limit operator and, then, the chr. topology is {\em of
first order} if $\sigma \rightarrow (P,F)$ implies $(P,F)\in
L(\sigma)$  (see Appendix \ref{sA1} for subtleties on this
definition). \end{definition}

The topology of $M$ is generated by the sets type $I^{\pm}(p),
p\in M$ (Alexandrov topology) as $M$ is strongly causal. In
principle, a first requirement for any admissible topology on
$\overline{M}$ seems to maintain  all the sets $I^\pm((P,F))$ as
open. This property is characterized in the following lemma.
\begin{lemma}\label{lea} Let $\overline{M}$ be an admissible
 completion as a point set for a strongly causal spacetime $M$.
If a topology on $\overline{M}$ satisfies that $I^{\pm}((P,F))$
are open for all $(P,F)\in \overline{M}$ then the following
property holds: if $(P_{n},F_{n})\rightarrow (P,F)$ then $P\subset
{\rm LI}(P_{n})$, $F\subset {\rm LI}(F_{n})$. Moreover, the
converse is true if the topology is  sequential (see Defn.
\ref{d6.1} and Remark \ref{r6.1}). In this case,
$I^\pm(p)(=I^{\pm}(p,M))$ is also open in $\overline{M}$ for all
$p\in M$.
\end{lemma}
{\it Proof.} ($\Rightarrow$). If $p\in P$ then $(P,F)\in
I^{+}(p,\overline{M})$, which is open. As
$\{(P_{n},F_{n})\}\rightarrow (P,F)$, necessarily
$(P_{n},F_{n})\in I^{+}(p, \overline{M})$
 for all $n$ big enough. Whence, $p\in P_{n}$
 for big $n$,
 and thus, $p\in {LI}(P_{n})$. In conclusion, $P\subset
{\rm LI}(P_{n})$. The inclusion $F\subset {\rm LI}(F_{n})$ is
analogous.

($\Leftarrow$).  Let us prove that $I^{+}((P_{0},F_{0}))^{c}$ is
sequentially closed for any $(P_{0},F_{0})\in\overline{M}$. Let
$(P,F)$ be in the limit of a sequence $\{(P_n,F_n)\}\subset
I^+((P_0,F_0))^c$ and so $F_0\cap P_n=\emptyset$  for all $n$. By
assumption, $P\subset {\rm LI}(P_n)$ and, therefore, $F_0\cap
P=\emptyset$. Thus, $(P,F)\in I^+((P_0,F_0))^c$. (This last
argument applied to points in $M$ also proves the last assertion.)
\cvd

\bdefi\label{j} The coarsest topology in $\overline M$ such that
$I^\pm((P,F))$ is open for all $(P,F)\in \overline M$ will be
called the {\em coarsely extended Alexandrov topology
(CEAT)}.\edefi

\begin{remark}\label{re1} (A) {\rm CEAT is characterized by (see Remark \ref{r6.2}(2)):
\be \label{cAt} (P_{n},F_{n})\rightarrow (P,F)
\quad \Leftrightarrow \quad P\subset {\rm LI}(P_{n}) \;\;
\hbox{and} \;\; F\subset {\rm LI}(F_{n}). \ee As pointed out by
Geroch et al. \cite{GKP} such a topology is not enough for the
completion, see Fig. \ref{fig1} (A) or (B). Thus, an admissible
topology must also contain other open subsets, but they must be
``as few as possible'' --otherwise, a topology such as the
discrete one (which regards all the subsets as open) could be
admitted, but this is not useful at all. More precisely,
properties as the convergence of sequences in $M$ or the density
of $M$ in $\overline M$, must be retained; so, not too many
subsets must be admitted as open.}

(B) {\rm  The sets $I^\pm((P,F))$ are always open for the chr.
topology. In fact, the chr. topology is sequential (see
Proposition \ref{p6.1}) and  Lemma \ref{lea} is applicable (recall
Remark \ref{r6.2}(C)).}
\end{remark}
Indeed, the following results will show that, for the points
$(P,F)$ with $P\neq \emptyset \neq F$, the CEAT will be enough.
Recall previously
the following general property (one only needs the first
countability of CEAT to prove it, see Remark \ref{r6.2}(1)):

\begin{figure}
\centering
\ifpdf
  \setlength{\unitlength}{1bp}%
  \begin{picture}(486.09, 185.20)(40,0)
  \put(0,0){\includegraphics{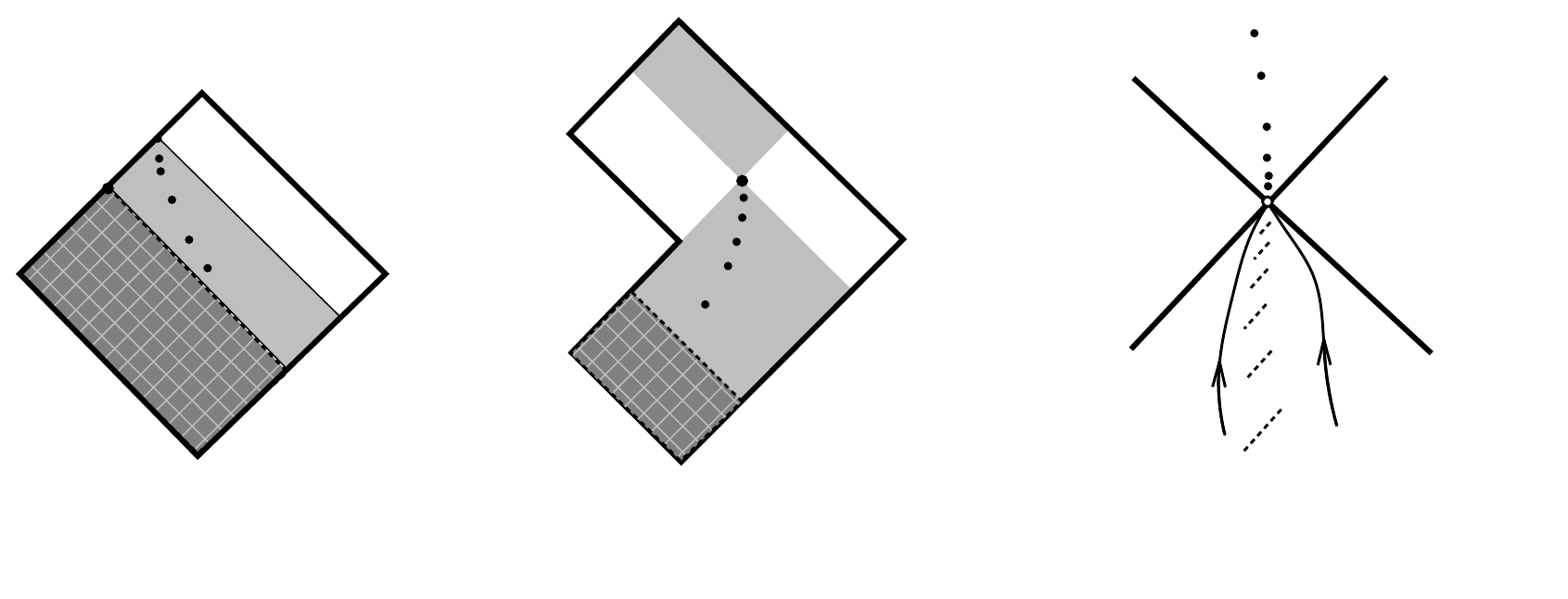}}
  \put(53.87,8.96){\fontsize{14.23}{17.07}\selectfont A}
  \put(10.98,129.68){\fontsize{5.69}{6.83}\selectfont $(P,\emptyset)$}
  \put(26.90,145.16){\fontsize{5.69}{6.83}\selectfont $(P',\emptyset)$}
  \put(58.36,121.68){\fontsize{5.69}{6.83}\selectfont $x_n$}
  \put(43.26,71.31){\fontsize{8.54}{10.24}\selectfont $P$}
  \put(76.42,90.77){\fontsize{8.54}{10.24}\selectfont $P'$}
  \put(200.48,61.36){\fontsize{8.54}{10.24}\selectfont $P$}
  \put(222.15,79.93){\fontsize{8.54}{10.24}\selectfont $P'$}
  \put(218.17,149.36){\fontsize{8.54}{10.24}\selectfont $F'$}
  \put(231.15,100.99){\fontsize{5.69}{6.83}\selectfont $x_n$}
  \put(192.52,132.55){\fontsize{8.54}{10.24}\selectfont $\uparrow P$}
  \put(205.35,8.73){\fontsize{14.23}{17.07}\selectfont B}
  \put(363.07,76.18){\fontsize{5.69}{6.83}\selectfont $\gamma'$}
  \put(416.86,75.64){\fontsize{5.69}{6.83}\selectfont $\gamma$}
  \put(325.93,53.63){\fontsize{8.54}{10.24}\selectfont $P'=I^-[\gamma']$}
  \put(416.58,54.07){\fontsize{8.54}{10.24}\selectfont $P=I^-[\gamma]$}
  \put(396.17,163.06){\fontsize{5.69}{6.83}\selectfont $x_n$}
  \put(395.00,9.45){\fontsize{14.23}{17.07}\selectfont C}
  \put(364.00,158.35){\fontsize{8.54}{10.24}\selectfont $F$}
  \end{picture}%
\else
  \setlength{\unitlength}{1bp}%
  \begin{picture}(486.09, 185.20)(40,0)
  \put(0,0){\includegraphics{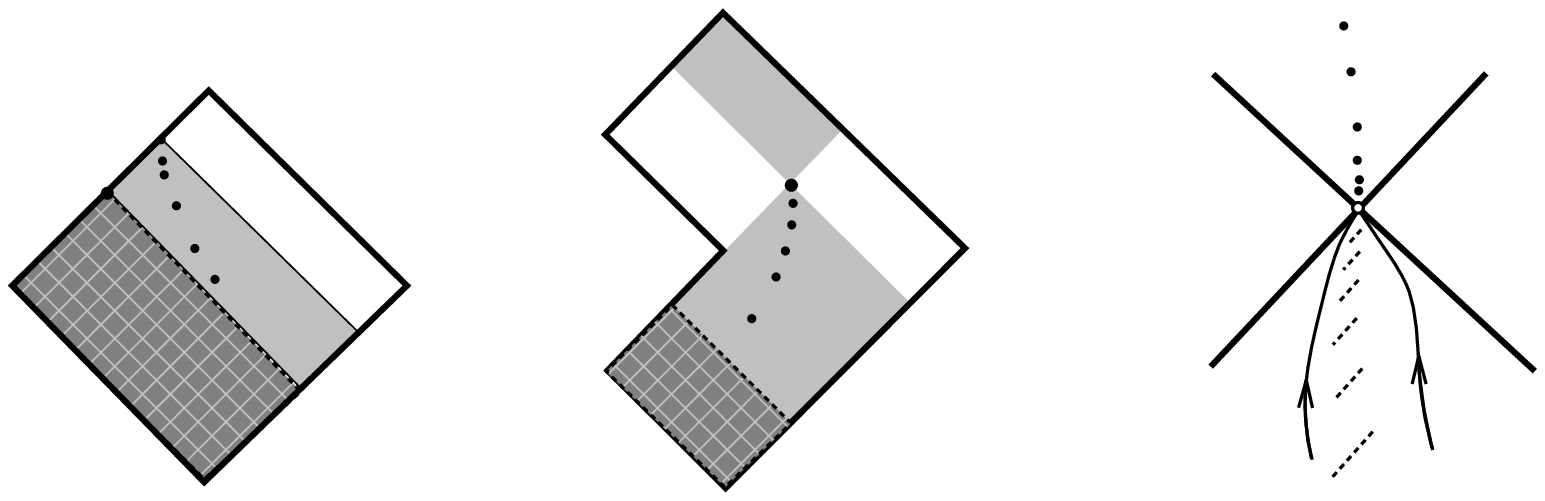}}
  \put(53.87,8.96){\fontsize{14.23}{17.07}\selectfont A}
  \put(10.98,129.68){\fontsize{5.69}{6.83}\selectfont $(P,\emptyset)$}
  \put(26.90,145.16){\fontsize{5.69}{6.83}\selectfont $(P',\emptyset)$}
  \put(58.36,121.68){\fontsize{5.69}{6.83}\selectfont $x_n$}
  \put(43.26,71.31){\fontsize{8.54}{10.24}\selectfont $P$}
  \put(76.42,90.77){\fontsize{8.54}{10.24}\selectfont $P'$}
  \put(200.48,61.36){\fontsize{8.54}{10.24}\selectfont $P$}
  \put(222.15,79.93){\fontsize{8.54}{10.24}\selectfont $P'$}
  \put(218.17,149.36){\fontsize{8.54}{10.24}\selectfont $F'$}
  \put(231.15,100.99){\fontsize{5.69}{6.83}\selectfont $x_n$}
  \put(192.52,132.55){\fontsize{8.54}{10.24}\selectfont $\uparrow P$}
  \put(205.35,8.73){\fontsize{14.23}{17.07}\selectfont B}
  \put(363.07,76.18){\fontsize{5.69}{6.83}\selectfont $\gamma$}
  \put(416.86,75.64){\fontsize{5.69}{6.83}\selectfont $\gamma'$}
  \put(325.93,53.63){\fontsize{8.54}{10.24}\selectfont $P=I^-[\gamma]$}
  \put(416.58,54.07){\fontsize{8.54}{10.24}\selectfont $P'=I^-[\gamma']$}
  \put(396.17,163.06){\fontsize{5.69}{6.83}\selectfont $x_n$}
  \put(395.00,9.45){\fontsize{14.23}{17.07}\selectfont C}
    \put(364.00,158.35){\fontsize{8.54}{10.24}\selectfont $F$}
  \end{picture}%
\fi \caption{ Some open subsets of Lorentz-Minkowski $\LL^2$. In
Figure (A) the sequence $\{x_{n}\}_n$ converges to both
$(P,\emptyset)$, $(P',\emptyset)$ with CEAT, but does not converge
to $(P,\emptyset)$ with any topology satisfying (A1), (A2) in
Defn. \ref{k} (requirement (A3) ensures the convergence to
$(P',\emptyset)$). Analogously, in Figure (B) the sequence
$\{x_{n}\}_n$ converges to both $(P,\emptyset)$, $(P',F')$ with
CEAT, but does not
converge to $(P,\emptyset)$ if (A1), (A2) are fulfilled. In Figure
(C), the sequence $\{x_n\}_n$ (as well as the sequence constantly
equal to $(P,F)$) converges to both $(P,F)$, $(P',\emptyset)$ with
the Marolf-Ross topology, but only converges to $(P,F)$ with the
chr. topology.\label{fig1}}
\end{figure}

\begin{proposition}\label{ll} For any  topology $\tau$ finer than
CEAT, they are equivalent:
\begin{itemize}
\item[(1)] $\tau$ and CEAT admit a common local basis of
neighborhoods of $(P,F)$. \item[(2)] If
$(P_{n},F_{n})\not\rightarrow (P,F)$ with $\tau$ then
$(P_{n},F_{n})\not\rightarrow (P,F)$ with CEAT. 
\end{itemize}
\end{proposition}
(Notice that the converse of (2) always holds  by the hypotheses
on $\tau$.)

\begin{lemma}\label{iu} Let $\{(P_{n},F_{n})\}$ be a sequence of
pairs in $\overline{M}$ and assume that $P\sim_{S}F$ with
$P\neq\emptyset\neq F$. If $P\subset {\rm LI}(P_{n})$, $F\subset
{\rm LI}(F_{n})$ then $P$ and $F$ are maximal into ${\rm
LS}(P_{n})$ and ${\rm LS}(F_{n})$, resp.
\end{lemma}
{\it Proof.} Assume by contradiction the existence of, say, some
$P'$ satisfying $P\varsubsetneq P'\subset {\rm LS}(P_{n})$. Then,
any point of $P'$ will lie in the past of infinitely many pairs
$(P_{n},F_{n})$ and, by transitivity, will lie also in the past of
any point of $F$ (recall that $F\subset {\rm LI}(F_{n})$).
Therefore, $P'$ will be included in $\downarrow F$, in
contradiction with $P\sim_{S} F$.  \cvd

\smallskip

\noindent  Now, we have:
\begin{proposition}\label{leb} Let $(P,F)\in \overline M$ with
$P\neq \emptyset \neq F$. Then
$\{(P_{n},F_{n})\}$ converges to $(P,F)$ with CEAT iff it
converges with the chr. topology, i.e
 $P\in \hat{L}(P_{n})$ and 
$F\in \check{L}(F_{n})$. 
\end{proposition}
{\em Proof.} The implication to the right follows from Lemma
\ref{iu} (see Defn. \ref{overline} and (\ref{cAt})), and to the
left from Remark \ref{re1} (B). \cvd

\begin{remark}\label{reb} (CEAT/chr. topology is the natural one around any $(P,F)$ with $P\neq\emptyset\neq F$).
{\em Previous proposition means that, around any pair $(P,F)$ with
$P\neq\emptyset\neq F$, CEAT has so many open sets (in the sense
of Prop. \ref{ll}) as the chronological topology. As this topology
will not suffer  the pathologies found in Fig. \ref{fig1} (A) (see
below),  it is is not necessary to add new open subsets to CEAT
around such a pair $(P,F)$. So, this topology (or, equivalently,
the chr. one) becomes the coarsest admissible possibility. {\rm
Moreover, as the CEAT is second countable (recall Remark
\ref{r6.2}(1)), each pair $(P,F)$ has a countable neighbourhood
basis}.

In particular, the topology induced on $M$ from the chr. topology
agrees with the manifold topology, as $M$ is strongly causal and,
thus, endowed with Alexandrov topology. }
\end{remark}
Previous remark covers points with $P\neq\emptyset\neq F$ and, now, we state
conditions for topological admissibility which cover all the points.
\begin{definition}\label{k} Let $\overline{M}$ be an admissible completion as a point set for some strongly causal spacetime $M$.
The {\em conditions of admissibility} for a  topology $\tau$  on
$\overline{M}$ are:
\begin{itemize}
\item[(A1)] $\tau$ is finer than CEAT, i.e. $I^{\pm}((P,F))$ is
open for all $(P,F)\in \overline M$. \item[(A2)] The limits for
$\tau$ are compatible with the empty set, i.e.: if
$\{(P_{n},F_{n})\}_{n}\rightarrow (P,\emptyset)$ (resp.
$(\emptyset,F)$) and it happened $P\subset P'\subset LI(P_{n})$
(resp. $F\subset F'\subset LI(F_{n})$) for some $(P',F')\in
\overline{M}$, then $(P',F')=(P,\emptyset)$ (resp.
$(P',F')=(\emptyset,F)$). \item[(A3)] $\tau$ is maximally coarse
among the topologies satisfying the other conditions, i.e. no
topology satisfying (A1) and (A2) (or eventually alternative --or additional-- conditions of admissibility)
is strictly coarser than $\tau$.
\end{itemize}
\end{definition}

\begin{remark}\label{re} (Discussion on the role of the conditions for
admissibility.) {\em Conditions (A1) and  (A3)  have been widely
justified above, so, we focus on condition (A2). This seems the
obvious choice from the viewpoint of set convergence in Figs.
\ref{fig1} (A), (B). However, it can be also understood from the
following more general viewpoint.

First, notice that, under the hypotheses in the formulation of
(A2), any topology which satisfies a minimum compatibility with
the set-point limit operator $L$ must ensure that the constant
sequence $\{(P',F')\}_n$ converges to $(P,\emptyset)$. In fact, by
hypothesis, $(P_{n},F_{n})\rightarrow (P,\emptyset)$. But $P'$ is
closer to $P$ that all $P_{n}$ (and also $P'$ is ``better
adapted'' as a point set limit to $P_{n}$ than $P$) because
$P\subset P'\subset LI(P_{n})$. Moreover, this inclusion also
implies that $F'$ is closer to $\uparrow P$ than all $F_{n}$ (as
$F'\subset \uparrow P'\subset \uparrow P$). In conclusion, the
convergence $\{(P',F')\}_n\rightarrow (P,\emptyset)$ is compelling
from the set-point approach.

But, now, it is also natural to impose $(P',F')=(P,\emptyset)$,
since, otherwise, we would admit an (apparently artificial) pair
of non-$T_1$ separated points. However, as Marolf-Ross topology is
not $T_1$, we will come back to this point in the next section
(Prop. \ref{pA2}, Remark \ref{rterminator}).}\end{remark}
Moreover, under (A1) the hypothesis (A2) has also a very simple
characterization, which can be also regarded as an argument in
favor of it.
\begin{proposition} Let $\tau$ be any topology which satisfies the condition of admissability (A1). Then, condition (A2) is equivalent to:
\begin{itemize}
\item[(*)] If $\{(P_{n},F_{n})\}_{n}\rightarrow (P,\emptyset)$
(resp. $(\emptyset,F)$) then $P\in\hat{L}(P_{n})$ (resp. $F\in
\check{L}(F_{n})$).
\end{itemize}
\end{proposition}
{\em Proof.} To check that (*) implies (A2), assume that
$\{(P_{n},F_{n})\}_{n}\rightarrow (P,\emptyset)$, and so, $P\in
\hat{L}(P_{n})$. If $P\subset P'\subset {\rm LI}(P_n)$ for some
$(P',F')\in\overline{V}$, then $P\subset P'\subset {\rm LS}(P_n)$.
Therefore, the maximality of $P$ into ${\rm LS}(P_n)$ ensures
$P=P'$, and thus, $F'=\emptyset$.

For the converse, assume again $\{(P_{n},F_{n})\}_{n}\rightarrow
(P,\emptyset)$. By (A1), $P\subset {\rm LI}(P_{n})$. Assume that
$P\subset P'\subset {\rm LS}(P_{n})$. Take some chain
$\{p'_{k}\}_{k}$ generating $P'$, and choose some sequence
$\{n_{k}\}_{k}$ so that $p'_{k}\in P_{n_{k}}$ for all $k$. Then,
$p'_{k}\in P_{n_{k'}}$ for all $k'\geq k$, and so, $P\subset
P'\subset {\rm LI}(P_{n_{k}})$. Moreover, notice that
$\{(P_{n_k},F_{n_k})\}_{k}\rightarrow (P,\emptyset)$. Hence, by
(A2), $P=P'$, $F'=\emptyset$. In conclusion, $P$ is maximal into
${\rm LS}(P_n)$, and thus, $P\in \hat{L}(P_n)$. \cvd

The following result shows that the T$_{1}$ character of the
topology is obtained by grant under the hypotheses (A1), (A2).

\bprop \label{pT1} Any topology $\tau$ on $\overline M$ which
satisfies the admissibility conditions (A1) and (A2) is T$_{1}$.
\eprop {\em Proof.}  Assume the existence of $(P,F), (P',F')\in
\overline{M}, (P',F')\neq (P,F)$ such that $\{(P',F')\}_n
\rightarrow (P,F)$. By condition (A1) and  Lemma \ref{lea},
$P\subset P'$, $F\subset F'$. Assume, for example, that the first
 inclusion is strict. In this case, either $F=\emptyset$ or
$P=\emptyset$, since, otherwise, $P'$ would violate the maximality
of $P$ in $\downarrow F$, in contradiction with $P\sim_{S} F$.
Therefore, the conclusion follows by applying condition (A2). \cvd


\smallskip


\noindent Finally, we show how these conditions do single out a
topology.


\begin{theorem}\label{spc} Let $\overline{M}$ be an admissible completion as a point set for a strongly causal
spacetime $M$,  and assume that the chr. topology on
$\overline{M}$ is of first order. Then, it is the unique topology
among the sequential ones which satisfies the conditions of
admissibility (A1), (A2), (A3) in Defn. \ref{k}.

\end{theorem}
{\it Proof.}  As pointed out in Remark \ref{re1} (B), the chr.
topology satisfies condition (A1). In order to prove (A2), assume
that $\{(P_{n},F_{n})\}\rightarrow (P,\emptyset)$.  By the first
order property, $(P,\emptyset)\in L((P_{n},F_{n}))$. Then, $P$ is
a maximal IP in ${\rm LS}(P_{n})$, and so, if $(P',F')\in
\overline{M}$ satisfies $P\subset P'\subset {\rm LI}(P_{n})$,
necessarily $P=P'$ (and therefore $F'=\emptyset$).

Let $\tau$ be any  sequential topology satisfying (A1) and (A2).
Condition (A3), plus the uniqueness, will hold if any sequentially
closed set $C$ for the chr. topology is also  sequentially closed
for $\tau$. Thus, it is enough to check that, whenever
$\{(P_{n},F_{n})\}_{n}\rightarrow (P,F)$ with $\tau$, the limit
also holds with the chr. topology. In the case $P\neq\emptyset\neq
F$, the result follows from Prop. \ref{leb}. Otherwise, assume,
for example, $P\neq\emptyset$, $F=\emptyset$. From Lemma
\ref{lea}, $P\subset {\rm LI}(P_{n})$. By contradiction, assume
$P\varsubsetneq P'\subset {\rm LS}(P_{n})$. There exists a
subsequence $\{(P_{n_{k}},F_{n_{k}})\}$ such that $P\varsubsetneq
P'\subset {\rm LI}(P_{n_{k}})$ (for every $k$ take $n_{k}$ such
that $p'_{k}\in P_{n_{k}}$, being $\{p'_{k}\}$ a future chain
generating $P'$). From the completeness condition in Defn.
\ref{def}, there exists some $F'$ such that $(P',F')\in
\overline{M}$. As the subsequence also converges to
$(P,\emptyset)$, the hypothesis (A2) yields the contradiction
$P=P'$. \cvd

\smallskip

\noindent Summing up:

\begin{quote}
Three conditions on admissibility 
for a topology have been stated:

Condition (A1) is just a minimum natural requirement on
compatibility with the chronology (even though it disregards some
topologies in the literature, see Remark \ref{3.27} below).

Condition (A2)  becomes natural for set point convergence (Remark
\ref{re}) and ensures the $T_1$ character of the topology.
However, it is not fulfilled by Marolf-Ross topology (this is
discussed further in Subsection \ref{sMR}).

Condition (A3) is a minimality requirement which, on one hand, is
necessary to ensure that not too many subsets are regarded as open
(Remark \ref{re1} (A)) and, on the other, can be claimed to ensure
uniqueness of the topology.

 In fact, when the chr. topology is of first order (recall Remark
\ref{r6.15}(2)), it becomes the unique admissible sequential
topology (Th. \ref{spc}), and so, it is canonically selected.
\end{quote}

\subsubsection{Summary on possible completions and our choice}

Taking into account the stated conditions on admissibility plus the uniqueness of the
elements singled out by these conditions, the following definitions are distinguished.
\begin{definition}\label{deff}
Let $M$ be a strongly causal spacetime. Then:

(A) An {\em admissible completion} $\overline{M}$ of $M$ is any
admissible completion as a point set (according to Defn.
\ref{def}) endowed with the extended chronological relation
$\overline{\ll}$ (Defn. \ref{def1'}) and  an admissible topology
(Defn. \ref{k}).

(B) The {\em Marolf-Ross completion} (introduced by these authors
in \cite{MR}) is the set $\overline{M}$ of all the S-related pairs
$(P,F)$ endowed with the extended chronological relation
$\overline{\ll}$ and the Marolf-Ross topology (Defn. \ref{de}).

(C) A {\em chronological completion} (introduced by Flores in
\cite{F}) is any completion $\overline{M}$ which is minimal as a
point set (in the sense of \cite[Defn. 7.3, 7.1]{F}), endowed with
the extended chronological relation $\overline{\ll}$ and the chr.
topology.

(D) The {\em c-completion} of $M$ (singled out by S\'anchez in
\cite{S}) is the completion $\overline{M}$ composed by all the
S-related pairs $(P,F)$, endowed with the extended chronological
relation $\overline{\ll}$ and the chr. topology.  A {\em standard
completion} is any generalization of the c-completion obtained by
using an admissible completion as a point set, instead of all the
S-related pairs $(P,F)$.

\end{definition}
 Admissible completions (A) constitute the general
framework.  They may vary, as a point set, between the minimal
option (C) and the maximum and canonically determined one in (B)
and (D) (notion of c-completion). Its topology may differ of the
chr. topology only if this topology is not of the first order or
if a non-sequential admissible topology were chosen
---such possibilities are open for future researchers on this
topic.

Marolf-Ross completion (B) is discussed below. It satisfies many
of the nice properties of these boundaries,
 even though  its failure in the condition of admissibility (A2)
suggests, in particular, that this completion is a non specially
privileged one among other conceivable possibilities.

The notion of c-completion in (D) avoid the lack of uniqueness of
the minimal completions for a single spacetime in (C). However,
the standard completions defined in (D) include both, the
c-completion and all the chronological completions in (C), as well
as any intermediate possibility (whenever it is admissible as a
point set). Even though the c-boundary will be our choice to be
used in Section \ref{s4}, we consider the general framework of
standard completions along the remainder of Section \ref{s3}. This
 stresses that most properties remain valid for all of them. Moreover, this viewpoint may be also useful for future
developments, in order to consider the inclusion of further
boundary points (see Section \ref{scausalcompletion}).

\subsection{MR topology vs chr. topology}\label{sMR}

Marolf and Ross \cite{MR} introduced two reasonable topologies for
their completion. One of them does not fulfill the condition of
admissibility (A1) (see \cite[Sect. 5.4, footnote 9]{S}) and will
not be taken into account here. The other one fulfills (A1), but
does not fulfill (A2) (for example, it may be non-$T_1$).
Of course this topology agrees the chr. one in most cases. Nevertheless, we will stress
here the diferences.
MR topology can be rewitten as follows \cite{F}:
\begin{definition}\label{de} Let $\overline{M}$ be an admissible completion as a point set for some strongly causal spacetime $M$. The {\em MR topology} on $\overline{M}$ is generated by using $L^{+}(S)$, $L^{-}(S)$, for any subset $S\subset \overline{M}$, as closed subsets of $\overline{M}$,
where
\[
\begin{array}{c}
L^{+}(S)=Cl_{FB}[S\cup L^{+}_{IF}(S)] \\ L^{-}(S)=Cl_{PB}[S\cup
L^{-}_{IP}(S)]
\end{array}
\]
with 
\[
\begin{array}{c}
Cl_{FB}(S)=S\cup\{(P,\emptyset)\in\overline{M}:
P=I^{-}({\rm LI}(P_n))\;\;\hbox{for some sequence}\;\; (P_{n},F_{n})\in S\} \\
Cl_{PB}(S)=S\cup\{(\emptyset,F)\in\overline{M}: F=I^{+}({\rm
LI}(F_n))\;\;\hbox{for some sequence}\;\; (P_{n},F_{n})\in S\}
\end{array}
\]
and
\[
\begin{array}{c}
L^{+}_{IF}(S)=\{(P,F)\in\overline{M}: F\neq\emptyset,
F\subset\cup_{(P',F')\in S}F'\} \\
L^{-}_{IP}(S)=\{(P,F)\in\overline{M}: P\neq\emptyset,
P\subset\cup_{(P',F')\in S}P'\}.
\end{array}
\]
\end{definition}
  Some caution must be taken into account,
for example, the equality
 $Cl_{FB}((Cl_{FB}(S)))=Cl_{FB}(S)$ must be checked.   Assuming that it holds
 (eventually,
under favorable hypotheses in the spirit of Proposition
\ref{p6.2}), one can find  properties such as:

\begin{itemize}
\item[(a)] The MR topology on any admissible completion
$\overline{M}$ satisfies (A1) (i.e., it includes CEAT). \item[(b)]
Let $(P,F)\in \overline{M}$ with $P\neq\emptyset\neq F$. A
sequence $\{(P_{n},F_{n})\}_{n}\subset \overline M$ converges to $
(P,F)$ with the MR topology iff it converges with the chr.
topology (and thus iff it converges with CEAT).
\end{itemize}

\smallskip

\noindent Therefore, the differences between both topologies
appear when condition $P\neq \emptyset\neq F$ does not hold. This
is related to the possible violation of (A2). In order to analyze
it, notice the following result, which is straightforward from
Prop. \ref{pT1}.

\begin{proposition}\label{pA2}
Let $\overline{M}$ be an admissible completion as a point set of a
strongly causal spacetime $M$ endowed with a topology  $\tau$
which satisfies (A1). Then, (A2) is equivalent to the following
pair:

\bit\item[(A2)'] If $\{(P_{n},F_{n})\}_{n}\rightarrow
(P,\emptyset)$ (resp. $(\emptyset,F)$) and it happened $P\subset
P'\subset LI(P_{n})$ (resp. $F\subset F'\subset LI(F_{n})$) for
some $(P',F')\in \overline{M}$, then the constant sequence
$\{(P',F')\}_n$ also converges to $(P,\emptyset)$ (resp.
$(\emptyset,F)$) \item[(A2)''] $\tau$ is $T_1$.\eit
\end{proposition}

\begin{remark}\label{rterminator} {\rm Our conclusion on MR topology in relation to (A2) is the following:

(i) The arguments in Remark \ref{re} show that condition (A2)' is
compelling for any useful topology.  Nevertheless, the cases when
MR topology  satisfies (A2)' do not seem clear.

(ii)  MR topology may not satisfy condition (A2)'', as Marolf and
Ross argued that  the example in Fig. \ref{fig1} (C)
 cannot be
 $T_{1}$. 
 Recall that this property not only seems harmless and desirable  but
 also
allows to fix univocally a topology. That is, even if (A2)' were
satisfied, the absence of (A2)'' makes  unclear that a topology
could be univocally selected by a requirement of coarseness such
as\footnote{MR topology may be non coarser than the chr. one, in
contraposition to what was stated in \cite[Th. 8.1]{F}. In fact,
in the proof of this result
 the following property was wrongly assumed: if $\{P_{n}\}_{n}$,
$\{P^{n}_{k}\}_{k}$ are sequences of IPs such that
$P\in\hat{L}(P_{n})$ and $P_{n}\in \hat{L}(P^{n}_{k})$ for all
$n$, then there exists some subsequence $\{k_{n}\}_{n}\subset
\{n\}_{n}$ such that $P\in \hat{L}(P^{n}_{k_{n}})$.} (A3) in Theorem \ref{spc}.

\smallskip
\noindent Summing up, one can state  a version of Theorem
\ref{spc} by considering only  conditions (A1), (A2)' and (A3),
which seem to include Marolf-Ross topology at least in favorable
cases. However, even in this case, $T_1$ separability must be
imposed to fix a topology, i.e., the full condition (A2) seems
unavoidable.

}\end{remark}

\subsection{Properties of the standard completions}\label{s3.6}

The properties fulfilled by the completions of a chronological
space $X$ were studied  in \cite{F}. Some of them were proved in a
more general setting of completions (Remark \ref{r2.2} (A)) and
others  for the chronological completions of strongly causal
spacetimes. At any case, they hold for any admissible completion
and, then, for the c-completion. So, we can prove\footnote{ In
\cite[Th. 7.9]{F} the following property was also claimed: {\em if
two points $(P,F), (P',F')\in \overline{M}$ are non-Hausdorff
related then both lie in $\partial M$}. However, this result is
not true in general (it was obtained as a consequence of the also
uncorrect result \cite[Prop. 7.8]{F}).}:

\begin{theorem}\label{rev}
For any   standard   completion $\overline{M}$ of a strongly
causal spacetime $M$:
\begin{itemize}

\item[(i)] Any chain in the completion has a limit. Moreover, any
inextensible timelike curve in $M$ has a limit in $\partial M$.

\item[(ii)] The inclusion $\mathbf{i}: M \rightarrow \overline{M}$
is a topological  embedding, and $\mathbf{i}(M)$ is dense in
$\overline{M}$.

 \item[(iii)] The boundary $\partial M$ is a
closed subset of $\overline{M}$.


\item[(iv)] $I^\pm((P,F))$ is open for any $(P,F)\in
\overline{M}$.

\item[(v)] Each single point is closed, i.e., the topology is
$T_1$.

\end{itemize}
\end{theorem}
{\it Proof.}  Statement (i)  follows directly from the {\em
completeness} and {\em S-relation} conditions satisfied by any
admissible completion as a point set (Defn. \ref{def} (1), (2)).
In particular, notice that any future-directed chain $\varsigma$
admits as a limit any $(P,F)\in\overline{M}$ with
$P=I^{-}[\varsigma]$ (Defn. \ref{overline},  Prop. \ref{p6.1}(1)).
This also proves the second assertion in statement (ii). For the
first one, recall   Remark \ref{reb}.  This remark also proves
that $i(M)$ is an open subset and, so, statement (iii) (see also
\cite[Th. 7.6]{F}, which is valid for any admissible completion,
non necessarily minimal). Assertion (iv) has been imposed for any
admissible completion, and have been proved for the chr. one in
Remark \ref{re1} (B). For (v), just apply Defn \ref{overline}.
\cvd

\begin{remark}\label{3.27} {\em  It is not clear the cases when (v) is
 fulfilled by MR topology,  and (iv) is not fulfilled
neither by Harris' topology (who necessarily uses a different
notion of chronological relation) nor by the alternative MR
topology (as pointed out  in \cite[Sect. 5.4, Fig. 10]{S}).}
\end{remark}

\subsection{Globally hyperbolic case and conformally flat
ends}\label{s3.7}

Put  $I(p,q):=I^{+}(p)\cap I^{-}(q)$, $J(p,q):=J^{+}(p)\cap
J^{-}(q)$. Next, we focus our attention on the case when $M$ is
globally hyperbolic, i.e. $M$ is causal with compact $J(p,q)$
\cite{BeSa2}. First, following the proof of \cite[Th. 9.1]{F} we
have:
\begin{theorem}\label{h} Let $\partial M$ be any
admissible boundary of a strongly causal spacetime $M$. The following properties are equivalent:
\begin{itemize}
\item[(i)] $M$ is globally hyperbolic. \item[(ii)] if $(P,F)\in
\partial M$ then either $P=\emptyset$ or $F=\emptyset$.
\item[(iii)] if $(P,F)\in
\partial M$ then either $\uparrow P=\emptyset$ or $\downarrow
F=\emptyset$.
\end{itemize}
In particular,  the c-completion is the unique standard
completion.
\end{theorem}
{\it Proof.} $(i)\Rightarrow (ii)$ Assume the existence of
$(P,F)\in\partial M$ with $P\neq\emptyset\neq F$. Choose points
$p\in P$, $q\in F$, and a chain $\varsigma\subset M$ generating
$P$, and thus, with endpoint $(P,F)$. Then, $\varsigma$ is
eventually contained in $I(p,q)$. Moreover, no subsequence of
$\varsigma$ converges in $M$, since, otherwise, we would
contradict the terminal character of $P$. Whence, $J(p,q)\subset
M$ is not compact, in contradiction with $(i)$.

$(ii)\Rightarrow (iii)$ Assume the existence of
$(P,\emptyset)\in\partial M$ with $\uparrow P\neq \emptyset$. Let
$\emptyset\neq F$ be a maximal IF into $\uparrow P$. Notice that
$F$ must be terminal, since, otherwise, $F=I^{+}(p)$ and any chain
$\varsigma$ generating $P$ would satisfy $\varsigma\rightarrow p$
with the topology of the manifold (in contradiction with the
terminal character of $P$). Let $\emptyset\neq P\subset P'$ be
some maximal (thus, necessarily terminal) IP into $\downarrow F$.
Then, $P'\sim_{S} F$,
in
contradiction with $(ii)$.

$(iii)\Rightarrow (i)$ By \cite[p. 409, Lemma 14]{O}, it suffices
to show that $J(p,q)$ is included in some compact set $K\subset
M$. By contradiction, assume that $K:=\overline{I(p',q')}$ (which
includes $J(p,q)$ if $p'\ll p$, $q\ll q'$) is not compact for some
points $p',q'\in M$. Then, there exists some $\sigma\subset
I(p',q')$ with no subsequences converging in $M$. From
\cite[Theorem 5.11]{FH} applied to $\hat{M}$, there exists some
subsequence $\sigma'\subset\sigma$ and some TIP
$P\in\hat{L}(\sigma')$ such that $\emptyset\neq I^{-}(p')\subset
P$. Moreover, $P$ constitutes some pair $(P,F)$ (recall the
completeness of $\overline{M}$, Defn. \ref{def} (1)) and $\uparrow
P\supset I^{+}(q')\neq \emptyset$, in contradiction with (iii).
\cvd

\smallskip

Next, our aim is to study the concept of asymptotic conformal flatness.
The following result  will be useful in order to discuss this notion,
but it may have interest in its own right.
\begin{proposition}\label{prr1} Any point of a standard   completion $\overline{M}$ of a globally
hyperbolic spacetime $M$, admits a neighborhood which is
sequentially compact.
\end{proposition}
{\it Proof.} We have to prove that any $(P,F)\in\overline{M}$
admits a neighborhood $V$ such that any sequence in $V$ admits a
converging subsequence. Since $M$ is locally compact and it is
open in $\overline{M}$, we can restrict our attention to some
$(P_{0},\emptyset)\in\partial M$. For some $p_{0}\in P_{0}$ take
$V=\{(P,F)\in\overline{M}: I^{-}(p_{0})\subset P\}$. Notice that
$V$ is a neighborhood of $(P_{0},\emptyset)$ because it contains
$I^{+}(p_{0},\overline{M})$. So, it suffices to show that $V$ is
sequentially compact. Let $\sigma$ be any sequence inside $V$.
From \cite[Theor. 5.11]{FH},  there exists some subsequence
$\sigma^{\infty}$ with $\hat{L}(\sigma^{\infty})\neq \emptyset$
and choose $P'\in\hat{L}(\sigma^{\infty})$, so that
$I^{-}(p_{0})\subset P'$. Now, if $P'$ is a TIP then
$(P',\emptyset)\in\partial M$ (recall Theorem \ref{h}), and so,
$(P',\emptyset)\in L(\sigma^{\infty})\cap V$. Otherwise,
$P'=I^{-}(p)$, for some $p\in M$, we can assume without
restriction that $\sigma^{\infty}\subset M$ (recall that $\partial
M$ is closed), and so, from \cite[Th. 2.3]{H2} or \cite[Prop.
2.5]{F}, we have $\sigma^{\infty}\rightarrow p\in V$. \cvd

\smallskip

\noindent The following example shows that the hypothesis of
global hyperbolicity in previous result is necessary.

\begin{figure}
\centering \ifpdf
  \setlength{\unitlength}{1bp}%
  \begin{picture}(172.49, 167.73)(0,0)
  \put(0,0){\includegraphics{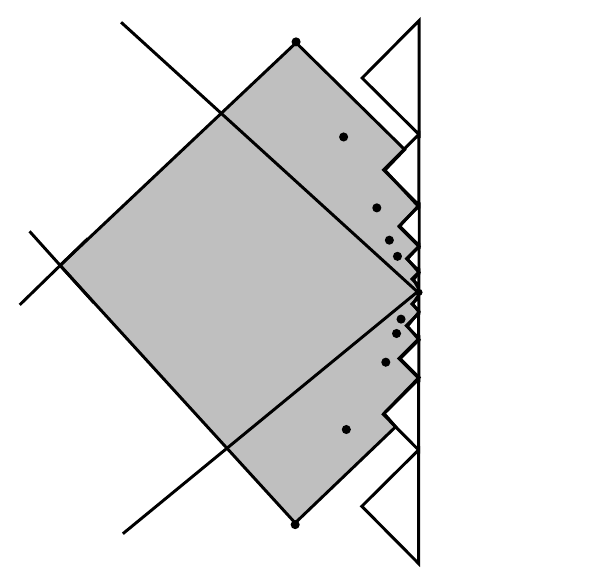}}
  \put(121.67,84.10){\fontsize{4.70}{5.64}\selectfont $(0,0)\equiv (P,F)$}
  \put(89.14,157.74){\fontsize{4.70}{5.64}\selectfont $q_n$}
  \put(83.97,13.47){\fontsize{4.70}{5.64}\selectfont $p_n$}
  \put(54.51,151.03){\fontsize{7.05}{8.47}\selectfont $F$}
  \put(52.19,14.78){\fontsize{7.05}{8.47}\selectfont $P$}
  \put(33.84,94.15){\fontsize{7.05}{8.47}\selectfont $U_n$}
  \put(124.80,140.15){\fontsize{4.70}{5.64}\selectfont $T_n^+$}
  \put(125.32,19.21){\fontsize{4.70}{5.64}\selectfont $T_n^-$}
  \end{picture}%
\else
  \setlength{\unitlength}{1bp}%
  \begin{picture}(172.49, 167.73)(0,0)
  \put(0,0){\includegraphics{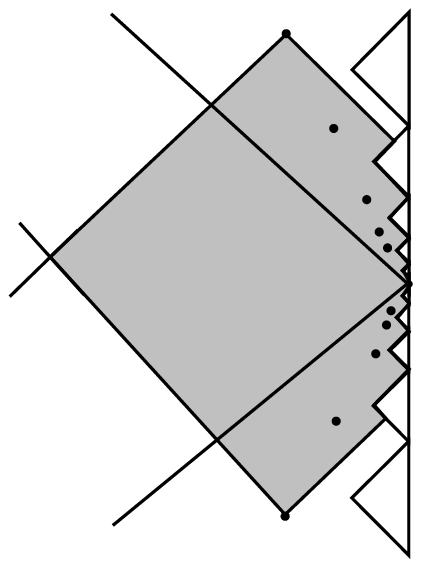}}
  \put(121.67,84.10){\fontsize{4.70}{5.64}\selectfont $(0,0)\equiv (P,F)$}
  \put(89.14,157.74){\fontsize{4.70}{5.64}\selectfont $q_n$}
  \put(83.97,13.47){\fontsize{4.70}{5.64}\selectfont $p_n$}
  \put(54.51,151.03){\fontsize{7.05}{8.47}\selectfont $F$}
  \put(52.19,14.78){\fontsize{7.05}{8.47}\selectfont $P$}
  \put(33.84,94.15){\fontsize{7.05}{8.47}\selectfont $U_n$}
  \put(124.80,140.15){\fontsize{4.70}{5.64}\selectfont $T_n^+$}
  \put(125.32,19.21){\fontsize{4.70}{5.64}\selectfont $T_n^-$}
  \end{picture}%
\fi \caption{The origin corresponds with a pair $(P,F)\in
\partial M$ which does not have a sequentially compact neighborhood (see Example \ref{esequent}). Any open neighborhood
$U_{n}=I^{+}(p_{n},\overline{M})\cap
I^{-}(q_{n},\overline{M})\subset\overline{M}$ of $(P,F)$ contains
a sequence with no limit in $\overline{M}$.\label{fig2}}
\end{figure}

\begin{example}\label{esequent} {\em The c-completion $\overline{M}$ of the spacetime $M$ in Fig.
\ref{fig2} does not satisfy the property in Prop. \ref{prr1}. In
fact, the spacetime $M$ is defined by removing from $\LL^{2}$ in
coordinates $(x,t)$ the region $x\geq 0$ and the interior regions
of the isosceles triangles $T^{+}_{n}$, $T^{-}_{n}$ whose bases
are determined, resp., by the pair of vertexes $(0,1/n)$,
$(0,1/(n+1))$ and $(0,-1/n)$, $(0,-1/(n+1))$ for all $n$, and the
angle between their sides and the $t$ axis are $\pm 45^0$. Notice
that no vertex on the $t$-axis correspond with a point in the
c-boundary, but $(0,0)$ does correspond, concretely, with $(P,F)$
where $P=\{(x,t)\in M: t<x<0\}$ and $F=\{(x,t)\in M: -t<x<0\}$.
Let $\{p_{n}=(x_n,-1/n)$\}, $\{q_{n}=(x'_n,1/n)\}$ be chains in
$M$ generating $P$, $F$, resp. It suffices to observe  (recall
Remark \ref{reb}) that any open neighborhood
$U_{n}=I^{+}(p_{n},\overline{M})\cap
I^{-}(q_{n},\overline{M})\subset\overline{M}$ of $(P,F)$ contains
a sequence $\{x^m_{n}\}_m\subset U_{n}$ composed by points with
$t$ components constantly equal to some $1/n'$, $n'>n$ and
approaching to the $t$-axis; in particular, $\{x^m_{n}\}_m$ has no
limit in $\overline{M}$.}
\end{example}

\smallskip

Next, we are going to apply the causal boundary approach to
formulate the notions of {\em conformally  flat end} and {\em
asymptotically conformally  flat end}.

In Riemannian Geometry, there is a natural way to say that a
(connected) Riemannian manifold $(M,g_R)$ has a {\em flat} or {\em
asymptotically flat end}. Namely, a flat end  appears when, for
some closed ball $\overline{B(p,R)}\subset M$ of sufficiently big
radius $R$, a connected component of $M\setminus
\overline{B(p,R)}$ is isometric to $\R^n$ minus some compact
subset. Here $\R^n$ is endowed with its natural Euclidean metric
$g_E$, but if we allow  a metric equal to $g_E$ plus terms which
decrease with some power of $1/r$, (where $r$ is the
$g_E$-distance to the origin) then an asymptotically flat end
appears. Notice that, if one considers the classical Cauchy
completion $M_C$ of $(M,g_R)$ (defined by using classes of
equivalence of Cauchy sequences), the expression ``some closed
ball $\overline{B(p,R)}\subset M$ of sufficiently big radius $R$''
in previous definition can be substituted by ``some compact subset
of the Cauchy completion\footnote{A subtlety appears here with the
topology. One can consider $M_C$ endowed with the metric topology
associated to the Riemannian distance on $M$ extended to $M_C$.
However, this topology may be non-locally compact and, so, the
Cauchy completion of a bounded region may be non-compact. However,
$M_C$ can be identified  as a point set with a subset of the
Gromov completion (which becomes the natural completion for causal
boundaries, see \cite{FHSst}). For Gromov's topology the previous
undesirable property cannot hold and, so, it can be used
rigourously along the present discussion.} $M_C$''.

Our aim is to point out that the viewpoint ``a flat/asymptotically
flat end appears when, up to a compact subset of the completion,
the manifold is isomorphic to some model space'', can be
transplantated directly to c-completions of spacetimes, even
though some small technicalities must be noticed.

Notice that the c-completion of $\LL^{4}$
 is equal to the usual conformal one obtained from
the natural Penrose embedding in Einstein static universe
$\LL^1\times \SSS^3$, but the so-called {\em spacelike infinity}
$i^0\in \overline{\LL^{4}}$
 (see more details in Remark \ref{rbclm} (1) below).
Assume that, for some compact subset $K$ of the c-completion
$\overline{M}$ of a spacetime, a connected component $E$ of
$M\setminus K$ is conformal to
$\LL^{4}\setminus {\cal K}$, where ${\cal K}\subset
\overline{\LL^{4}}$ is also compact (say, ${\cal K}$ eventually
contains the future and past timelike infinities $i^\pm$). To
regard
this connected component $E$ 
as a {\em conformally flat end} of the spacetime is then natural.
If  $\LL^4$ is replaced by an ambient spacetime which behaves as
an asymptotically flat one (say, which satisfies the axioms in
\cite[Chapter 11]{Wald}) then $E$ can be regarded as an {\em
asymptotically conformally  flat end}. From a formal viewpoint, it
is also convenient to give a more abstract notion of end, because
of the dependence of the previous one with the ``concrete
details'' of the removed $K$. Summing up, let $\overline{M}$ be
the c-completion of a strongly causal spacetime $M$, and recall
the following definitions (for simplicity, they are stated just
for conformal flatness):

\begin{itemize}
\item[(1)] Let $K\subset \overline{M}$ compact. A connected
component $E$ of $M\setminus K$ is a {\em  conformally flat end}
if it is conformal to $\LL^{4} \setminus {\cal K}$, where ${\cal
K}$ is some compact subset of the c-completion\footnote{Notice
that this map is then extensible to a unique continuous map from
$\overline{M}\setminus K$ to $\overline{\LL^{4}} \setminus {\cal
K}$, which can be also called {\em conformal} as it preserves the
chronological relations.} $\overline{\LL^{4}}$ which includes
$i^\pm$.

\item[(2)] Two conformally flat ends $E,E'$ are {\em equivalent},
$E\sim E'$,
if there exists some compact
$\tilde{K}\subset \overline{M}$ such that\footnote{Alternatively,
if $(E\cap E')\setminus\tilde{K}$ is an end. Recall that even in
this version some subset $\tilde{K}$ must be removed because, even
though $E$ and $E'$ are connected, it might happen that $E\cap E'$
is not.} $E\setminus \tilde{K}=E'\setminus\tilde{K}$. Then,
each class of the relation of equivalence
 $\sim$ is called an {\em abstract conformally flat end}.
\end{itemize}
Notice that in non-locally compact cases as  in Fig. \ref{fig2},
there are no asymptotically flat ends according to previous
definition. Nevertheless, one might like that the vertexes of each
triangle $T^\pm_n$ in the $t$-axis corresponded to an asymptotic
end. One can enlarge our definition, say, admitting the existence
of ends when, in our previous definition,
 $J^+(K)\cup J^-(K)$ is removed, instead of $K$. In principle, we will
 not worry about this possibility, as it seems important only
 when the c-boundary is
not locally sequentially compact, and this seems of limited
interest (recall Prop. \ref{prr1}).

Analogous definitions and considerations can be stated for the
notion of asymptotically conformally  flat spacetimes --which,
obviously, include all the classical  asymptotically flat ones,
such as ${\mathbb L}^{4}$, Schwarzschild,  the stationary part of
Kerr (all with a unique end), Kruskal (two ends) or
Reissner-Nordstrom (infinitely many ends).


\subsection{Endpoints for causal curves}
\label{scausalcompletion}

Up to now, we have dealt only with chronological relations,
timelike curves etc. However, one can wonder what would happen if
we also consider causal elements such as extended causal
relations, endpoints of causal curves, etc.
 To this aim, recall first:


\bprop\label{ppp} Let $\gamma:[a,b)\rightarrow M$ be a
future-directed inextensible causal curve. Then:

(i) $I^-[\gamma]$ is a TIP.

(ii) The inclusion $\uparrow \gamma \subset \uparrow I^-[\gamma]$
always holds, and the equality holds if $\gamma\subset
I^-[\gamma]$.\eprop {\it Proof.}  (i) To check that $I^-[\gamma]$
is an IP, notice that from its definition it is enough to prove
that, for any two points $p_{1}, p_{2}\in I^{-}[\gamma]$ there
exists another point $p_{3}\in I^-[\gamma]\cap I^{+}(p_{1})\cap
I^{+}(p_{2})$. As $p_i\ll \gamma(t_i)$ $i=1,2$ for some $t_i$, it
is enough $p_3=\gamma(t_3)$ for any $t_3\in (t_1,b)\cap (t_2,b)$.
The terminal character of $I^{-}[\gamma]$ follows from the
inextensibility of $\gamma$ by the same argument as in the
timelike case.

(ii) The first assertion follows directly from the definitions and
the second one from the fact that $\gamma\subset I^{-}[\gamma]$
implies $\uparrow \gamma\supset \uparrow I^{-}[\gamma]$. \cvd

\smallskip

\noindent The part (i) means that no more generality is obtained
if chronological past and futures of causal curves are considered
instead of TIP's and TIF's  --and, then, for all the construction
of the boundary, which depends only on these sets. Nevertheless,
the part (ii) suggests that inextensible causal curves may have no
endpoint in the c-boundary.

\begin{definition} A  standard completion  is {\em properly causal} if any inextensible causal curve has an endpoint.
\end{definition}

\begin{figure}
\centering
\ifpdf
  \setlength{\unitlength}{1bp}%
  \begin{picture}(221.39, 147.64)(0,0)
  \put(0,0){\includegraphics{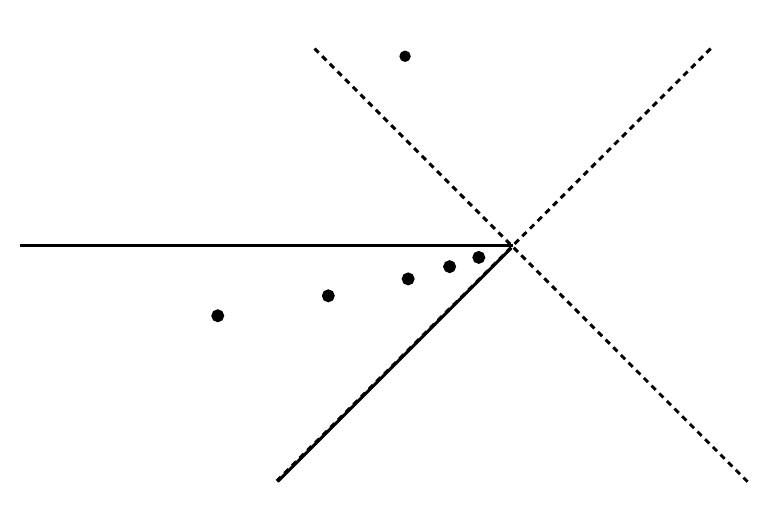}}
  \put(133.91,16.38){\fontsize{14.23}{17.07}\selectfont $P$}
  \put(135.68,121.18){\fontsize{14.23}{17.07}\selectfont $F$}
  \put(59.62,47.77){\fontsize{8.54}{10.24}\selectfont $x_n$}
  \put(120.11,135.33){\fontsize{8.54}{10.24}\selectfont $q$}
  \put(86.17,7.53){\fontsize{8.54}{10.24}\selectfont $\gamma$}
  \end{picture}%
\else
  \setlength{\unitlength}{1bp}%
  \begin{picture}(221.39, 147.64)(0,0)
  \put(0,0){\includegraphics{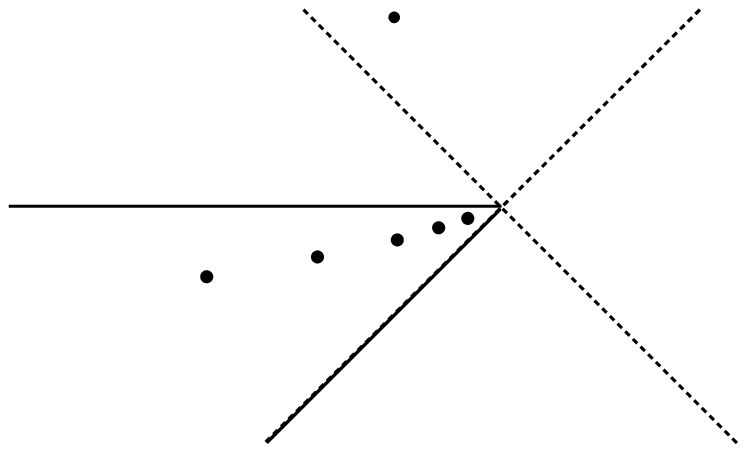}}
  \put(133.91,16.38){\fontsize{14.23}{17.07}\selectfont $P$}
  \put(135.68,121.18){\fontsize{14.23}{17.07}\selectfont $F$}
  \put(59.62,47.77){\fontsize{8.54}{10.24}\selectfont $x_n$}
  \put(120.11,135.33){\fontsize{8.54}{10.24}\selectfont $q$}
  \put(86.17,7.53){\fontsize{8.54}{10.24}\selectfont $\gamma$}
  \end{picture}%
\fi \caption{Non-properly causal completion for any admissible
boundary.\label{fig2'}}
\end{figure}
The following example shows a non-properly causal c-completion.
\bexam {\em Let $M=\LL^2\setminus \{(x,0): x\leq 0\}$ and put
$\gamma(s)=(s,s)$ for $s\in [-1,0)$ (see Fig. \ref{fig2'}). Then
$\uparrow \gamma \varsubsetneq \uparrow I^-[\gamma]$ and $\gamma$
has no endpoint in the c-completion $\overline{M}$. This example
(emphasized in \cite{S}) stresses our main choices for the
c-boundary: (i) $I^{-}(q)\cap F\neq\emptyset$, $I^{-}(q)\cap
I^{+}(x_{n})=\emptyset$, so $(P,F)\overline{\ll} q$, but
$x_{n}\not\ll q$; (ii) as we imposed that $I^{-}(q)$ is open, this
implies $x_{n}\not\rightarrow (P,F)$ (in fact,
$F\not\in\check{L}(I^{+}(x_{n}))$; (iii) analogously, the causal
curve $\gamma$ will not have an endpoint. So, in order to ensure
the convergence of causal curves, one should include new boundary
points (namely, $(P,\emptyset)$, $(\emptyset,F)$, but recall that
their components are not S-related). Properties (i), (ii), (iii)
also happen in any admissible topology as well as in the MR
topology (Defn. \ref{de}).}\eexam If one is interested in defining
a properly causal boundary for any spacetime,
the framework of the already  constructed  admissible boundaries
is useful. As suggested in previous example, the c-boundary would
be enlarged 
in order to obtain a properly causal one. Nevertheless, in order
to get this larger  boundary as a full general construction, one
must restart all the process above and, in particular, the causal
relation should be extended to the completion. Some difficulties
of this extension were pointed out by Marolf and Ross in
\cite[Sect. 3.2]{MR}, and some of the choices may be
non-indisputable at this moment.

Therefore, instead of considering such a general construction, we
give some simple and quite general criteria which ensures that the
c-boundary is properly causal. Presumably, they hold in most cases
of interest.

\btheo\label{ge} A standard completion $\overline{M}$ is properly
causal if one of the following properties hold:

(i) $\overline{M}$ is compact.

(ii) $M$ is globally hyperbolic.

(iii) For any future-directed (resp. past-directed) inextensible
lightlike pregeodesic $\rho: [a,b)\rightarrow M$ with no cut
points\footnote{Recall that these points, as well as conjugate
points, are conformally invariant for lightlike geodesics.}  it is
$\uparrow \rho \supset \uparrow I^-[\rho]$ (resp. $\downarrow \rho
\supset \downarrow I^+[\rho]$). \etheo {\em Proof}. For case (i),
take some inextensible future-directed causal curve $\gamma$. By
the compactness of $\overline{M}$, there exists some sequence
$\varsigma_{0}=\{\gamma(t_{n})\}_{n}$ exhausting $\gamma$ such
that $\varsigma_{0}\rightarrow (P,F)$ for some $(P,F)\in
\overline{M}$. As $L(\varsigma_0) \cup \varsigma_0$ is closed (any
$(P',F')\in L(\varsigma_0)$ satisfies $P'=I^-(\gamma)$; then, $F'$
is  maximal  in $\uparrow I^-(\gamma)$ and $L(L(\varsigma_0) \cup
\varsigma_0)=L(\varsigma_0) \cup \varsigma_0$), necessarily
$(P,F)\in L(\varsigma_0)$, i.e.
\begin{equation}\label{er}
\left\{\begin{array}{l} P\subset LI(I^{-}(\gamma(t_{n}))) \\
P\;\;\hbox{maximal}\;\; IP\;\;\hbox{in}\;\;
LS(I^{-}(\gamma(t_{n}))),\end{array}\right. \quad
\left\{\begin{array}{l} F\subset LI(I^{+}(\gamma(t_{n}))) \\
F\;\;\hbox{maximal}\;\; IF\;\;\hbox{in}\;\;
LS(I^{+}(\gamma(t_{n}))),\end{array}\right.
\end{equation}
(the second lines required only if $P\neq\emptyset$ or
$F\neq\emptyset$).  Moreover, as the points on $\gamma$ are
causally related, conditions (\ref{er}) also hold for any other
sequence exhausting $\gamma$. Hence, $(P,F)$ is an endpoint of
$\gamma$.

For case (ii), just observe that Prop. \ref{ppp} (i) and Theorem
\ref{h} (ii) ensure that any inextensible future-directed causal
curve $\gamma$ has endpoint $(I^{-}[\gamma],\emptyset)\in
\overline{M}$.

Finally, for case (iii) let $\gamma:[a,b)\rightarrow M$ be some
inextensible future-directed causal curve. Assume first that some
restriction  $\rho = \gamma|_{[c,b)}$ lies under condition (iii).
From this condition, $\rho$ has endpoint any $(P,F)\in
\overline{M}$ with $P=I^{-}[\gamma]$, and so has $\gamma$. Now,
assume that previous hypothesis does not hold for any restriction
of $\gamma$. Then, there exists a sequence of
 points $\{t_n\} \rightarrow b$ such that each $\gamma(t_{n+1})$ is the cut point of
 $\gamma|_{[t_n,b)}$. Connecting the even consecutive
points $\gamma(t_{2n})$
 by timelike curves (and eventually smoothing it),
an inextensible timelike curve $\rho$ through all $\gamma(t_{2n})$
is constructed. Then, the curve $\rho$ has some endpoint, and so
has the curve $\gamma$. \cvd

\begin{remark}\label{rge} {\em The criteria in Theorem \ref{ge} can be combined. So,
 if (i) holds up to a globally hyperbolic subset, its c-boundary
becomes also properly causal. In fact, in this case, either the
inextensible causal curve $\gamma$ crosses infinitely many times a
compact subset $K$ (thus, one can apply (i)) or a sequence of
points exhausting $\gamma$ remains in some globally hyperbolic
connected component of $M\setminus K$ (and thus, one can apply
(ii)). In particular, this observation is applicable to spacetimes
with asymptotically flat ends (recall
Section \ref{s3.7}).}
\end{remark}

\subsection{Appendix: $L$-operator and convergence in
chr. topology}\label{sA1}

Notice that the chr. topology, as well as Harris' one \cite{H2},
is defined by starting at a limit operator $L$, which defines the
closed sets (Defn. \ref{overline}). It is easy to check that these
sets satisfy the axioms of a topology, just by taking into account
that, if $\tilde \sigma$ is a subsequence of $\sigma$, then
$L(\sigma) \subset L(\tilde \sigma)$. Nevertheless, the name
``limit operator'' suggests that $L$ should map each sequence
$\sigma$ in the set composed by all the points to which $\sigma$
converges. It is not clear if this property might fail for the
operator $L$ in the chr. chronology. At any case, the simple
requirement of being ``of first order'' in Defn. \ref{overline},
would prevent it.

Let us see this question from a more general viewpoint (see also
\cite[p. 562]{H2}). Let $X$ be any set, ${\cal S}(X)$ the set of
all the sequences in $X$ and ${\cal P}(X)$ the set of parts of
$X$. We will mean by a limit operator any map ${\cal L}:{\cal
S}(X)\rightarrow {\cal P}(X)$ such that if $\sigma \in {\cal
S}(X)$ and $\sigma'$ is a subsequence of $\sigma$ then ${\cal
L}(\sigma) \subset {\cal L}(\sigma')$. Its {\em derived topology}
$\tau_{\cal L}$ is the one such that a subset $C\subset X$ is
closed if and only if ${\cal L}(\sigma) \subset C$ for any
sequence $\sigma$ of $C$. The limit operator is {\em of first
order} if the convergence of $\sigma$ to $x\in {\cal P}(X)$ with
the $\tau_{\cal L}$ topology implies that $x\in {\cal L}(\sigma)$.
For example, if ${\cal L}$ is the limit operator which maps all
${\cal S}(X)$ in the empty set, then $\tau_{\cal L}$ is the
discrete topology and ${\cal L}$ {\em is not} of first order.

\begin{definition}\label{d6.1}
A subset $C$ of a topological  space $X$ is {\em sequentially
closed} if for any sequence in $C$ which converges to some point
$x\in X$, then $x\in C$. The topological space $X$ is a {\em
sequential space} if every sequentially closed subset of $X$ is
closed.
\end{definition}

\begin{remark}\label{r6.1} {\em In general, closed subsets are
sequentially closed, but the converse does not hold. Observe that,
in a sequential space, any non-closed subset $A$ contains a
sequence converging to a point in $\overline{A}\setminus A$. If,
in addition, for any point $x\in \overline{A}\setminus A$ there
exists a sequence in $A$  which converges to $x$, then the
topological space $X$ is called  {\em Fr\'echet-Urysohn}. Any
first countable space is Fr\'echet-Urysohn and, then, sequential
(see \cite{goreham} for a systematic study). }\end{remark}

In the case of a topology defined by a limit operator ${\cal L}$,
the following result holds.

\begin{proposition} \label{p6.1} Let $X$ be a set endowed with a topology $\tau_{\cal L}$ derived from a limit operator ${\cal L}$. Then, the following assertions hold:
\begin{itemize}
\item[(i)] If $x\in {\cal L}(\sigma)$ with $\sigma=\{x_n\}\subset
X$, then $\sigma$ converges to $x$.  \item[(ii)] $X$ is a
sequential space. 
\end{itemize}

\end{proposition}
{\it Proof.} (i) By contradiction, assume that $\sigma$ does not
converge to $x$ with the topology $\tau_{\cal L}$.
Then, there exists some neighborhood $U$ of $x$ and some
subsequence $\tilde{\sigma}=\{x_{n_k}\}_{k}$ such
that $x_{n_k}\not\in U$ for any $k$. Since $x\in
{\cal L}(\sigma)$, it is also $x\in {\cal L}(\tilde{\sigma})$,
and thus, $x$ must belong to any closed set containing
$\tilde{\sigma}$. In particular, $x$ belongs to the closed set
$U^{c}$, in contradiction to the initial hypothesis.

(ii) Let $C$ be a sequentially closed subset of $X$. Let
$\sigma\subset C$ be a sequence, and assume $x\in {\cal
L}(\sigma)$. From (i), $\sigma$ converges to $x$. Since $C$ is
sequentially closed, necessarily $x\in C$. Thus, $C$ is closed,
and so, $X$ is a sequential space. \cvd

\begin{remark}\label{r6.15} {\em (1)
This proposition proves, in particular, that {\em the chr.
topology is sequential}. Nevertheless, there exist examples
showing that it may not satisfy the  first axiom of countability,
i.e., some points may not admit a countable topological basis.

In fact, the pair $(P',\emptyset)$ in the example illustrated by
Fig. 1(C) shows this situation. To see it, consider the natural
coordinates $(x,t)$ (such that the central removed point is the
origin of $\LL^2$) and the sequence of points $\{(-1/n,1/n)\}$,
which converges to the pair $(P',\emptyset)$, but does not
converge to $(P,F)$. By contradiction, assume that
$(P',\emptyset)$ admits a countable basis of neighborhoods $U_k$.
For each $k$, the points $(-1/n,1/n)$ must enter in $U_k$ for all
$n$ big enough. In particular, $(-1/n_k,1/n_k)$ belongs to $U_k$
for some $n_k$. Thus, one can construct a sequence of points
$(y_k,t_k)$ in the spacetime such that $t_k>|y_k|$, with $y_k$,
$t_k$ converging to $0$, and $(y_k,t_k)$ belonging to $U_k$ for
each $k$. In particular, $(y_k,t_k)$ converges to $(P,F)$, and
does not converge to $(P',\emptyset)$, in contradiction to the
fact that $U_k$ is a basis of neighborhoods. Notice also that $F$,
regarded as a subset of $\overline{M}$, also shows explicitly
that, in general, the closure of a set does not coincide with the
set of limits of sequences there, i.e. the chr. topology is not
Fr\'echet-Urysohn.

(2) However, the possibilities for the chr. topology of being
non-first countable, or non-first order,  are very pathological
(the last possibility is clearly conceivable, but we do not have
any non-first order explicit example). In fact, both possibilities
are forbidden under mild hypotheses, as the property of being {\em
separating} explained next.
}
\end{remark}

\begin{lemma}
For any two pairs $(P,F)$, $(P',\emptyset)$, in the chr.
completion with $P'\varsubsetneq P$, the set $P\setminus
\overline{P'}$ is not empty.
\end{lemma}

\noindent {\em Proof.} Assume, by contradiction, $P\subset \overline{P'}$ under $P'\varsubsetneq
P$. Then,
 any chain $\{p_n\}$ generating $P$ satisfies $\{p_n\}\subset \overline{P'}$
but it is not
included in $P'$ for $n$ big enough. So, for such a $n$, the open subset
$I^+(p_n)\cap I^-(p_{n+1})$ (included in $P$,
and, thus, in $\overline{P'}$), does not intersect $P'$, a contradiction. \cvd


\begin{definition} The chr. topology of a standard completion
is {\em separating} if the following property  for pairs type
$(P',\emptyset)$, plus the analogous one for $(\emptyset, F')$,
holds: for any two pairs $(P,F)$, $(P',\emptyset)$, with
$P'\varsubsetneq P$, there exists some $x\in P\setminus
\overline{P'}$ such that the open sets $I^+(x,\overline{M})$,
$\overline{I^+(x,\overline{M})}^{c}$ separate them (i.e. $(P,F)\in
I^{+}(x,\overline{M})$, $(P',\emptyset)\in
\overline{I^{+}(x,\overline{M})}^{c}$).
\end{definition}

\begin{example} {\em The chr. topology of the example in Fig. 1(C) is not separating.
A simple case where the separating property holds non-trivially,
is $M=\LL^2\setminus \{(x,t): -x\leq t\leq 0\}$ (think, for
example, in $P,F,P'$ as the following subsets computed on $\LL^2$:
$P=I^-((0,0)), F=I^+((0,0)), P'=I^-((1,-1))$). }\end{example}

\begin{proposition}\label{p6.2}
Assume that the chr. topology of a standard completion
$\overline{M}$ is separating. Then, the chr. topology is of first
order and second countable. Moreover, for any $D=\{x_{n}\}\subset
M$  which is dense in $M$ (and thus in $\overline{M}$) and
chronologically dense in $M$, the set
\[
\Omega=\{I^{+}(x_{m},\overline{M})\cap I^{-}(x_{n},\overline{M}):
n,m\in \N\}\cup \{I^{\pm}(x_{n},\overline{M})\cap
\overline{\cup_{i=1}^{k} I^{\pm}( x_{j_{i}},\overline{M})}^{c}: n,
k, j_{i}\in {\mathbb N}\},
\]
is a  topological basis.
\end{proposition}

\noindent {\it Proof.} First, let us prove that $\Omega$ is a
topological basis and, thus, the chr. topology is second
countable. Clearly, we can restrict our attention to
$(P',F')\in\partial M$.

First, assume $P'\neq\emptyset\neq F'$, and let
$\{p'_{n}\},\{q'_{n}\}\in D$ be chains generating $P'$, $F'$,
resp. Consider the nested neighborhoods $U_{n}\in \Omega$ of
$(P',F')$ given by $U_{n}=I^{+}(p'_{n},\overline{M})\cap
I^{-}(q'_{n},\overline{M})\subset\overline{M}$.  Let $U$ be an
arbitrary neighborhood of $(P',F')$ in $\overline{M}$. We have to
prove that $U_{n}\subset U$ for some (and then all the following)
$n$. Assume by contradiction the existence of $r_{n}\in
U_{n}\setminus U$. Since each $r_{n}\in U_{n}$, necessarily
$P'\subset {\rm LI}(I^{-}(r_{n}))$, $F'\subset {\rm
LI}(I^{+}(r_{n}))$. Moreover, since $P'\sim_{S}F'$,  Lemma 3.15
implies that $P'$ is a maximal IP into ${\rm LS}(I^{-}(r_{n}))$,
and $F'$ is a maximal IF into ${\rm LS}(I^{+}(r_{n}))$, i.e., this
completes $(P',F')\in L(r_{n})$. Then, $r_{n}\rightarrow (P,F)$
with the chr-topology, in contradiction with $r_{n}\not\in U$ for
any $n$.

Next, assume that $(P',F')\in\partial M$ satisfies, say,
$P'\neq\emptyset$, $F'=\emptyset $, and fix a chain
$\{p'_{n}\}\subset D$ generating $P'$. Let $\{s_{i}\}\subset D$ be
the (countable) subset of $\overline{P'}^{c}$ composed by those
points in $D$ which separate $(P',\emptyset)$ and $(P,F)$, for
some $(P,F)\in\overline{M}$. Recall about this subset that, from
the separation property assumed in the hypotheses, for any $(P,F)$
as above there exists some $x\in P\setminus \overline{P'}$ which
separates $(P',\emptyset)$ and $(P,F)$. Thus,  any $s\in D$
satisfying $x\ll s\in P$ provides an element of $\{s_i\}$
associated to $(P,F)$, and the full $\{s_i\}$ is constructed by
taking all such $s$ for all the pairs $(P,F)$. In particular, the
chronological density of $D$ ensures that, if at least one $(P,F)$
as above exists, then  $\{s_i\}$ is infinite (otherwise,
$\{s_i\}$ is empty). Now, define the open subsets $U_{n}\in
\Omega$ of $(P',\emptyset)$ as
$U_{n}=I^{+}(p'_{n},\overline{M})\cap A_n$, where
$A_n=\overline{\cup_{i=1}^{n} I^{+}( s_{i},\overline{M})}^{c}$
 (if  $\{s_i\}$
is
 empty, put $A_n=\overline{M}$),
for all $n$. Since $(P',\emptyset)\not\in
\overline{I^{+}(s_i,\overline{M})}$ for all $i$, the open subsets
$U_n$ contain $(P',\emptyset)$. Let $U$ be an arbitrary
neighborhood of $(P',\emptyset)$ in $\overline{M}$, and let us
check $U_{n}\subset U$ for some $n$. Assuming by contradiction the
existence of $r_{n}\in U_{n}\setminus U$ for all $n$ as above, one
deduces again $P'\subset {\rm LI}(I^{-}(r_{n}))$. Moreover, by the
property defining the subset $\{s_i\}$, if $P'\varsubsetneq
P\subset {\rm LS}(I^{-}(r_{n}))$, then $P\setminus \overline{P'}$
must contain some $s_{i_{0}}\in \{s_{i}\}$. Hence $(P,F)\in
I^{+}(s_{i_{0}},\overline{M})$, and thus $s_{i_{0}}\in P\subset
{\rm LS}(I^-(r_n))$. In conclusion, $s_{i_{0}}\in I^{-}(r_{n})$
for infinitely many $n$, in contradiction with the hypothesis
$r_{n}\in U_{n}$ (which implies $r_{n}\not\in I^{+}(s_{i_{0}})$
for all $n$ big enough). Therefore, $P'$ is maximal in ${\rm
LS}(I^{-}(r_{n}))$, and so, $(P',\emptyset)\in L(r_{n})$. Thus,
$r_{n}\rightarrow (P',\emptyset)$ with the chr. topology, in
contradiction with $r_{n}\not\in U$ for any $n$.

To check  the first order property, assume that
$\sigma=\{(P_n,F_n)\}$ converges to $(P',F')$ with the
chr-topology. Then, it is not a restriction to assume that
$(P_{n},F_{n})\in U_{n}$ for every $n$, being $\{U_{n}\}_{n}$ a
countable basis of neighborhoods for $(P,F)$ as above. The proof
above shows that any sequence $\{r_n\}$ such that $r_n\in U_n$
satisfies $(P',F')\in L(\{r_n\})$ and, thus, the required
inclusion $(P',F')\in L(\sigma)$. \cvd

\begin{remark}\label{r6.2} {\em  (1) CEAT  (Definition \ref{j}) is always
second countable, as it admits as a topological subbasis: $
\{I^{\pm}(x_{n},\overline{M}): n\in \N\}$, with $D=\{x_{n}\}$ as
in Proposition \ref{p6.2}. In fact, for any $(P,F)$ and any
$(P_0,F_0)$ in, say, $I^+(P,F)$, the density of $D$ implies the
existence of some $x_n\in F\cap P_0$ and, thus, $(P_0,F_0)\in
I^+(x_{n},\overline{M})\subset I^+(P,F)$.

(2) CEAT can be characterized in terms of the following limit
operator $L^0$: for any sequence $\sigma=\{(P_{n},F_{n})\}$, we
define $(P,F)\in L^0(\sigma)$ iff $P\subset {\rm LI}(P_{n})$ and
$F\subset {\rm LI}(F_{n})$. In fact, denote by $\tau_{L^0}$ the
topology derived from $L^{0}$. CEAT is finer than $\tau_{L^0}$
from the implication to the right of Lemma \ref{lea} (recall that,
as CEAT is second countable,
 the continuity of the inclusion $(M,$CEAT$)\hookrightarrow
(M,\tau_{L_0})$ is characterizable by sequences). Analogously,
$\tau_{L^0}$ is finer than CEAT from the implication to the left
of Lemma \ref{lea}, taking into account that, if $\{(P_n,F_n)\}$
converges to $(P,F)$ with $\tau_{L^0}$, then $P\subset {\rm
LI}(P_n)$ and $F\subset {\rm LI}(F_n)$. To check this last
property just argue by contradiction: if, say, there exists some
$p_0\in P$ with $p_0\not\in P_n$ for infinitely many $n$, then the
set of pairs $(P',F')$ with $p_0\not\in P'$ is closed for
$\tau_{L^0}$ and does not contain $(P,F)$, an absurd. Recall that
this property is equivalent to the first-order character of $L^0$,
because of the particular expression of this operator.

(C) The same argument shows that if $\{(P_n,F_n)\}$ converges to
$(P,F)$ with the chr. topology, then $P\subset {\rm LI}(P_n)$ and
$F\subset {\rm LI}(F_n)$. However, this does not imply that the
chr. operator $L$ is of first order.}
\end{remark}

\section{Conformal boundary vs c-boundary}\label{s4}

\subsection{First definitions on the conformal boundary}\label{s4.1}

As above, $M$ will be a strongly causal spacetime and $\partial M$
and $\overline{M}$ will denote the causal boundary and
completion, resp. Recall that a smooth map  $i: M\hookrightarrow
M_0$ between two Lorentzian manifolds $(M,g), (M_0,g_0)$
 is called an {\em open  embedding} if $i(M)$
is an open subset in $M_0$ and $i$ is a diffeomorphism between $M$
and its image $i(M)$. The map $i$ is {\em conformal} if the
pullback metric $i^*g_0$ satisfies $i^*g_0 = \Omega g$ for some
function $\Omega>0$ on $M$; for spacetimes, we also assume that
the time-orientations are preserved. Eventually,
$\{\Omega(p_n)\}_n$ may tend to 0 or $\infty$ when the sequence
$\{i(p_n)\}_n$ converges to the topological boundary of $i(M)$ in
$M_0$. However, we will not care on this possibility, as we will
consider only properties which are conformally invariant. So, once
$i$ is defined, one is free to re-scale $g$ so that $i$ becomes an
isometry.

\bdefi\label{d4.1} (Envelopment.) A {\em (conformal) envelopment}
of $M$ is an open conformal embedding $i:M\hookrightarrow M_{0}$
in some ({\em ``aphysical''}) strongly causal spacetime $M_{0}$.

Then, the {\em conformal completion} of $M$ (w.r.t. $i$) is  the
closure $\overline{M}_{i}:=\overline{i(M)}\subset M_{0}$, and the
{\em conformal boundary} is the topological one
$\partial_{i}M:=\overline{i(M)}\setminus i(M)$. \edefi

\begin{remark} \label{rconfchr}
{\em This definition considers the conformal completion as a point
set, but it is natural to endow it  with a topology and a
chronology too. The problems to define these elements will be
discussed later, but we point out now  the natural choices. The
topology on $\overline M_i$ is just the induced one from $M_0$.
For the chronological relation, there are some candidates. As
natural options, we can consider the following two definitions:
\bit \item[$W)$]  $p\ll_i^W q$ (or just $p\ll_i q$) iff there
exists some continuous curve $\gamma:[a,b]\rightarrow \overline
M_i$ with $\gamma(a)=p, \gamma(b)=q$ such that
$\gamma\mid_{(a,b)}$ is future-directed smooth timelike and
contained in $M$. \item[$S)$] $p\ll_i^S q$ iff there exists some
curve $\gamma:[a,b]\rightarrow \overline M_i$ which is smooth and
future-directed timelike in $M_0$, with $\gamma(a)=p, \gamma(b)=q$
and $\gamma\mid_{(a,b)}$  contained in $M$. \eit

These binary relations represent the weakest and the strongest
reasonable choices for the chronology in the completion
$\overline{M}_i$. They are different in general (Figs.
\ref{fig2''}, \ref{f7}) and may be non-transitive
(Figs. \ref{f1}, \ref{f2}).
Notice that the chronological relation $\ll_i^W$ is intrinsic to
$\overline{M}_i$ and, so, it will be the natural candidate to
agree with the relation $\overline{\ll}$ in the c-completion (as
the c-boundary only has a topology but not a differentiable or
metric structure). For this reason, $\ll^W_i$ will be preferred to
$\ll^S_i$. Analogously, the following causal relations are
defined:

\bit \item[$W)$] $p\leq_i^W q $ (or just $p\leq_i q$) iff either
$p=q$ or there exists some continuous curve
$\gamma:[a,b]\rightarrow \overline{M}_i$ with $\gamma(a)=p,
\gamma(b)=q$ such that $\gamma$ is future-directed and
continuous-causal (Remark \ref{r2.2o}) when regarded as a curve in
$M_0$. \item[$S)$] $p\leq_i^S q$ iff there exists some
 curve $\gamma:[a,b]\rightarrow
\overline{M}_i$ which is smooth and future-directed causal in
$M_0$, with $\gamma(a)=p, \gamma(b)=q$. \eit About the possible
causal relations in $\overline{M}_i$, recall that the connecting
causal curves may naturally go along the boundary. As this
boundary may be non-smooth, the curve $\gamma$ which is considered
for $\leq_i^W$ is only continuous-causal (as a curve in $M_0$).
Nevertheless, even if $\gamma$ is a smooth curve in $M$ with just
an endpoint in the boundary $\partial_iM$, the causal relation in
$M_0$ becomes essential (see Fig. \ref{fig2''}). So, both causal
relations $\leq_i^W, \leq_i^S$ are extrinsic, i.e., they require
the causality in $M_0$ and, in principle, depend on the conformal
embedding. But $\leq_i^W$ uses less structure on the boundary
(which may be non-smooth) and so, it will be preferred below.
}\erema

\begin{figure}
\centering
\ifpdf
  \setlength{\unitlength}{1bp}%
  \begin{picture}(294.44, 173.11)(0,0)
  \put(0,0){\includegraphics{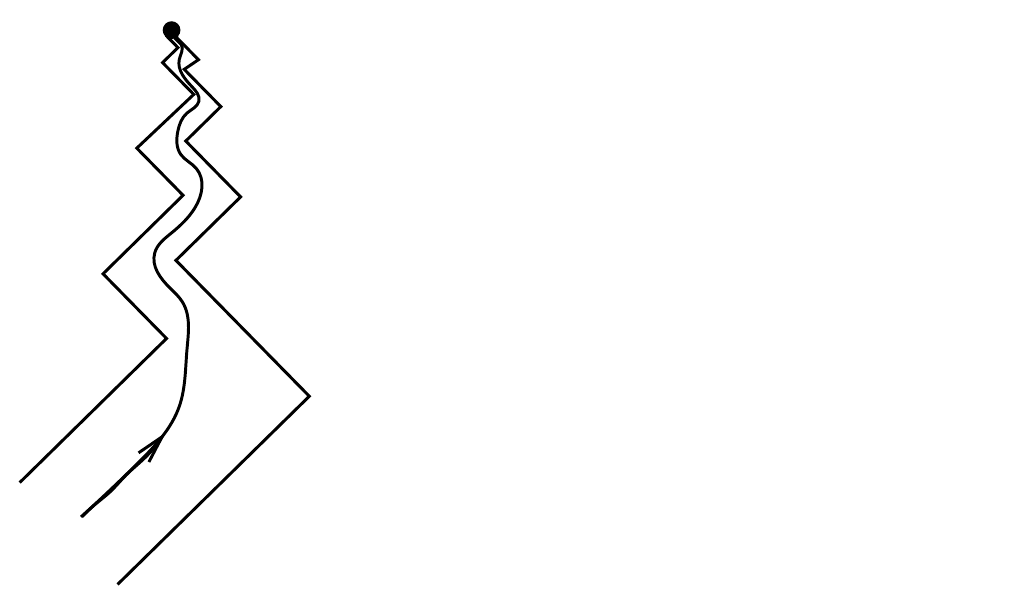}}
  \put(43.14,172.03){\fontsize{8.54}{10.24}\selectfont $z$}
  \put(28.14,23.03){\fontsize{8.54}{10.24}\selectfont $\gamma$}
  \put(108.73,98.70){\fontsize{14.23}{17.07}\selectfont $\overline{M}_i^*\subset \LL^2$}
  \end{picture}%
\else
  \setlength{\unitlength}{1bp}%
  \begin{picture}(294.44, 173.11)(0,0)
  \put(0,0){\includegraphics{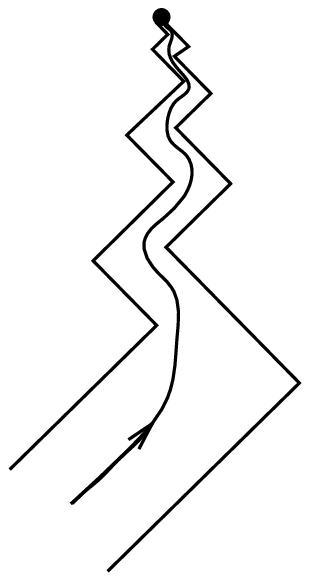}}
  \put(43.14,172.03){\fontsize{8.54}{10.24}\selectfont $z$}
  \put(28.14,23.03){\fontsize{8.54}{10.24}\selectfont $\gamma$}
  \put(108.73,98.70){\fontsize{14.23}{17.07}\selectfont $\overline{M}_i^*\subset \LL^2$}
  \end{picture}%
\fi \caption{Because of the infinity breaks of $\partial^{*}_i M$,
the causal curve $\gamma$, which is smooth in $M$, cannot be
smoothly extended to $z\in\partial^{*}_{i}M$. The extended curve
is causal continuous because there are close smooth causal curves
in $M_0$ which connect any point of $\gamma$ with $z$.
Nevertheless, no such curves are contained in
$\overline{M}^{*}_i$. Any point $p$ on $\gamma$ satisfies
$p\ll_{i}^W z$, but $p\not\ll_{i}^S z$. \label{fig2''}}
\end{figure}

\begin{convention} \label{convention2} {\em
When there is no possibility of confusion, the conformal embedding
$i$ will be dropped, and $M$ will be regarded as an open subset of
$M_0$. However, the subscript $i$ is retained in $\partial_iM,
\overline{M}_i$ --and so these spaces are distinguished from the
causal ones $\partial M, \overline{M}$.

The preferred relations $\ll^W_i, \leq ^W_i$ will be denoted just
$\ll_i, \leq_i$.

The elements in $M$ such as the chronological futures or past will
be denoted as above (say, $I^\pm(p):=I^\pm(p,M)$), and we will add
the subscript 0 if these elements are taken in $M_0$
($I_0^\pm(p):=I^\pm(p,M_0)$). Other elements will be written more
explicitly. 
For example, $I^+(p,\overline{M}_i) (:= I^+(p,\overline{M}_i,
\ll_i)$) denotes the chronological future of $p$ in the completion
$\overline{M}_i$ computed with the preferred relation $\ll_i$.

As an abuse of notation, if $z\in
\partial_iM$ then $I^{\pm}(z)$ will represent $I^\pm(z,\overline{M}_i)\cap M$, which turns out the set
of points in $M$ which can be connected to $z$ by means of a
continuous curve $\gamma:[a,b]\rightarrow \overline M_i$,
$\gamma(b)=z$, which is smooth, future or past-directed timelike, and
contained in $M$, on $[a,b)$. }
\end{convention}

\bdefi \label{dacc} (Accessibility). A future-directed (resp.
past-directed) timelike curve $\gamma:[a,b)\rightarrow M$ has an
{\em $i$-endpoint} $p\in \overline M_i$ if $p=\lim_{t\rightarrow
b}\gamma(t)$.

If $p\in \overline{M}_i$ is the $i$-endpoint of a future-directed
(resp. past-directed) timelike curve, then $p$ is said {\em future
(resp. past) accessible}, and $p$ is called {\em accessible} if it
is either future or past accessible. The set of all the accessible
points of the conformal boundary $\partial_iM$ will be denoted by
$\partial_{i}^{*}M$ and the set of all the accessible points of
the conformal completion by $\overline{M}^*_i (=M\cup
\partial_{i}^{*}M)$. \edefi

In order to have a useful envelopment we must ensure that it
yields a true completion, that is, no part which should correspond
to the boundary is missing. This is easy to formalize if one only
cares on Causality.

\bdefi \label{dcc} (Chr-completeness.) An envelopment
$i:M\hookrightarrow M_{0}$ is {\em chronologically complete} if
any timelike curve $\gamma:[a,b)\rightarrow M$ which is
inextensible in $M$ (and, thus, generates a TIP or a TIF) has an
$i$-endpoint $p$ in the conformal boundary. \edefi That is,
chronological completeness means that the accessible boundary
$\partial_i^*M$ contains $i$-endpoints for all inextensible
timelike curves in $M$. For example, if $M$ is the half-space of
$\LL^N$ obtained as $t>0$, its inclusion in $M_0=\LL^N$ is an
envelopment which is {\em not} chronologically complete, as no
inextensible future-directed curves in $M$ will have an
$i$-endpoint in $M_0$. Before going further, notice the following
technicality which will be frequently claimed:

\begin{lemma}\label{piwdef} Let $P$ be a TIP of $M$ and $\gamma_1, \gamma_2$
two future-directed timelike curves which generate them, i.e.,
$P=I^-[\gamma_1]=I^-[\gamma_2]$. If $\gamma_1$ admits $z\in
\partial^*_iM$ as its  $i$-endpoint, then $z$ is  also the $i$-endpoint of
$\gamma_2$.
\end{lemma}
{\it Proof.} Under these hypotheses, $\gamma_{1}\subset
I^{-}[\gamma_{2}]$ and $\gamma_{2}\subset I^{-}[\gamma_{1}]$,
thus, $I_{0}^{-}[\gamma_{1}]=I_{0}^{-}[\gamma_{2}]$. As $z$ is the
$i$-endpoint of $\gamma_1$, $I_{0}^{-}[\gamma_{1}]$ is a PIP in
$M_0$, i.e., $I_0^-(z)=I^-_0[\gamma_1]=I^-_0[\gamma_2]$ and, by
Proposition \ref{paux} (ii) applied to $M_0$, $z$ is also the
$i$-endpoint of $\gamma_2$.
 \cvd

\smallskip

So, one trivially has the following characterization of
chronological completeness, which also defines the projection maps
$\hat\pi, \check\pi$. \btheo  \label{tproy} An envelopment is
chronologically complete if and only if the natural projections of
the future and past causal preboundaries,
$\hat{\partial}M\rightarrow
\partial^{*}_{i}M, 
\check{\partial}M\rightarrow
\partial^{*}_{i}M,$
 which map each TIP $P=I^-[\gamma]$ or TIF $F=I^+[\gamma]$ in
the $i$-endpoint of $\gamma$, are well-defined on all
$\hat{\partial} M$ and $\check{\partial} M$. In this case,  the
natural projections
of the precompletions \be\label{proy} \hat{\pi}:\hat{M}\rightarrow
\overline{M}_{i},\qquad \check{\pi}:\check{M}\rightarrow
\overline{M}_{i} \ee are obviously defined. \etheo When, say, a
TIP $P$ satisfies $\hat{\pi}(P)=z\in \partial_i^*M$, we say that $P$
projects or {\em is associated} to $z$.

 The simplest
criterion to ensure chronological completeness is compactness
(even though this may be too restrictive and the boundary may
include non-accessible points).
\begin{proposition} \label{pcompact} An envelopment is chronologically
complete if the conformal completion $\overline{M}_i$ is compact.
\end{proposition}
{\it Proof.} Take $P\in \hat{\partial}M$  and let $\gamma:
[a,b)\rightarrow M$ be some inextensible future-directed timelike
curve in $M$ such that $I^{-}[\gamma]=P$. Since $\overline{M}_{i}$
is compact, there exists some sequence $t_n\nearrow b$ such that
$\{\gamma(t_n)\}_n$ converges to some $p_0\in \overline{M}_i$. As
$M_0$ is strongly causal, $\gamma$ is continuously extensible to
$p_0$ (Proposition \ref{paux}(i)) and, as $\gamma$ is inextensible
in $M$, $p_0\in
\partial_i^*M$.  \cvd

\smallskip

Along all our study, we will assume that the envelopment is
chronologically complete. Moreover, we will focus on the
accessible part $\partial_i^*M$ of the conformal boundary
$\partial_i M$. About these hypotheses notice: \ben \item {\em
Chr-completeness.} If it were dropped, one may think that the
completion could be ``still completed''. Moreover, this property
allows to relate the conformal boundary with the c-boundary
(Theorem \ref{tproy}).

\item {\em Restriction to $\partial_i^*M$}. One might think that
the accessible part of $\partial_iM$ may not be ``complete
enough'' as one would like to consider also ``completions of
spacelike directions'' --say, as in the case of the point $i^0$ in
the standard compactification of $\LL^4$. Nevertheless, we do not
worry on non-accessible points because of the following: \bit
\item One seems to be forced to impose compactness for $\overline
M_i$ if spacelike directions were also completed in some sense. In
fact, spacelike directions behave in a rather uncontrolled way,
and we are considering a conformally invariant construction. Even
in the (positive-definite) Riemannian case, a conformally
invariant boundary cannot rely on the canonically associated
distance, as all non-compact Riemannian manifolds are conformal to
an incomplete one \cite{NO}. So, a Riemannian conformal completion
becomes a compactification, as in the case of the Riemann sphere
completing the complex plane. Nevertheless, the requirement to
have a compact boundary may be very restrictive and, even if this
is fulfilled, the properties of the points in
$\partial_iM\setminus\partial_i^*M$ may be very irregular (recall
examples such as Fig. \ref{f6}, the point $(0,1/2)$ in Fig.
\ref{f3}, or the case of $i^0$ itself, emphasized in  Remark
\ref{r13}(2)).

\item Even an envelopment with compact $\overline{M}_i$ may have
``non conformally invariant properties'', as shown in Fig.
\ref{f2} and \ref{f6}. That is, a compactification by itself will
not be enough, as boundary points in a compactification may
``disappear'' in a different one.\end{itemize}
\end{enumerate} 

\subsection{Requirements to relate the conformal and c-boundary}\label{s4.2}

One can try to define a natural correspondence between the causal
and conformal completions by using the projections (\ref{proy}) as
follows:

\be \pi:\overline{M}\rightarrow \overline{M}_{i}^{*},\qquad
\pi((P,F))=\left\{\begin{array}{lll} \hat{\pi}(P) & \hbox{if} & P\neq \emptyset\\
\check{\pi}(F) & \hbox{if} & F \neq \emptyset.
\end{array}\right.
\ee Nevertheless, this map is well-defined only if the following
consistency  holds: \be \label{estarc}
(P,F) \in
\partial M \;\; \hbox{and} \;\; P\neq \emptyset \neq F
\quad \Rightarrow \quad  \hat{\pi}(P) = \check{\pi}(F). \ee

Now, remark:

\ben \item Condition (\ref{estarc}) does not hold in general. In
fact, it is not only possible to find counterexamples for a
general envelopment (Fig. \ref{f3}), but also for an envelopment
with a $C^0$ boundary (Fig. \ref{f4}).

\item Assume that (\ref{estarc}) holds and, so, the projection
$\pi$ is well-defined. As we are considering only the accessible
part $\partial_{i}^*M$, the map $\pi$ is automatically onto. It
would be desirable that $\pi$ were also one to one. However, these
do not happen in general. This is not surprising for general
envelopments, as arbitrary identifications can be introduced
(Figs. \ref{f1}, \ref{f2}). Nevertheless, other examples show that
this is inherent to the particular approach. For example, when a
``non-regular'' point (in the sense of chronological sets, below
Defn. \ref{chronrel}) for $\ll_i$ appears as in Fig. \ref{f5}, we
have two non-Hausdorff related points of the c-boundary which
project necessarily onto a single point of $\partial_{i}^*M$.

\item Assume that the  projection $\pi$ is well-defined and
bijective. As emphasized in Remark \ref{rconfchr}, the completion
$\overline{M}_i$ must be endowed with a topology and a chronology,
and it would be desirable that $\pi$ were a homeomorphism and a
chronological isomorphism (the latter meaning that two points in
$\overline{M}$ are chronologically related if and only if so are
their images in $\overline{M}_{i}^*$). Nevertheless, Fig. \ref{f6}
shows\footnote{In this example $\pi$ is not bijective.
Nevertheless, this does not affect the essence of the example: if
all the removed vertical segments (but the limit one) are widened
a bit, a bijective example is obtained.} that the topology of the
conformal boundary may depend strongly on the conformal embedding
(even if there are no artificial identifications as those in Figs.
\ref{f1},\ref{f2}).

\item Subtler problems appear for the chronology and its interplay
with the topology. For example, Fig. \ref{f7} shows an open domain
of $\LL^2$ with $C^\infty$ boundary and the following undesirable
property: the chronological past of a point $q$ (computed in
$\overline M_i$ with $\ll_i$) is not open (with the induced
topology from $M_0$).

\een Summing up, previous discussion shows that some additional
conditions must be imposed on the envelopments in order to obtain
a satisfactory conformal boundary. The smoothability of the
boundary will play an important role (and will be discussed
specifically in Section \ref{s4.4}). Nevertheless, the examples
cited above show that
 even this condition is not enough. Moreover, it may be also too
restrictive and, so, more general conditions will be introduced
first.

\begin{remark}\label{rlars} {\em As a more speculative issue, one could
try to define a more general notion of conformal boundary by
considering a conformal embedding $i: M\hookrightarrow M_0$ not
necessarily open (i.e., possibly ${\rm dim}M<{\rm dim}M_0$). Many
of the notions and results for conformal envelopments could be
extended directly to this more general case. In this case, to
ensure the uniqueness of the conformal boundary would be more
complicated, as there are more possibilities for the embedding.
However, the existence for all stably causal spacetimes is
guaranteed by a result in \cite{MuSa}. In this reference, any
stably causal spacetime is shown to admit a conformal embedding in
Lorentz-Minkowski $\LL^{N_0+1}$, for some big $N_0$. By composing
this embedding with the natural open conformal embedding of
$\LL^{N_0+1}$ in Einstein static universe $\LL^1\times {\mathbb
S}^{N_0}$, one finds a conformal embedding $i: M\hookrightarrow
\LL^1\times {\mathbb S}^{N_0}$ with compact closure and, then,
chronologically complete. However, we focus only on the classical
case of conformal boundary through open embeddings.}
\end{remark}

\subsection{Main result}\label{s4.3}

Next, let us state some general conditions so that the boundary
will have a well-defined $\pi$ with satisfactory properties.
Of course, such  conditions will not be intrinsic to $M$, but
depends on how $M$ lies in $M_0$.
\begin{definition}\label{d1}
Let $\gamma:[a,b]\rightarrow \overline{M}_i$ be a continuous curve
such that $\gamma\mid_{[a,b)}$ is a future-directed (resp.
past-directed) smooth timelike curve contained in $M$ and
$\gamma(b)\in \overline{M}_i$. Then, $\gamma$ is future (resp.
past) {\em deformably timelike (in its $i$-endpoint)} if there
exists a neighborhood $U=U_{0}\cap \overline{M}_i$ of $\gamma(b)$
(where $U_{0}$ is an open set of $M_{0}$) such that $\gamma
(a)\ll_i w$ (resp. $w\ll_i\gamma(a)$) for all $w\in U$.

A point $z\in \partial_i^*M$ is {\em  \deform} if all the TIPs and
TIFs associated to $z$ (i.e, TIPs and TIFs defined by timelike
curves with $i$-endpoint $z$) can be constructed as the
chronological past or future of {\deform}
curves.
\end{definition}
Trivially, timelike curves with i-endpoint in the spacetime $M$
are
\deform. The role of timelike deformability is stressed by the
following result.
\begin{proposition}\label{l2}
(1) Let $z\in \partial_i^* M$ be \deform. Then, all the
future-directed timelike curves in $M$ with $i$-endpoint $z$ have
the same chronological past in $M$, namely, $I^-(z)$ (according to
Convention \ref{convention2}).

(2) All the  points in $\partial_i^*M$ are {\deform} if and only if
 the open sets in $\overline{M}_i$ for the CEAT associated to $\ll_i$ are
open sets for the  topology induced from $M_0$. In this case, any
curve which is smooth and timelike in $M$ and has an endpoint in
$\partial_iM$ is
\deform.
\end{proposition}
{\it Proof.} (1)  Let $\gamma,\tilde\gamma$ be two future-directed
timelike curves with $i$-endpoint $z$, and define
$P=I^{-}[\gamma], \tilde P=I^{-}[\tilde\gamma]$. Now, consider two
future-directed {\deform} curves $\sigma,\tilde{\sigma}$ such that
$P=I^{-}[\sigma]$, $\tilde P=I^{-}[\tilde\sigma]$. By Lemma
\ref{piwdef}, $\sigma,\tilde\sigma$ converge to the same point
that $\gamma, \tilde\gamma$, resp., that is, they converge to $z$.
By timelike deformability, $\sigma \subset I^-(\tilde \sigma)$ and
viceversa and, thus,  $P=\tilde P$.

(2) For the implication to the right, consider an open set $U$ for
the CEAT and suppose without restriction that $U:=I^+(p,
\overline{M}_i)$. As the chronological relation $\ll$ is open in
$M$, for any point in $U\cap M$ there exists an open set for the
induced topology containing that point and contained in $U$. On
the other hand, if $z\in U\cap
\partial_i^*M$, there exists a future-directed deformably timelike curve from $p$
to $z$ (just consider the TIP associated to $z$, which contains
$p$, and apply that $z$ is \deform). Therefore, there exists an
open set $V$ for the induced topology such that $z\in V\subset U$.

For the implication to the left, observe that any timelike curve
$\gamma:[a,b]\rightarrow \overline{M}_i^*$ with endpoint
$z\in\partial^{*}_{i}M$ is \deform. In fact, by hypothesis,
$U:=I^+(\gamma(a), \overline{M}_i)$ is an open set for the induced
topology which contains $z$, and so, $\gamma$ is \deform. \cvd

\begin{remark}
{\em If $z$ is {\deform} for all $z\in \partial_i^*M$, the CEAT is
coarser than the  topology on $\overline{M}_i$ induced from $M_0$.
In particular, the condition of admissibility (A1) (see Defn.
\ref{k}) holds for this induced topology.}
\end{remark}

\begin{definition}\label{d2}
A point $z\in \partial_i^*M$ is  {\em  (locally) timelike
transitive} if it admits a neighborhood $V=\overline{M}_i\cap V_0$
($V_0$ open in $M_0$) such that  for any $x,x'\in V$ :\bit \item
$x\ll_iz\leq_ix'\Rightarrow x\ll_ix'$. \item
$x\leq_iz\ll_ix'\Rightarrow x\ll_ix'$. \eit
\end{definition}

\begin{figure}
\centering
\ifpdf
  \setlength{\unitlength}{1bp}%
  \begin{picture}(313.54, 146.14)(0,0)
  \put(0,0){\includegraphics{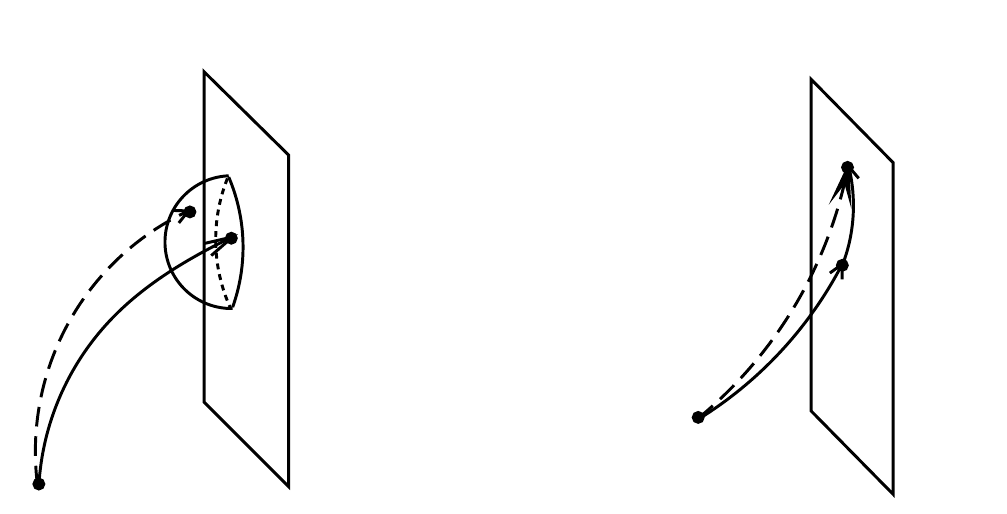}}
  \put(23.92,37.06){\fontsize{7.20}{8.64}\selectfont $\gamma$}
  \put(14.92,7.06){\fontsize{7.20}{8.64}\selectfont $\gamma(a)$}
  \put(68.50,69.65){\fontsize{7.20}{8.64}\selectfont $\gamma(b)$}
  \put(7.62,132.40){\fontsize{11.99}{14.39}\selectfont Deformably timelike}
  \put(204.39,24.78){\fontsize{7.20}{8.64}\selectfont $x$}
  \put(240.43,62.72){\fontsize{7.20}{8.64}\selectfont $z$}
  \put(210.93,132.27){\fontsize{11.99}{14.39}\selectfont Timelike transitive}
  \put(5.67,110.94){\fontsize{8.54}{10.24}\selectfont $M$}
  \put(247.54,93.16){\fontsize{7.20}{8.64}\selectfont $x'$}
  \put(43.27,90.20){\fontsize{7.20}{8.64}\selectfont $w$}
  \put(41.50,57.60){\fontsize{8.54}{10.24}\selectfont $U$}
  \put(210.68,85.89){\fontsize{8.54}{10.24}\selectfont $V$}
  \end{picture}%
\else
  \setlength{\unitlength}{1bp}%
  \begin{picture}(313.54, 146.14)(0,0)
  \put(0,0){\includegraphics{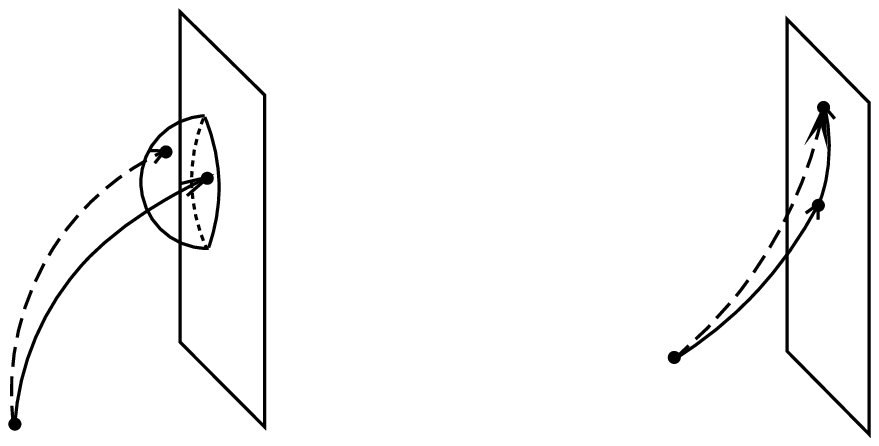}}
  \put(23.92,37.06){\fontsize{7.20}{8.64}\selectfont $\gamma$}
  \put(14.92,7.06){\fontsize{7.20}{8.64}\selectfont $\gamma(a)$}
  \put(68.50,69.65){\fontsize{7.20}{8.64}\selectfont $\gamma(b)$}
  \put(7.62,132.40){\fontsize{11.99}{14.39}\selectfont Deformably timelike}
  \put(204.39,24.78){\fontsize{7.20}{8.64}\selectfont $x$}
  \put(240.43,62.72){\fontsize{7.20}{8.64}\selectfont $z$}
  \put(210.93,132.27){\fontsize{11.99}{14.39}\selectfont Timelike transitive}
  \put(5.67,110.94){\fontsize{8.54}{10.24}\selectfont $M$}
  \put(247.54,93.16){\fontsize{7.20}{8.64}\selectfont $x'$}
  \put(43.27,90.20){\fontsize{7.20}{8.64}\selectfont $w$}
  \put(41.50,57.60){\fontsize{8.54}{10.24}\selectfont $U$}
  \put(210.68,85.89){\fontsize{8.54}{10.24}\selectfont $V$}
  \end{picture}%
\fi \caption{Examples of deformably timelike curve (left) and timelike transitive point (right).}
\end{figure}

\begin{definition}\label{ta}
Let $i:M\hookrightarrow M_0$ be an envelopment. A point $z\in
\partial_i^* M$ is {\em regularly accessible} if it is both {\deform} and timelike transitive.

The boundary is {\em regularly accessible} if all its points in
$\partial_{i}^* M$ are regularly accessible.
\end{definition}

The following proposition will be useful:
\begin{lemma}\label{plim}
Consider a point $p\in \overline{M}_i$ and suppose that there
exists a sequence of future-directed causal curves
$\gamma_{n}:[0,a_{n}]\rightarrow M$ such that $p$ is an
accumulation point of $\{\gamma_{n}(0)\}_n$. Moreover, suppose
that the sequence $\{\gamma_{n}(a_{n})\}_n$ does not converge to
$p$. Then, there exists a future-directed causal curve
$\gamma:[0,A]\rightarrow \overline{M}_i$, $A>0$, which is limit
curve\footnote{See \cite[Sect. 3]{BEE} for details.} of
$\{\gamma_{n}\}_{n}$, continuous causal in $M_{0}$ and satisfying
$\gamma(0)=p$.
\end{lemma}
{\it Proof.} It is a consequence of the proof of \cite[Prop.
3.31]{BEE} by taking into account that $\gamma$ must lie in
$\overline{M}_i$ as $\{\gamma_{n}\}_{n}\subset M$. \cvd

\smallskip

\smallskip

Now, we are in conditions to establish the main result of this
section.
\begin{theorem}\label{t1}
Let $i:M\hookrightarrow M_{0}$ be a chronologically complete
envelopment. If its boundary is regularly accessible then the
conformal and c-completion are equivalent, in the sense:
\begin{itemize}
\item[a)] The map $\pi:\overline{M}\rightarrow
\overline{M_{i}}^{*}$ is well-defined and bijective. \item[b)]
$\pi$ is an homeomorphism. \item[c)] $\pi$ is a chronological
isomorphism.
\end{itemize}
\end{theorem}
{\it Proof. }
a) Consider a pair $(P,F)\in \partial M$, i.e., $P\sim_{S}F$, and
let $\{p_{n}\}_n,\{q_{n}\}_n$ be future and past-directed chains
generating $P,F$ resp. By contradiction, suppose that
$p=\hat{\pi}(P)\neq \check{\pi}(F)=q$. As $P\subset \downarrow F$,
up to a subsequence, there exists a future-directed timelike curve
$\gamma_{n}$ joining each $p_{n}$ with $q_{n}$. As $p,q$ are
accumulation points of $\gamma_{n}$,  Lemma \ref{plim} implies the
existence of a limit continuous causal curve in $M_0$  such that
$\gamma:[0,A]\rightarrow \overline{M}_i$, $A>0$, with
$\gamma(0)=p$.

As $\partial_{i}^* M$ is regularly accessible, $P=I^{-}[\alpha]$
for some  future-directed timelike curve, necessarily {\deform}
(Prop. \ref{l2}). Now, let $r\in Im(\gamma)\backslash\{p\}$ close
enough to $p$, i.e., such that $r\in V$, where $V$ is the
neighborhood associated
to the timelike transitivity of $p$. 
Then, $r$ is reachable by a future-directed
{\deform} curve $c$ starting at $\alpha (0)$, and $I^-[c]=I^-(r)$ is
an IP by Prop. \ref{l2}. Now:
\begin{itemize}
\item $P \subsetneq I^{-}[c]$. In fact, choose any $p_{n}$. By
transitivity in $V$, there exists a timelike curve
$\tilde{c_n}$ from $p_n$ to $r$ 
and, by Prop. \ref{l2},  $p_{n}\in
I^-[\tilde{c}_n]=I^-(r)=I^{-}[c]$, which shows the required
inclusion. Moreover, $P\neq I^{-}[c]$ because otherwise Lemma
\ref{piwdef} would contradict $p\neq r$.\item $I^{-}[c]\subset
\downarrow F$. Let $r_{m}:=c(t_{m})$ be a timelike chain
generating $I^{-}[c]$. By using that $c$ is deformably timelike,
for every $r_{m}$ there exists a neighborhood $U_{m}$ of $r$, such
that $r_m\ll_i w$ for all $w\in U_{m}$. Moreover, because of the
convergence of $\{\gamma_n\}$, there exist a subsequence
$\{s_{n_k}\}$ such that $\gamma_{n_k}(s_{n_k})\in U_m$ for all
$k$. This last property implies  $r_{m}\ll_{i}
\gamma_{n_k}(s_{n_k})\ll_{i} q_{n_k}$, and thus, $r_{m}\ll_{i}
q_n$ for all large $n$. As  $m$ is arbitrary, the required
inclusion follows.
\end{itemize}
This contradicts the maximality of $P$ in $\downarrow F$ and,
necessarily, $\hat{\pi}(P)= \check{\pi}(F)$.

As the map $\pi$ is onto (its codomain was restricted to
$\partial_i^*M$), we only have to check the injectivity. Notice
that if  $\pi(P,F)=\pi(P',F')$ and $P\neq \emptyset \neq F$ then
 implies $(P',F')=(P,F)$ (if, say, $P'\neq \emptyset$ Prop. \ref{l2}(1) implies $P'=P$, thus $P'\nsim \emptyset$ and analogously
 $F'=F$). So, reasoning by
contradiction, it is necessary to check only the case when
$(\emptyset,F),(P,\emptyset)\in
\partial M$ are projected onto the same point $z\in \partial^*_iM$.
Let $\gamma, \rho$ be deformably timelike curves such that
 $P=I^-(\gamma)$ and $F=I^+(\rho)$.
 As $\gamma$ is  timelike deformable
and $\gamma,\rho$ converge necessarily to $z$, any point of the
curve $\gamma$ can be joined with a point of $\rho$ sufficiently
close to $z$, so $P\subset \downarrow F$ (and analogously
$F\subset \uparrow P$).  But we are assuming  $P\nsim_S F$ in $M$
and, then, there exists $P',F'$ with $P\subsetneq P'$,
$F\subsetneq F'$ and $P'\sim_SF'$ (notice that if, say,
$P\subsetneq P'$, then $P'\nsim_SF$ as $\emptyset\sim_S F$).
Then $z':=\pi(P',F')$ is well-defined and, as $P'\neq \emptyset
\neq F'$ necessarily $z\neq z'$. But this violates the strong
causality of $M_0$, as the following almost closed timelike curves
can be constructed. Let $P, F, P', F'$ be generated by, resp. the
chains $\{x_n\}_n, \{y_n\}_n, \{x'_n\}_n, \{y'_n\}_n$. As
$P\subset P'\subseteq \downarrow F'$ and $F\subset F'$, choosing
$x_{n_1}\in P$, $y_{n_1}\in F$ arbitrarily close to $x$, then we
can take $x'_{n_2}\in P'$ and $y'_{n_2}\in F'$ arbitrarily close
to $y$ such that $x_{n_1}\ll_i x'_{n_2}\ll_i y'_{n_2}\ll_i
y_{n_1}$.


\smallskip

b) Let us prove the continuity of $\pi$ first.  It suffices to
prove that for any sequence $\{(P_{n},F_{n})\}_{n}$ such that
$(P_0,F_0)\in L((P_n,F_n))$ then 
$p_{n}:=\pi((P_{n},F_{n})) \rightarrow
p_0:=\pi((P_0,F_0))$.\footnote{This is a general property of
sequential spaces (see for example, \cite[Lemma 3.1]{goreham}). In
fact, assume that this property holds. Let $C$ be any closed
subset in $\overline{M}^*_i$, and let us prove that $\pi^{-1}(C)$
is chr.-closed  on $\overline{M}$. Let $\sigma$ be any sequence in
$\pi^{-1}(C)$ such that $(P,F)\in L(\sigma)$. Then, $\pi(\sigma)$
is contained in $C$, and $\pi(\sigma)$ converges to $\pi(P,F)$. As
$C$ is closed in $\overline{M}^*_i$, necessarily $\pi(P,F)\in C$,
i.e., $(P,F)\in \pi^{-1}(C)$, as required.} So, by contradiction,
suppose that $p_n\not\rightarrow p_0$. Consider also an auxiliary
complete Riemannian metric $h$ in $M_0$ with associated distance
$d_h$. As the part a) of the proof ensures that $\hat \pi$ and
$\check \pi$ agree, we will prove just the continuity of $\hat
\pi$ and assume without loss of generality $P_0\neq\emptyset$ and
$P_{n}\neq\emptyset$.

Denote by $\{p_{n}^{m}\}_{m=1}^{\infty}$ a chain generating
$P_{n}$ (and so converging to $p_{n}$) and  by $\{p_0^{m}\}_m$ a
chain generating $P_0$ (and converging to $p_0$). As $P_0\subset
LI(P_n)$, we can assume  $p_0^n\in P_n$, up to a subsequence. For
each $n$ take $m_{n}$ big enough such that
$p_0^n\ll_{i}p_n^{m_{n}}$ and 
$d_h(p_n^{m_{n}},p_n)<\frac{1}{n}$. Let
$\gamma_n:[0,a_n]\rightarrow M$ a sequence of timelike curves
joining $p_0^n,p_n^{m_{n}}$. As $p_n\not\rightarrow p_0$ by
assumption, necessarily  $p_{n}^{m_{n}}\not\rightarrow p_0$ and
Lemma \ref{plim} implies the existence of a limit curve
$\gamma:[0,A]\rightarrow \overline{M}_i.$

Observe that $p_0$ is reachable by a future-directed curve which
is necessarily \deform. In particular, the point $q=\gamma(t)$ for
small $t$  is reachable by a future-directed (deformably timelike)
curve $c_q$. If we denote $P_q=I^-[c_q]$ then the inclusions
$P_0\subsetneq P_q$ and $P_q\subset LI(P_{n})$ follow by reasoning
as in the
 two itemized arguments in the part a). 
Summing up, $P_0\subsetneq P_q \subset LI(P_n)$ and, so, $P_0$ is
not maximal in $LS(P_n)$, in contradiction with $P_0\in
\hat{L}(P_n)$.

For the continuity of $\pi^{-1}:\overline{M}_i^{*}\rightarrow
\overline{M}$, it is enough  to prove that, for any sequence
$\{p_{n}\}_n\subset\overline{M}_i^*$ which converges to a point
$p\in \overline{M}_i^*$, then $(P,F)\in L((P_n,F_n))$, where
$(P_n,F_n)= \pi^{-1}(p_n)$, $(P,F)= \pi^{-1}(p)$ (recall that
$\overline{M}_i^{*}$ is metrizable -in particular, sequential- and
Prop. \ref{p6.1}(i)). In order to check  $P\in \hat{L}(P_n)$ (the
proof of $F\in \check{L}(P_n)$ would be analogous), we only need
to consider the case $P\neq \emptyset$ (recall Defn.
\ref{overline}) and work with the map $\hat \pi$. Consider a curve
$\gamma$ such that $P=I^{-}[\gamma]$. As $\gamma$ is deformably
timelike and $p_n\rightarrow p, \gamma(t)\ll_ip_n$ for $n$ large
enough and, thus, $\gamma(t)\in P_n$ for $n$ large enough, so
$P\subset LI(P_n)$. In order to conclude that $P\in \hat{L}(P_n)$,
assume by contradiction that the maximality of $P$ is violated,
i.e, there exists $P'\supsetneq P$ maximal in $LS(P_n)$ and so
$p':=\hat{\pi}(P') \neq p$ (recall
Prop \ref{l2}). Take some chain $\{y'_{m}\}_{m}$ generating $P'$,
and choose a sequence $\{n_{m}\}_{m}$ so that $y'_{m}\in
P_{n_{m}}$ for all $m$. Then, $y'_{m}\in P_{n_{m'}}$ for all
$m'\geq m$ and $P'$ is included in $LI(P_{n_m})$, that is, $P'\in
\hat{L}(P_{n_m})$. Thus,  $\{p_{n_m}\}_m$ converges to both, $p$
(by hypothesis) and $p'$ (by applying the proved continuity of
$\hat\pi$ on $\{P_{n_m}\}_m$), in contradiction with the
Hausdorffness of $M_0$.

\smallskip

c) Consider two points $(P,F)\overline{\ll} (P',F')$, that is,
$F\cap P'\neq \emptyset$. Denote by $\gamma_{F}, \gamma_{P'}$ two
timelike curves such that $F=I^{+}[\gamma_{F}],
P'=I^{-}[\gamma_{P'}]$, and let $p:=\pi((P,F)), p':=\pi((P',F'))$.
Choose any $r\in P'\cap F$. There exists $t_{0},t_{1}$ such that
$p\ll_i \gamma_{F}(t_{0})\ll r \ll \gamma_{P'}(t_{1})\ll_i p'$.
Thus, one can construct a curve connecting $p, p'$ which is
(piecewise) smooth and timelike in $M$, and shows $p\ll_i p'$.

Conversely, let $p\ll_i p'$ in $\overline{M}_i^*$, and consider a
connecting curve $\gamma:[0,1]\rightarrow \overline{M}_i^*$ as in
the definition of $\ll_i$ (Remark \ref{rconfchr}). We have to
check that $P'\cap F\neq \emptyset$, for the pairs $(P,F) =
\pi^{-1}(p)$ and  $(P',F') = \pi^{-1}(p')$. But $P'=I^{-}[\gamma]$
and $F=I^+[\tilde{\gamma}]$, for $\tilde{\gamma}(t)=\gamma(1-t)$
(recall Prop. \ref{l2}(1)) and, thus, $\gamma(t)\in P'\cap F$ for
all $t\in (0,1)$. \cvd


\subsection{Envelopments with $C^1$ boundary}\label{s4.4}

The regular accessibility imposed in the main result (Theorem
\ref{t1}) comprises two technical requirements, timelike
deformability and transitivity. In this subsection, we  explore
some  assumptions which imply regularity with a double aim: to
make clearer and more easily computable the notion of regularity.
A natural simplifying hypothesis for the (accessible) completion
is to have a structure of {\em manifold with boundary}. About the
order of smoothability, examples such as Fig. \ref{f4} show that
$C^0$ is too weak, but a differentiability greater than 1 might be
too restrictive to produce non-trivial results.

\subsubsection{Notion of $C^r$ conformal boundary}
In order to make clear some notation and  subtleties, we recall
explicitly the following notion.

\bdefi \label{decb0} ($C^r$ point). An envelopment
$i:M\hookrightarrow M_{0}$ has a {\em $C^r$ ($r=0,1$) boundary at
$z\in
\partial_{i}M$} if $z$ admits a {\em $C^r$ chart of $M_{0}$ adapted to the boundary}, i.e.,
there exists an open neighborhood $V_0$ of $z$ in $M_0$ and a
$C^r$
 coordinate chart $\varphi: V_0 \rightarrow \R^N$ such
 that\[
\varphi(\overline M_{i}\cap V_0)=\varphi(V_0)\cap \R^N_+,\quad
\hbox{where}\;\; \R^N_+=\{(x^1, x^2,\dots ,x^{N})\in \R^N:
x^N\geq 0\}.
\]
If the envelopment $i$ has a $C^r$ ($r=0,1$) boundary for all
$z\in\partial_{i}^* M$ then it is called {\em envelopment with
$C^r$ ($r=0,1$) boundary}.

\edefi In the case of an envelopment of $C^{1}$ boundary at some
$z\in \partial_iM$, one says that a vector $v$ tangent to $M_{0}$
at $z$ {\em points out inwards} (resp. {\em points out outwards})
if $v^{N}>0$ (resp. $v^{N}<0$), where $(v^{1},\ldots,v^{N})\in
T_{0}{\mathbb R}^{N}$ is the expression of $v$ in terms of a chart
$(V_{0},\varphi)$ adapted to the boundary such that
$\varphi(z)=0$.

Notice that when  $i:M\hookrightarrow M_0$ is an envelopment with
$C^r$ boundary, only the accessible part of the conformal
completion $\overline{M}_{i}^*=M\cup
\partial_{i}^*M$ is required to be a $C^r$ manifold with boundary, and $\partial_{i}^*
M$ is then a hypersurface embedded in $M_0$.

\brema \label{r13}{\em

(1) The canonical conformal embedding of Lorentz-Minkowski in
Einstein static universe (see for example \cite{HE, Wald}), $i:
\LL^4\hookrightarrow \LL^1 \times \SSS^3$, is an envelopment with
$C^0$ boundary according to our definition. It is not $C^1$
because the points $i^\pm$ are included in
$\partial_{i}^{*}M=\partial_{i}^{*}\LL^4$. The spacelike infinity
$i^0$ is included in $\overline{M}_i$ (which is {\em not} a $C^0$
manifold with boundary) but it is not
included in $\overline{M}^*_i$. 

 (2) Assume that not only $\overline{M}^*_i$ but also all $\overline{M}_i$ is a $C^1$ manifold with
 boundary. Then, a simple argument based on partitions of the unity shows the existence of
 a smooth function $\Omega$ on all
 $M_0$ such that
 \be
 \Omega(p)>0 \Leftrightarrow p\in M, \quad \partial_iM= \Omega^{-1}(0), \quad d\Omega_p\neq 0 , \; \forall p\in\partial_iM .
 \ee
This  viewpoint in the definition of conformal completion appears
in, for example, \cite{Fr} and \cite{Wald}. Of course, in
favorable cases one can then restrict the function $\Omega$ on
$\overline{M}_{i}^{*}$ and consider this intrinsically as a $C^1$
manifold with boundary.
}
 \erema

\subsubsection{Regular and strong accessibility of a $C^1$
boundary} As discussed above, the differences between the strong
and weak versions of the chronological 
and causal 
relations defined in Remark \ref{rconfchr} yield problems in order
to identify the conformal boundary with the c-boundary. On one
hand, the differences between the two causal relations $\le_i,
\le_i^S$ are connected with the loss of timelike transitivity of
$\partial^*_i M$. On the other, the differences between the two
chronological relations $\ll_i, \ll_i^S$ may cause the loss of
timelike deformability, as we analyze now. In fact, the main
difficulty to ensure this last property is that some point of
$\partial^*_i M$ may be the $\hat \pi$ or $\check \pi$ projection
for different TIPs or TIFs. This will happen only if the timelike
curve which generates the TIP or TIF cannot be  extended as a
smooth timelike curve to $\partial^{*}_{i}M$.

\begin{definition}\label{s-a} (Strong accessibility)
Let $\gamma:[a,b)\rightarrow M$ be a future-directed (resp.
past-directed) timelike curve
 with $i$-endpoint $z\in
\overline{M}_i$. If, up to a reparametrization, $\gamma$ can be
smoothly extended to $b$ with  timelike velocity $\gamma'(b)\in
T_zM_0$, then $\gamma$ is called a future (resp. past) strongly
timelike curve. 

If $z$ is the $i$-endpoint of a future (resp. past) strongly
timelike curve, then $z$ is said {\em future (resp. past) strongly
accessible}.

A future (past) accessible point $z\in \overline{M}_{i}$ is {\em
strongly accessible} if it is future (resp. past) strongly
accessible.

The completion $\overline{M}_{i}$ is {\em strongly accessible} if
all the accessible points are strongly accessible.

%

\end{definition}
Recall that
 $z\in
\partial_iM$ is called strongly accessible when it is
an i-endpoint just for  one strongly timelike curve --or two, one
future-directed and one past-directed, in the case that $z$ is
future and past accessible. This is simpler than timelike
deformability, as will be stressed in Remark \ref{rbclm} (2).
\begin{figure}
\centering
\ifpdf
  \setlength{\unitlength}{1bp}%
  \begin{picture}(413.80, 125.46)(0,0)
  \put(0,0){\includegraphics{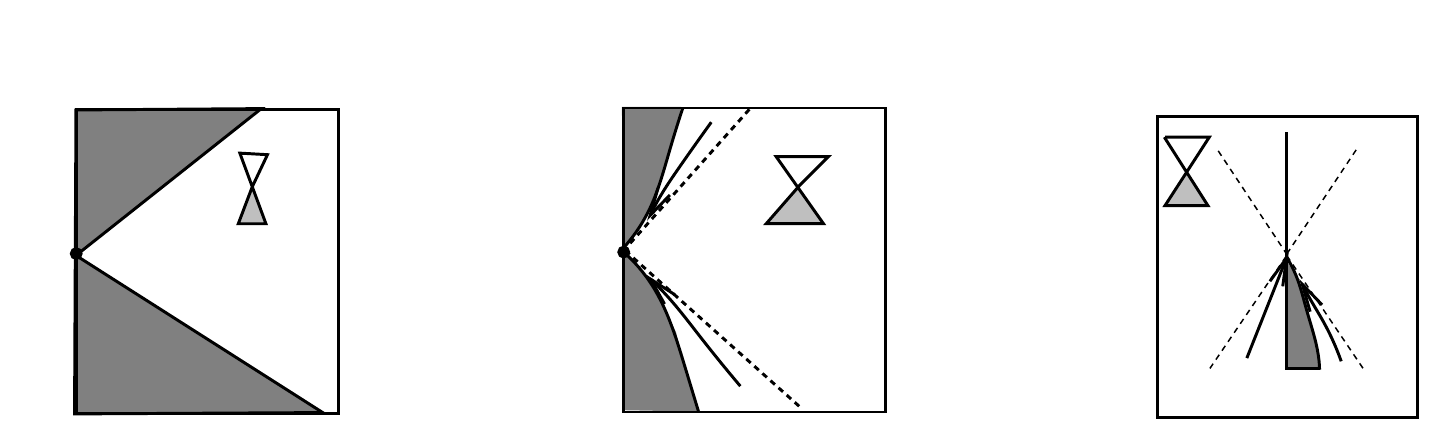}}
  \put(375.42,51.46){\rotatebox{0.00}{\fontsize{8.54}{10.24}\selectfont \smash{\makebox[0pt][l]{$z$}}}}
  \put(169.35,49.98){\fontsize{8.54}{10.24}\selectfont $z$}
  \put(11.67,49.54){\fontsize{8.54}{10.24}\selectfont $z$}
  \put(56.73,108.23){\fontsize{14.23}{17.07}\selectfont A}
  \put(214.41,108.67){\fontsize{14.23}{17.07}\selectfont B}
  \put(365.77,108.38){\fontsize{14.23}{17.07}\selectfont C}
  \end{picture}%
\else
  \setlength{\unitlength}{1bp}%
  \begin{picture}(413.80, 125.46)(0,0)
  \put(0,0){\includegraphics{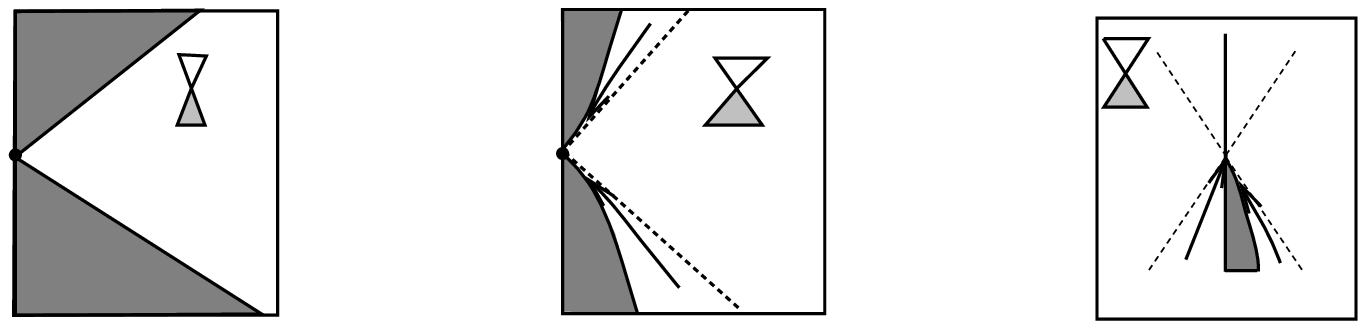}}
  \put(375.42,51.46){\rotatebox{0.00}{\fontsize{8.54}{10.24}\selectfont \smash{\makebox[0pt][l]{$z$}}}}
  \put(169.35,49.98){\fontsize{8.54}{10.24}\selectfont $z$}
  \put(11.67,49.54){\fontsize{8.54}{10.24}\selectfont $z$}
  \put(56.73,108.23){\fontsize{14.23}{17.07}\selectfont A}
  \put(214.41,108.67){\fontsize{14.23}{17.07}\selectfont B}
  \put(365.77,108.38){\fontsize{14.23}{17.07}\selectfont C}
  \end{picture}%
\fi \caption{\label{accesibilidad}In Figure (A), point $z$ is not
accessible, because it is not reachable by a timelike curve from
$M$. In Figure (B), any timelike curve reaching $z$ must be null
at $z$, i.e. $z$ is future and past accessible, but not strongly
accessible. In Figure (C), point $z$ is the endpoint of two
future-directed timelike curves, one of them is strongly timelike
and
the other one is not.}
\end{figure}

For $C^1$ boundaries, the conditions of accessibility and at least
future or past strong accessibility are not restrictive:

\begin{proposition}\label{l3} Assume that the conformal boundary is $C^1$ at some
$z\in\partial_{i} M$. Then $z$ is either future or past strongly
accessible.

In particular, if the envelopment has a $C^{1}$ boundary then all
the boundary is accessible, i.e.
$\partial_{i}M=\partial^{*}_{i}M$.
\end{proposition}
{\it Proof.}
Let $v \in T_zM_0$ be any timelike vector. 
If $v$ points out inwards, the required strongly timelike curve
can be chosen  as the geodesic starting at $z$ with initial
velocity $v$. If $v$ points out outwards,  the timelike vector
$-v$ lies in previous case. If $v\in T_{z}\partial_{i}M$, as the
set of all timelike vectors at $T_zM$ constitutes an open subset,
a small perturbation of $v$ reduces the problem to previous cases. \cvd

\begin{remark} {\em As non-accessible points cannot be $C^1$,
either they represent a very particular situation (the case of
$i^0$ for $\LL^4\hookrightarrow \LL^1 \times \SSS^3$) or become
rather irregular (see Figs. \ref{f6}, \ref{f3}). So, essentially,
the non-accessible part $\partial_iM\setminus\partial_i^*M$  not
only is a non-causal ingredient, but also its  definition would
depend on the convenience for the particular phenomenon considered
(asymptotic flatness, AdS/CFT correspondence), and would require a
specific study in the corresponding theory.}
\end{remark}

The first of the following two lemmas will be useful  to relate
the strong and weak chronological relations under strong
accessibility, and the second lemma to relate the causal ones.

\begin{lemma}\label{lpertu}
Let $\gamma:[a,b]\rightarrow \overline{M}^*_i$ be a
future-directed strongly timelike curve with $\gamma (a)=q\in M$
and $\gamma (b)=z\in \overline{M}^*_i$. If $\overline{M}_i$ is
$C^1$ at $z$, then there exists a neighborhood $U_{0}$ of $z$ in
$M_0$ such that $q$ can be joined with any $r\in U:=U_0\cap
\overline{M}_i$ by means of a future-directed strongly timelike
curve.

In particular, $\gamma$ is {\deform} and all the points in $U\cap
\partial_iM$ are future strongly accessible.
\end{lemma}
{\it Proof.} The non-trivial case appears when $z\in \partial_i
M$, so, consider a chart $(V_0 ,\varphi)$ adapted to the boundary
with M on the $x^N>0$ side and  $z=\varphi^{-1}(0)$. Up to a
reparametrization, write $\varphi(\gamma(t))=(x_1(t),...,x^N(t)),
t\in [0,1]$. As the velocity satisfies $g(\dot\gamma,
\dot\gamma)<0$ and $\dot\gamma([0,1])$ is a compact subset of
$TV_0 (\equiv \varphi(V_0)\times \R^N)$, there exists $\epsilon>0$
such that, for any $(x,y)\in \varphi(V_0)\times \R^N$ with
$|x^i-x^i(t)|< \epsilon$ and $|y^i-\dot x^i(t)|< \epsilon$ (for
some $t\in [0,1]$ and all $i=1,\dots,N$) one has
$\varphi_*g((x,y),(x,y))<0$. Now, consider any $r\in
U:=\varphi^{-1}([0,\epsilon)^N)$, and put $\varphi(r)=(\epsilon^1,
\dots , \epsilon^N)$ (where $0\leq \epsilon^i<\epsilon$). Then,
the curve $\gamma_r(t)=\varphi^{-1}((x^1(t)+t\epsilon^1, \dots ,
x^N(t)+t\epsilon^N))$ connects $q$ and $r$ fulfilling the required
properties. \cvd

\begin{lemma}\label{lcaus}
Assume that $z\in \overline{M}^*_i$ is future strongly accessible
and $C^1$. Then there exists a neighborhood $U_0\subset M_0$ of
$z$ such that, for all future-directed causal piecewise smooth
curve $\sigma:[0,c]\rightarrow U=U_0\cap \overline{M}_{i}$, there
exists a variation $\sigma_{\epsilon}:[0,c]\rightarrow U$ such
that $\sigma(c)=\sigma_{\epsilon}(c)$
and $\sigma_{\epsilon}$ is future-directed strongly timelike for all $\epsilon$.
\end{lemma}
{\it Proof.} Consider a past-directed timelike vector field $T$
defined on some neighborhood of $\sigma$ in $\overline{M}_i$ which
points out inwards $M$ on the points of the boundary (say, $T$ is
obtained by extending an inward pointing timelike vector of $T_z
M$, whose existence is ensured in the proof of Prop. \ref{l3}).
Let $-\sigma(s):=\sigma(c-s)$  the reversed curve of $\sigma$. Any
variational field $V=\mu T$ on $-\sigma$ with $\mu=0$ at
$-\sigma(0)$ and $\mu'>0$ everywhere, will yield curves contained
in $M$ except at the $i$-endpoint $\sigma(c)$ (which remains
fixed). According to \cite[Lemma 10.45]{O}, if the covariant
derivative $V'$ of $V$ along $-\sigma$ satisfies
$g_0(V',(-\sigma)') <0$, the longitudinal curves of the variation
close to $\sigma$ are strongly timelike. In order to ensure this
property, notice that, $g(T,(-\sigma)')<-\epsilon$ and
$|g(T',(-\sigma)')|<1/\epsilon$ on $[0,c]$ for some $\epsilon>0$
small enough. As $V'=\mu' T+\mu T'$,
\[
g(V',(-\sigma)')=\mu' g(T,(-\sigma)')+\mu g(T',(-\sigma)') <
-\epsilon\mu' +\mu/\epsilon .
\]
Choosing $\mu(s)=e^{s/\epsilon^2}-1$, we have $\mu(0)=0$ and
$\mu'> \mu/ \epsilon^2$, 
i.e., the required inequality follows. \cvd


\begin{definition}
An envelopment $i:M\hookrightarrow M_{0}$ with $C^1$ boundary is
{\em \tamech} (resp. {\em  \tameca}) if
 $\ll_i\equiv\ll_i^S$
(resp. $\leq_i\equiv\leq_i^S$).
\end{definition}

\begin{remark}\label{rrrr1}
{\em Even though these definitions are stated from a global
viewpoint, it is enough to check them from a local one, which may
be easier in practice. That is, the envelopment is {\em  \tamech}
(resp. {\em \tameca}) when
 for all
$z\in \overline{M}_i$, there exists a neighborhood
$V=\overline{M}_i\cap V_0$ such that $V$, regarded as an
envelopment of $M\cap V_0$ in its own right, is  {\tamech} (resp.
\tameca). This is obvious for chronological tameness, and can be
checked for the causal one as follows. Assume that $x\leq_i y$.
From local causal tameness one can construct a piecewise smooth
future-directed causal curve $\sigma$ in $\overline{M}_i$ from $x$
to $y$. Let $\sigma(t_0)$ be a break of $\sigma$. A standard
argument shows that it can be smoothed if $\sigma(t_0)\in M$ or if
one of the tangent vectors at the break $\sigma'(t_0^+),
\sigma'(t_0^-)$ is timelike (see for example the proof of Prop.
10.46 in \cite[p. 235]{O}). If both vectors are lightlike and
$\sigma(t_0)\in
\partial_iM$, the argument also works by taking into account
that either $w=\sigma'(t_0^+)-\sigma'(t_0^-)$ points out inwards
$M$ or it is tangent to $\partial_iM$ (in the latter case, a close
vector $\overline{w}$ points out inwards and satisfies
$g(\overline{w},\sigma'(t_0^-))<0,
g(\overline{w},\sigma'(t_0^+))>0$, which will be enough for the
job). }
\end{remark}

\begin{theorem}\label{t10}
A chronologically complete envelopment $i:M\hookrightarrow M_0$
with $C^1$ boundary, {\tamech} and {\tameca} is regularly
accessible. So, its conformal completion and c-completion are
equivalent (in the sense of Theorem \ref{t1}).
\end{theorem}
{\it Proof:} Let $z\in \partial_i^*M$. To check  that $z$ is
\deform, consider, say, a TIP $P$ asociated to $z$, and let $\gamma$ be a future-directed timelike curve in $M$ with endpoint $z$
such that $P=I^-[\gamma]$. Notice that, for all  $x=\gamma(t)$
chronological tameness implies
  $x\ll_i^Sz$, that is, there exists a connecting  strongly timelike curve $\alpha$. Then,  Lemma \ref{lpertu} can be applied to this curve,
showing that  $\gamma$ fulfills the definition of timelike
deformability (Defn. \ref{d1}).

 To check that  $z$ is timelike transitive, let $U$ be a neighborhood as in Lemma \ref{lcaus}
 and take any $x,y\in U$ such that
$x\ll_iz\leq_iy$ (the other case is analogous). By  tameness
$x\ll_i^Sz\leq_i^Sy$ and, so, there exists a piecewise smooth
causal curve $\sigma: [0,2]\rightarrow \overline{M}_i$ from $x$ to
$y$ such that $\sigma(1)=z$ and $\sigma|_{[0,1]}$ is timelike. By
this last property, some timelike curve $\sigma_{\epsilon}$ close
to $\sigma\mid_{[1,2]}$, and provided by Lemma $\ref{lcaus}$, can
be modified close to $0$ in order to ensure $x\ll_i y$. \cvd

\smallskip

 Fig. \ref{f7}
is an example of envelopment with conformal boundary different to
its c-boundary (at the topological level) because of the lack of
chronological tameness. In order to apply Theorem \ref{t10}, we
are going to show that chronological tameness can be replaced by
strong accessibility, which is more easily computable.
\begin{lemma}\label{cregu} Assume that $i$ is causally tame.
If  $z\in \partial_{i}^*M$ is a future strongly accessible $C^1$
boundary point, then all future-directed timelike curves
$\sigma:[a,b)\rightarrow M$ with
 $i$-endpoint $z$, have the same chronological past.
Moreover, if $\partial_{i}^*M$ is $C^1$ and strongly accessible
then $i$ is chronologically tame.
\end{lemma}
{\it Proof.} Fix a future-directed  strongly timelike curve
$\hat\sigma:[a,b)\rightarrow M$ with  $i$-endpoint $z$, and let
$P=I^-(\sigma),\hat P=I^-(\hat\sigma)$.  From Lemma \ref{lpertu},
$I^+(\hat\sigma(t))$ contains a neighborhood of $z$ for each $t$,
and thus, it also contains some point of $\sigma$. So,
$\hat\sigma(t) \in I^-[\sigma]$ and $\hat P\subset P$.

In order to prove the reversed inclusion, take an arbitrary $q\in
P$. By causal tameness, $q$ can be joined with $z$ by means of a
smooth future-directed causal curve $\gamma:[a,b)\rightarrow M$
which is smoothly extensible to $b$. Then, consider two cases:

(a) $\gamma$ is extensible to a strongly timelike curve. In this
case, the roles of $\gamma$ and $\hat \sigma$ are interchangeable,
and $q\in I^{-}[\gamma]=I^{-}[\hat{\sigma}]=\hat P$, as required.

(b) $\gamma$ is only extensible as a smooth  curve at $b$
(necessarily causal in $M_0$). Lemma \ref{lcaus} allows to find
arbitrarily close strongly timelike curves $\gamma_\epsilon$
depending on the parameter $\epsilon$ and, from the previous case,
$I^-[\gamma_\epsilon]=\hat P$. Now, the open set $I^+(q)$ contains
 some point
in $\gamma_\epsilon$ for $\epsilon$ small enough. Therefore, $q\in
I^-[\gamma_\epsilon]=\hat P$, as required.

In order to check the last assertion, let $p,q\in \overline{M}_i$
such that $p\ll_iq$.
There exists a future-directed timelike curve $\sigma$ joining
them. The previous proof shows $I^-[\sigma]=I^-[\gamma]$ for any
future-directed strongly timelike curve $\gamma$ with $i$-endpoint
$q$. This property ensures the existence of a strongly timelike
curve joining $p$ and $q$, i.e, $p\ll_i^Sq$. \cvd

\smallskip

Summing up, from Theorem \ref{t10} and Lemma \ref{cregu}:

\begin{corollary}\label{t12}
The conformal and c-completions are equivalent  if the conformal
boundary is $C^1$, strongly accessible and causally tame.

\end{corollary}

\begin{remark} \label{rbclm}
{\em 
(1) Theorem \ref{t10} and Corollary \ref{t12} yield sufficient
conditions to ensure regular accessibility, which become
reasonably general and easy to compute. In fact, the sufficient
conditions ($C^1$ boundary, causal tameness and either
chronological tameness or just strong accessibility) are local,
recall Remark \ref{rrrr1}. So, one can determine first  the points
were these conditions are fulfilled (typically,  most of them) and
then concentrate on the boundary points were they are not.
For example, in the canonical envelopment of $\LL^4$ in Einstein
static universe, the accessible boundary $\partial^*_{i}\LL^4$ is
$C^1$ and satisfies the hypotheses of Corollary \ref{t12} at all
the points but $i^\pm$ (see also Theorem \ref{ttt1} below). And,
easily, $i^\pm$ satisfy the definition of regular accessibility.

(2) About the hypotheses of Corollary \ref{t12},  we emphasize
first that they are very easy to check in practice. Typically,
$C^1$ smoothability can be checked by inspection. As a difference
with regular accessibility, the strong accessibility of some $z\in
\partial_i M$ must be checked only when $z$ is accessible and, then,
one must check it {\em at most for two cases}, the future and past
direction (if $z$ is both, future and past accessible). This is an
important simplification in comparison with {\em timelike
deformability} (one of the two conditions for regular
accessibility), as this condition has to be checked for all the
TIPs and TIFs which project onto $z$. For example,  Fig.
\ref{accesibilidad} (C) shows a strongly accessible
$z\in\partial_i^*M$ which is not timelike deformable, as none of
its associated TIPs can be defined by means of a deformably
timelike curve (this is possible because $z$ is not $C^1$). Notice
also that both conditions, $C^1$ and strong accessibility, are
independent and necessary for Corollary \ref{t12}, as Figs.
\ref{f5} and \ref{f7} show easily.

About causal tameness, notice that it may fail only when there
exists some curve  $\gamma:[a,b]\rightarrow \overline{M}_i$
satisfying: (a) $\gamma$ is not smooth at some point (perhaps one
endpoint), (b)  $\gamma$ is continuous causal, regarded  as a
curve in $M_0$, and (c) no smooth causal curve contained in
$\overline M_i$ joins its endpoints. It is easy to construct an
example of this situation if the boundary is not $C^1$ (recall,
for instance, Fig. \ref{fig2''}). Under our additional hypotheses
($C^1$ and strong accessibility), it is conceivable the existence
of such an example. Nevertheless, to construct explicitly one
seems really difficult, and its existence should be regarded as a
very pathological case which would not happen in physical
examples. In fact, it is not necessary to check causal tameness in
the cases  to be studied below (globally hyperbolic spacetimes and
asymptotic conformally flat ends). This also happens in the two
dimensional case and, so, when Penrose-Carter diagrams are studied
(as one focuses on a Lorentzian surface, dropping other two
dimensions).}
\end{remark}

\subsubsection{$C^1$ conformal boundary with no timelike points}

Next, our aim is to exploit the causal character of the tangent
hyperplane to $\partial_iM (=\partial_i^*M)$ for a $C^1$ boundary.

\begin{definition}\label{dtne} Let $i:M\hookrightarrow M_{0}$ be
an envelopment with $C^{1}$ boundary. A point $z\in \partial_iM$
is spacelike (resp. null; timelike) if $T_z(\partial_{i}M)$ is
spacelike (resp. degenerate; timelike).
\end{definition}
\begin{proposition}\label{prop}
Any timelike (resp. spacelike) point of a $C^1$ conformal boundary
is past and future (resp. either past or future) accessible and,
then, strongly accessible.
\end{proposition}
{\it Proof.} Let $z\in \partial_iM$. If $z$ is timelike, there
exist both, a future-directed  and a past-directed timelike vector
in $T_zM_0$ which  points out inwards $M$. Therefore,  reasoning
as in the proof of Prop. \ref{l3}, we find that $z$ is not only
both, future and past accessible, but also  future and past
strongly accessible. If $z$ is spacelike, any causal vector at
$T_zM_0$ points out inwards for one time-orientation and outwards
for the reversed one. This implies that $z$ is only future or past
accessible --and then strongly accessible, notice also Prop.
\ref{l3}. \cvd


\smallskip

Next, our aim is to prove the following theorem.
\begin{theorem}\label{ttt1}
Let $i:M\hookrightarrow M_0$ be a chronologically complete
envelopment with $C^1$ boundary. If the boundary  $\partial_iM
(=\partial_i^*M)$ has no timelike points, then it is causally tame
and strongly accessible. Therefore,  $\partial_iM$ is regularly
accessible and the conformal and c-completions are equivalent.
\end{theorem}

Before starting the proof, recall that if $z\in  \partial_iM $ is
null or spacelike, one of the time-cones (future or past) is
pointing inwards  $M$. Moreover, due to the continuity of the
cones and the fact that there are no timelike points,  we can
choose a neighborhood $W_0 (\subset M_0)$ of $p$ such that
$W_0\backslash
\partial_iM$ has two connected components (one of them $M\cap W_0$), and
all the cones with the same time-orientation at any $q\in W_0\cap
\partial_iM$ point out the same
direction (outwards or inwards). 
\begin{lemma} \label{lll1}
Under previous assumptions, any smooth timelike curve
$\gamma:[a,b]\rightarrow W_0\subset M_0$  touches the boundary
$\partial_iM$ at most once.
\end{lemma}
{\it Proof:} Assume that $\gamma$ touches $\partial_{i}M$ at least
twice. As $\partial_iM$ has no timelike points, it is crossed
transversally by $\gamma$ and,  without loss of
generality, we can assume   that $\gamma$  
touches the boundary exactly at $\gamma(a)$ and $\gamma(b)$. Then,
$\gamma'(a),\gamma'(b)$ has the same time-orientation, but point
out different directions (inwards and outwards), in contradiction
with the definition of $W_0$.\cvd

\smallskip

{\it Proof of Theorem \ref{ttt1}: } Choose $W_0$ as above and
assume without lost of generality that the past cones point out
inwards at any $q\in W_0\cap
\partial_iM$.

In order to prove strong accessibility, notice that all the points
in $\partial_iM\cap W_0$ are future strongly accessible. So, it is
enough to prove that they are not past accessible,  even at any
null point of $\partial_iM\cap W_0$. Assume by contradiction that
$\sigma$
is a past-directed timelike curve in $M\cap W_0$ 
with $i$-endpoint $p\in\partial_iM\cap W_0$. Let $V_0$ be a
globally hyperbolic neighborhood of $p$ in $M_0$ contained in
$W_0$ (for their existence, see for example \cite[Th. 2.14]{MS})
and consider $t_0$ such that $\sigma(t_0)$ is contained in $V_0$.
By the globally hyperbolic character of $V_0$, there exists a
smooth past-directed timelike curve $\gamma:[0,1]\rightarrow V_0$
joining $\sigma(t_0)$ and $p$. As the previous lemma is applicable
to $\gamma$, $\gamma([0,1))\subset V_0\cap M$ and, then, the
timelike vector $\gamma'(1)$ points out outwards. This contradicts
 that the past time-cone in $p$ points out inwards.

In order to prove that the envelopment is causally tame, consider
again $z\in \partial_iM$ and a neighborhood $V_0\subset M_0$ as
above. We can also assume that all the causal  geodesics  in
$\overline{V_0}$ maximize the (Lorentzian) distance for $g_0$.
Consider a point $q\in V_0\cap \overline{M}_i$ such that $q\leq_i
p$  (the case $p\leq_i q$ is analogous). This implies the
existence of a causal past-directed maximizing geodesic
$\gamma:[0,1]\rightarrow V_0$. If $\gamma$ is lightlike, then the
only possible continuous causal curve in $M_0$ joining $p$ and $q$
is (up to a reparametrization) $\gamma$ and, from the fact that
$p\leq_iq$, necessarily $\gamma\subset \overline{M}_i$, i.e.,
$p\leq_i^Sq$. If $\gamma$ is timelike, previous lemma ensures that
$\gamma\subset \overline{M}_i$ (notice that the past cone of $p$
points out inwards) and again $p\leq_i^S q$ (moreover, $p\ll_i^S
q$). \cvd


\smallskip

As  straightforward  consequences of these results, we can obtain
the following two assertions, which are more or less implicitly
assumed by researchers. About the first one, the implication to
the right is well-known, but we are not aware of a rigorous proof
for the implication to the left.
\begin{corollary}\label{cfinal} A spacetime which
admits a chronologically complete envelopment with $C^1$ boundary
$\partial_iM (=\partial_i^*M)$ is globally hyperbolic if and only
if $\partial_iM$ does not have timelike points.
\end{corollary}
{\it Proof.} ($\Rightarrow$). Take a timelike point $z$ of the
boundary. From Prop. \ref{prop}, it is future and past (strongly)
accessible. Take a future and a past-directed timelike curve in
$M$ with $i$-endpoint $z$, and fix points $q_{1}$, $q_{2}$ at each
curve. By construction, $J^+(q_{1})\cap J^-(q_{2})$ is not
compact.

($\Leftarrow$). Assume by contradiction that $M$ is not globally
hyperbolic and, so, the $c$-boundary admits a pair $(P,F)\in
\partial M$ with $P\neq \emptyset\neq F$ (Theorem \ref{h}). By Theorem \ref{ttt1},
$z=\pi((P,F))$ is well-defined and it is the $i$-endpoint of both,
a past and a future-directed strongly timelike curve. These curves
yield both, a future and a past-directed timelike vector at $z$
pointing outwards, in contradiction with the inexistence of
timelike points. \cvd

\smallskip

The second consequence of Theorem \ref{ttt1}  concerns the
boundary of conformally flat and asymptotically conformally flat
ends (see Section \ref{s3.7}). Obviously, there are no timelike
points in the null part of the canonical conformal boundary
$\partial_i\LL^4$. So, this part agrees with a part of its
c-boundary (recall Remark \ref{rbclm}), and can be regarded also
as intrinsic for any conformally flat end. This intrinsic
character also holds for any asymptotically conformally flat end
$E$. More precisely,  the existence of some conformal envelopment
without timelike points is assumed in the axioms for asymptotic
flatness at null infinity (and spacelike infinity), see
\cite{Wald, AH}. The corresponding part $\partial_i^{{\rm asym}}E$
of the accessible conformal boundary $\partial^*_iM$ will agree
with a part of the causal boundary $\partial M$. From a formal
viewpoint, $\partial_i^{{\rm asym}}E$ can be more precisely
defined. Let $E$ be an asymptotically conformally flat end, i.e.,
$E$ is a suitable connected component of $M\backslash K$, for some
compact $K\subset\overline{M}$ (Section \ref{s3.7}). In principle,
the part $\partial_i^{{\rm asym}}E$ of the boundary of $E$ can be
distinguished from the remainder of $\partial_i^{*}E$ (which
involves points in the boundary of $K$, and must not be regarded
as asymptotic) because the axioms for asymptotic flatness
distinguish a null boundary. At any case, $\partial_i^{{\rm
asym}}E$ is equal to
$\lim_{n\rightarrow\infty}\partial_i(M\backslash K_n)$, where
$\{K_{n}\}_{n}$ is a sequence of compact subsets such that each
$K_{n+1}$ is included in the interior of $K_{n}$, and $K=\cap_{n}
K_{n}$. Here, $\partial^*_i(M\backslash K_n)$ denotes the
conformal boundary of $M\backslash K_n$ for the fixed conformal
envelopment $i$ which defines asymptotic conformal flatness, and a
point $z\in M_0$ belongs to the limit of the boundaries if
$z\in\partial^*_i(M\backslash K_n)$ for all $n$ greater than some
$n_z$. Summing up:

\begin{corollary} Let $E$ be an asymptotically conformally flat end.
The points in $\partial_i^{{\rm asym}}E$  are regularly accessible
and, then, they correspond with a subset of the c-boundary
$\partial M$.
\end{corollary}

\subsection{Appendix: Conformal boundary fauna}\label{sA2}

Previous results on the conformal boundary simplify the
computation of the c-boundary in some particular cases
(conformally flat, those which admit a natural Penrose-Carter
diagram). Nevertheless, more specific techniques are needed for
general classes of spacetimes, where a conformal embedding is not
expected. This is the case of plane wave type \cite{FS}, static
\cite{AF, FH} or stationary \cite{FHSst} spacetimes (as well as
spacetimes conformal to them, which include other well-known
families as, for example, Generalized Robertson-Walker
spacetimes).

However, when the conditions which ensure the equivalence between
the conformal and the c-boundary are not fulfilled, the conformal
boundary may present a very erratic behavior. Here, we collect
some examples, which are instructive in order to understand our
previous results. They are quite simple, as we consider always an
envelopment $i:M\hookrightarrow M_0$ where the aphysical space
$M_0$ is always $\LL^2$, $\LL^3$ or some trivial quotient, with
the timecones oriented as depicted in the figures.
\begin{itemize}
\item[Figure \ref{f1}:] Consider $M=\LL^2\setminus\{(x,t):x=0,t\in
[-1,1]\}$ and $M_0=\LL^2$. The depicted point $q$ corresponds with
two points of the c-boundary, $(P_1,F_1),
(P_2,F_2)$. Trivially, $p\ll_i q\ll_i r$, but $p\not\ll_i r$.

\item[Figure \ref{f2}:] The artificial character of the
identifications of the two points $(P_j,F_j)$, $j=1,2$, in
previous example is stressed in the following one. Consider the
spacetime $(M,g)$ given by $M=(0,2\pi)\times \R$,
$g=dx^{2}-dt^{2}$. If we take the natural inclusion
$i:M\hookrightarrow \LL^{2}$,  the conformal boundary
$\partial_{i}M$ is obviously $M=\{0,2\pi\}\times\R$,
and it can be identified with the c-boundary. Nevertheless,
regarding $x$ as an angular coordinate, one finds a conformal
embedding $\tilde{\i}$ of $M$ into the cylinder $C=(\SSS^{1}\times\R,d\theta^{2}-dt^{2})$ such that $\tilde{\i}(M)=(\SSS^{1}\setminus \{\theta=0\})\times\R$. So, the conformal boundary is
now a quotient of previous one, $\partial_{\tilde{\i}}M \equiv
\partial_{i}M/\sim$, where $(0,t)\sim (2\pi,t)$ for all $t\in
\R$, which shows the non-intrinsic character of this boundary.
Notice that $\partial_iM$ is a smooth hypersurface of $M_0$, and
strongly accessible. Nevertheless, $\tilde{\i}:\overline{M}_{\tilde{\i}}\rightarrow
M_0$ is not an envelopment with $C^0$ boundary $\partial_{\tilde{\i}}M$,
according to Defn. \ref{decb0}. As in previous figure,
$\ll_{\tilde{\i}}$ is not transitive.

\item[Figure \ref{f5}:] In the two previous figures, some points of the c-boundary
were artificially identified to a single point in the conformal
boundary. In the following one,  there is also an identification,
but it seems much more natural, as it yields a Hausdorff boundary.
Let $M_0=\LL^{2}$ and define $M$ just removing the subset
$C=\{(x,t):t\geq |2x|\}$. Now, consider the following two
past-directed strongly timelike curves:
\[
\gamma_{1}:[-1,0)\rightarrow M,\,
\gamma_{1}(x)=(x,-\frac{3}{2}x)\quad \gamma_{2}:[0,1)\rightarrow
M,\, \gamma_{2}(x)=(1-x,\frac{3}{2}(1-x)).
\]
Obviously, these two curves converge in $\overline{M}_i$ to the
single point $(0,0)$. However, $\gamma_i$ determine two different
c-boundary points, namely $(P_0,F_1), (P_0,F_2)$ where
$F_j=I^+[\gamma_j], j=1,2$ and $P_0=I^-[\gamma_0]$. These two
points are non-Hausdorff related in the c-completion
$\overline{M}$. According to Harris \cite{H2}, $P_0$ is not a {\em
regular} point of the future chronological completion $\hat M$ of
$M$, because the future of $P_0$ is not an indecomposable future
set. In the original construction by Geroch, Kronheimer and
Penrose \cite{GKP}, the Hausdorffness of the c-boundary was
imposed a priori. However, this imposition implied very
undesirable consequences (see, for example, \cite{S}).

\item[Figure \ref{f6}:] The lack of uniqueness and the ambiguities
of a general conformal boundary are stressed here. In fact, the
open subset $M$ of $M_0=\LL^2$ inherits a conformal boundary
$\partial_{i   }M$ through its natural inclusion $i$. The behavior
of $\partial_{i   }M$ is very different to the behavior of the
c-boundary $\partial M$. This boundary seems to be very
unappropriate, especially from the topological viewpoint, because
the sequence of points $\{p_n\}_n$ converges to $q$ in $\LL^2$,
but the corresponding sequence $\{P_n=(I^-(p_n),\emptyset)\}_n$ in
the c-boundary converges to $P=(I^-(p),\emptyset)$. Nevertheless,
one can find a second envelopment $\tilde{\i}: M\rightarrow
M_0=\LL^2$. Under this envelopment, the behavior of $\partial M$
agrees with the expected one from the conformal boundary
$\partial_{\tilde{\i}}M$, namely, ${\tilde p_n:=
\tilde{\i}(p_n)}\rightarrow \tilde p:=\tilde{\i}(p)$. In fact,
$\partial_{\tilde{\i}}M$ and $\partial M$ are essentially
equivalent (the differences between them appears in elements which
are not relevant for the example: some identifications in the
removed vertical lines, similar to those in Fig. \ref{f1}, and the
non-accessibility of the vertexes $(2,0), (0,2)$).

More precisely, $M$ is constructed from $\LL^2$ in null
coordinates as:
\[
\begin{array}{l}
M=\{(u,v)\in (0,2)\times (0,2)\}\setminus (\cup_{n}L_{n}),\quad
L_{n}:=\{(2-\frac{1}{n},v),\, v\in (1,2)\}\;\;\forall n\in \N,
\end{array}
\]
so that $(P_n,\emptyset),(P,\emptyset)\in \partial M$  project on
$p_n,p\in \overline{M}_i$. The convergence with the chr. topology
of $\{(P_{n},\emptyset)\}_n$  to $(P,\emptyset)$ is
straightforward from Defn. \ref{overline}. As
$\{p_{n}=\pi((P_n,\emptyset))\}_n\rightarrow q \in \partial
M\backslash
\partial^*M$, the projection $\pi:\overline{M} \rightarrow \overline{M}^*_i$ is well-defined but
non-continuous.


Now, consider a second envelopment
$\tilde{\i}:M\rightarrow\LL^{2}$ given by
$\tilde{\i}(u,v)=(u,h_{n(u)}(v))$. Here, the nonnegative integer
$n(u)$ is determined by $2-\frac{1}{n(u)}<u\leq
2-\frac{1}{n(u)+1}$ and $h_n:[0,2]\rightarrow [0,1+\frac{2}{n+2}]$
is a smooth function with $h'_n>0$ and $h_{n}(v)=v$ for all $v\in
[0,1+\frac{1}{n+2}]$. Clearly, $\tilde{\i}$ is a conformal map,
because it maps lightlike lines in lightlike lines. Putting
$\tilde{p}= \tilde{\i}(p)$, $\tilde{p_n}= \tilde{\i}(p_n)$ $\tilde
P=\tilde{\i}(P)$, $\tilde P_n=\tilde{\i}(P)$ one has $\{(\tilde
P_{n},\emptyset)\}_n \rightarrow (\tilde P,\emptyset)$ and the
projection onto de new conformal boundary $\tilde\pi:\partial
M\rightarrow
\partial^*_{\tilde{\i}}M$ becomes continuous:
$\{\tilde \pi((\tilde P_n,\emptyset))\}_n= \{\tilde
p_{n}\}_n\rightarrow \tilde p= \{\tilde \pi((\tilde
P,\emptyset))\}$.

 \item[Figure \ref{f3}:] In this example one can see easily that  the projection $\pi: \partial M \rightarrow \partial_iM$
 is not always well-defined, even for a chronologically complete
 envelopment. 
Notice that, when $(P,F)\in \partial M$ satisfy $P\neq
\emptyset\neq F$, $\pi$ is defined as $\pi ((P,F))=\hat \pi (P)=
\check \pi (F)$. But this last equality does not hold for the
depicted TIP $P$ and TIF $F$, as $\hat \pi (P)=(0,0), \check \pi
(F)=(0,1)$.

Formally $M_0=\LL^2$ in null coordinates $(u,v)$ and
$M=\R^{2}\setminus (\cup_{n} L_{n}\cup H)$, where
$L_{n}=\{(-\frac{1}{n},v):0\leq v\leq 1+\frac{1}{n}\}$ and
$H=\{(u,v):u\geq 0, v\leq 1\}$. The points $(0,0),(0,1)\in M_0$
are associated to the TIP $P=I^-((0,0))$ and the TIF
$F=I^+((0,1))$, resp. Notice that $P \sim_S F$ in $M$ (in fact,
$P=\downarrow F, F=\uparrow P$), i.e., $(P,F)\in \partial M$
because of the removed elements of $M_0$.

\item[Figure \ref{f4}:] As in previous figure, this example shows
that $\pi$ is not always well-defined. Nevertheless, this case is
more sophisticated because the envelopment has now a $C^0$
boundary.

Let $M_0=\LL^{3}$ in natural coordinates, and consider only the
region $x>0$. In this region, $M$ is obtained by removing the
following two subsets:
$$\begin{array}{l}\{(x,y,t):(x,y,t)\in J^+((0,0,0))\cap \{(x,y,t):
y\leq 1\}
\},\\
 \{(x,y,t):(x,y,t)\in J^-((0,1,1))\cap \{(x,y,t): y\geq 0\}.\end{array}$$
 Here, the points $(0,0,0),
(0,1,1) \in
\partial_iM$ are reachable by a future-directed and past-directed
timelike curve resp. They determine, resp. a TIP $P$ and a TIF $F$
(not drawn). Taking into account that the coordinate $u=t+y$ is
null, one sees that $P\subset \downarrow F$ and $F\subset \uparrow
P$. Moreover, the suppressed subsets remove the TIPs and TIFs,
which violate the maximality of $P$ in $\downarrow F$, and
viceversa, that is, $P\sim_S F$. Summing up, $(P,F) \in \partial
M$. But $\check{\pi}(F)=(0,1,1)\neq (0,0,0)=\hat{\pi}(P)$.

\item[Figure \ref{f7}:] This very simple example stresses the
differences between the topologies of the conformal and
c-boundaries, even for a $C^1$ envelopment (when strong
accessibility is not ensured).

Consider the following open subset of $\LL^2$,
\[
M=\{(u,v)\in \R^{2}:u<0\}\cup \{(u,v)\in\R^{2}:u<v^2, u\geq
0\},
\]
with constant null cones depicted in the figure. The natural
inclusion in $M_0=\LL^2$ is an envelopment with $C^{1}$ boundary.
The points $(0,v)$ with $v<0$ correspond to pairs of the form
$(P,\emptyset)$. The point $(0,0)$ corresponds to a pair
$(P_{0},F_{0})$ (observe that $(0,0)$ is reachable by a
past-directed timelike curve, but it is not strongly accessible).

Consider the sequence $\{(0,v_{n})\}_{n}$ where $v_{n}\nearrow 0$.
This sequence in $\overline{M}_i$ is the projection of a sequence
$\{(P_{n},\emptyset)\}_n$ in $\overline{M}$. Obviously,
$\{(0,v_{n})\}_n \rightarrow (0,0)$ with the topology induced from
$\LL^{2}$, but the sequence $\{(P_{n},\emptyset)\}_n$ does not
converge to $(P_{0},F_{0})$ with the chronological topology.
Moreover, choose $q\in F_0$ and recall that $(P_0,F_0)\in
I^-(q,\overline{M}_i)$ but $(P_n,\emptyset)\not\in
I^-(q,\overline{M}_i).$

In fact, this last behavior is a consequence of the following
properties of $\partial M$: (i) $(P_0,F_0)\overline{\ll}q$ and
(ii) $I^-(q,\overline{M})$ is open. So, one must drop one of the
following elements in the conformal completion: (a) the relation
$\ll_i$ as the chronology of  $\overline{M}_i$, (b) the induced
topology from $M_0$ as the topology of  $\overline{M}_i$, or (c)
the open character of chronological futures and pasts in
$\overline{M}_i$. Elements (a) and (c) are intrinsic to $M$, but
(b) is extrinsic.

 \end{itemize}

\begin{figure}
\centering
\ifpdf
  \setlength{\unitlength}{1bp}%
  \begin{picture}(137.02, 161.38)(0,0)
  \put(0,0){\includegraphics{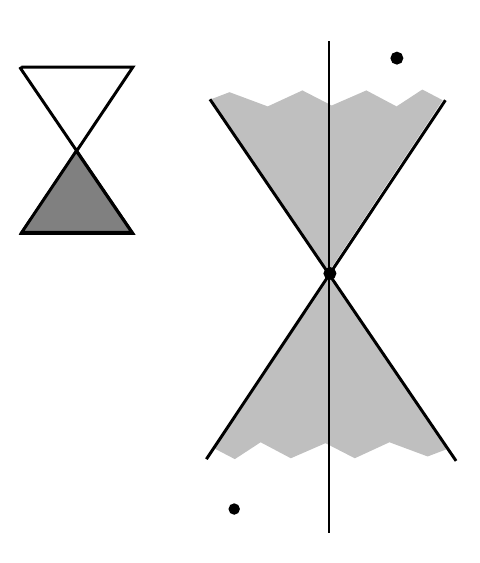}}
  \put(53.64,7.57){\fontsize{8.83}{10.60}\selectfont $p$}
  \put(99.62,80.94){\fontsize{8.83}{10.60}\selectfont $q$}
  \put(115.64,148.81){\fontsize{8.83}{10.60}\selectfont $r$}
  \put(73.78,39.49){\fontsize{8.54}{10.24}\selectfont $P_1$}
  \put(77.36,116.00){\fontsize{8.54}{10.24}\selectfont $F_1$}
  \put(97.32,116.26){\fontsize{8.54}{10.24}\selectfont $F_2$}
  \put(96.81,39.75){\fontsize{8.54}{10.24}\selectfont $P_2$}
  \end{picture}%
\else
  \setlength{\unitlength}{1bp}%
  \begin{picture}(137.02, 161.38)(0,0)
  \put(0,0){\includegraphics{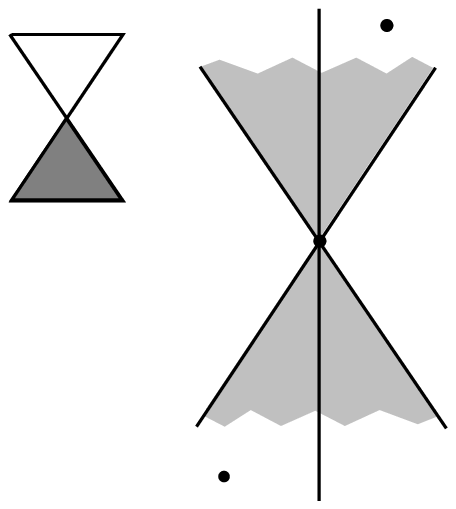}}
  \put(53.64,7.57){\fontsize{8.83}{10.60}\selectfont $p$}
  \put(99.62,80.94){\fontsize{8.83}{10.60}\selectfont $q$}
  \put(115.64,148.81){\fontsize{8.83}{10.60}\selectfont $r$}
  \put(73.78,39.49){\fontsize{8.54}{10.24}\selectfont $P_1$}
  \put(77.36,116.00){\fontsize{8.54}{10.24}\selectfont $F_1$}
  \put(97.32,116.26){\fontsize{8.54}{10.24}\selectfont $F_2$}
  \put(96.81,39.75){\fontsize{8.54}{10.24}\selectfont $P_2$}
  \end{picture}%
\fi \caption{\label{f1}}
\end{figure}

\begin{figure}
\centering \ifpdf
  \setlength{\unitlength}{1bp}%
  \begin{picture}(308.90, 130.82)(0,0)
  \put(0,0){\includegraphics{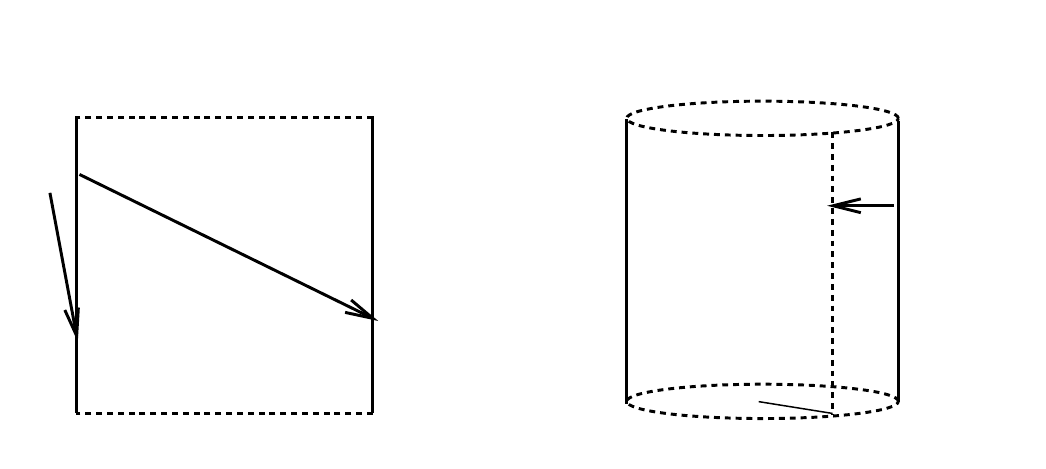}}
  \put(42.45,117.97){\fontsize{10.69}{12.83}\selectfont $(M,g)$}
  \put(197.94,117.50){\fontsize{10.69}{12.83}\selectfont $(M',g')$}
  \put(0.67,80.92){\fontsize{8.54}{10.24}\selectfont $\partial_{i}M$}
  \put(227.65,6.51){\fontsize{5.69}{6.83}\selectfont $\theta=0=2\pi$}
  \put(260.72,71.55){\fontsize{8.54}{10.24}\selectfont $\partial_{i}M/\sim$}
  \put(20.98,7.51){\fontsize{8.54}{10.24}\selectfont 0}
  \put(98.50,8.29){\fontsize{8.54}{10.24}\selectfont $2\pi$}
  \end{picture}%
\else
  \setlength{\unitlength}{1bp}%
  \begin{picture}(308.90, 130.82)(0,0)
  \put(0,0){\includegraphics{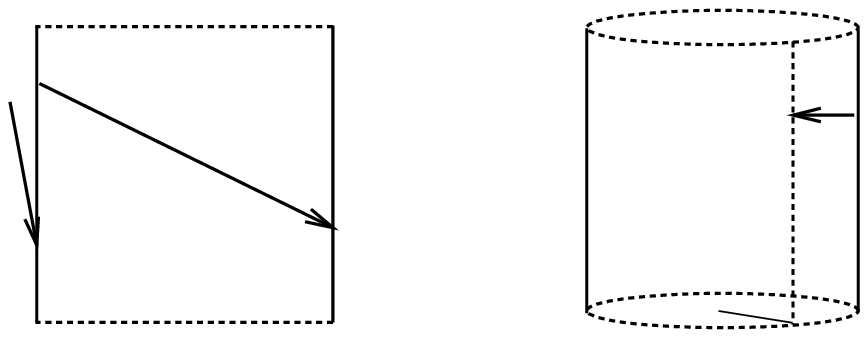}}
  \put(42.45,117.97){\fontsize{10.69}{12.83}\selectfont $(M,g)$}
  \put(197.94,117.50){\fontsize{10.69}{12.83}\selectfont $(M',g')$}
  \put(0.67,80.92){\fontsize{8.54}{10.24}\selectfont $\partial_{i}M$}
  \put(227.65,6.51){\fontsize{5.69}{6.83}\selectfont $\theta=0=2\pi$}
  \put(260.72,71.55){\fontsize{8.54}{10.24}\selectfont $\partial_{i}M/\sim$}
  \put(20.98,7.51){\fontsize{8.54}{10.24}\selectfont 0}
  \put(98.50,8.29){\fontsize{8.54}{10.24}\selectfont $2\pi$}
  \end{picture}%
\fi \caption{\label{f2}}
\end{figure}

\begin{figure}
\centering \ifpdf
  \setlength{\unitlength}{1bp}%
  \begin{picture}(185.37, 102.35)(0,0)
  \put(0,0){\includegraphics{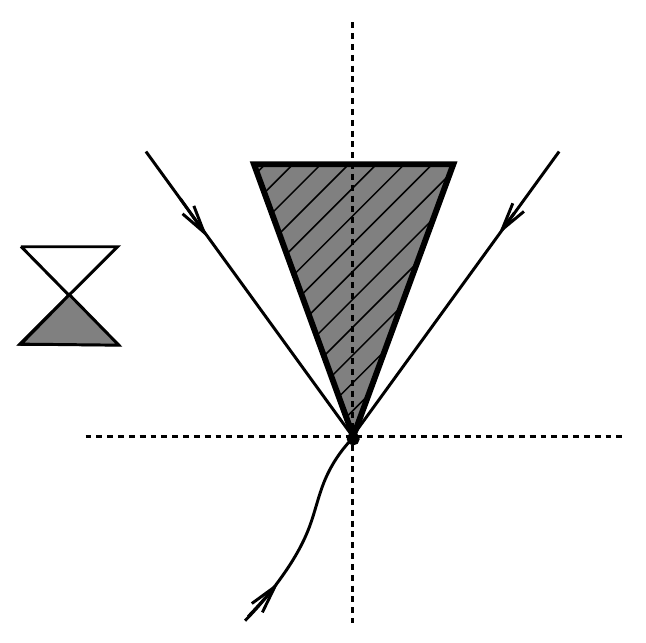}}
  \put(52.08,109.99){\fontsize{5.69}{6.83}\selectfont $\gamma_1$}
  \put(136.46,100.15){\fontsize{5.69}{6.83}\selectfont $\gamma_2$}
  \put(105.63,139.61){\fontsize{8.54}{10.24}\selectfont $C$}
  \put(103.23,48.51){\fontsize{10.69}{11.83}\selectfont $(0,0)$}
  \put(180.23,55.51){\fontsize{8.69}{9.83}\selectfont $x$}
  \put(93.23,170.51){\fontsize{8.69}{9.83}\selectfont $t$}
  \put(79.73,7.74){\fontsize{5.69}{6.83}\selectfont $\gamma_0$}
  \end{picture}%
\else
  \setlength{\unitlength}{1bp}%
  \begin{picture}(185.37, 102.35)(0,0)
  \put(0,0){\includegraphics{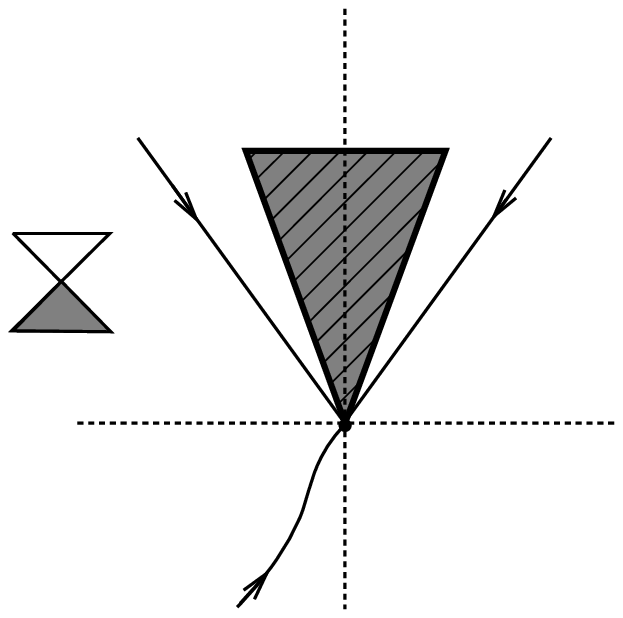}}
  \put(52.08,109.99){\fontsize{5.69}{6.83}\selectfont $\gamma_1$}
  \put(136.46,100.15){\fontsize{5.69}{6.83}\selectfont $\gamma_2$}
  \put(105.63,139.61){\fontsize{8.54}{10.24}\selectfont $C$}
  \put(103.23,48.51){\fontsize{10.69}{11.83}\selectfont $(0,0)$}
  \put(180.23,55.51){\fontsize{8.69}{9.83}\selectfont $x$}
  \put(93.23,170.51){\fontsize{8.69}{9.83}\selectfont $t$}
  \put(79.73,7.74){\fontsize{5.69}{6.83}\selectfont $\gamma_0$}
  \end{picture}%
\fi \caption{\label{f5}}
\end{figure}
\newpage
\begin{figure}
\centering
\ifpdf
  \setlength{\unitlength}{1bp}%
  \begin{picture}(489.32, 109.59)(0,0)
  \put(0,0){\includegraphics{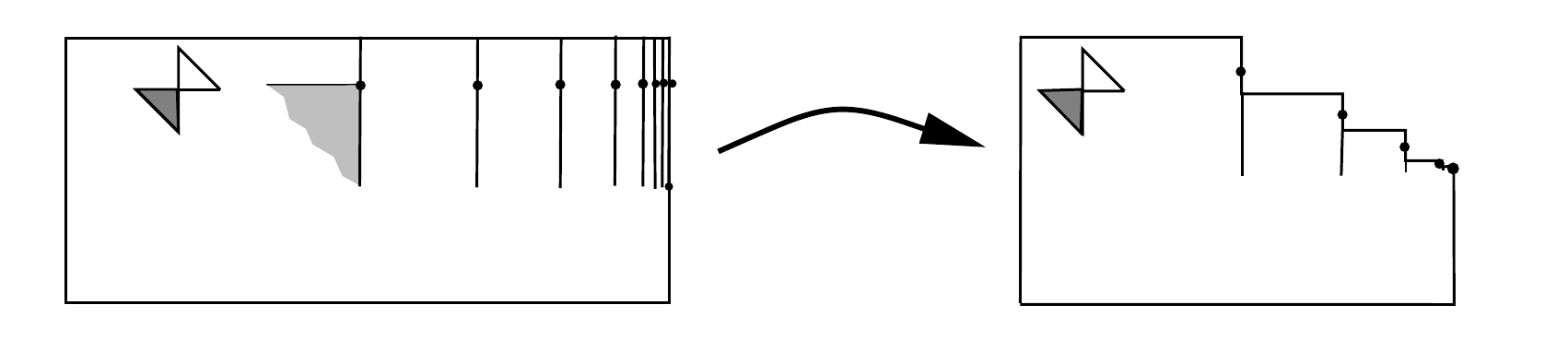}}
  \put(252.25,84.28){\fontsize{14.45}{17.35}\selectfont $\tilde{\i}$}
  \put(115.60,82.37){\fontsize{5.69}{6.83}\selectfont $p_1$}
  \put(153.60,82.37){\fontsize{5.69}{6.83}\selectfont $p_2$}
  \put(211.72,81.62){\fontsize{5.69}{6.83}\selectfont $q$}
\put(211.72,51.62){\fontsize{5.69}{6.83}\selectfont $p$}
  \put(112.85,44.67){\fontsize{8.54}{10.24}\selectfont $L_1$}
  \put(390.95,86.73){\fontsize{5.69}{6.83}\selectfont $\tilde{p}_1$}
  \put(422.95,73.73){\fontsize{5.69}{6.83}\selectfont $\tilde{p}_2$}
  \put(458.04,55.45){\fontsize{5.69}{6.83}\selectfont $\tilde{p}$}
  \put(90.33,66.45){\fontsize{8.54}{10.24}\selectfont $P_1$}
  \put(193.03,77.91){\fontsize{5.69}{6.83}\selectfont $p_n$}
  \put(441.16,62.38){\fontsize{5.69}{6.83}\selectfont $\tilde{p}_n$}
  \put(13.49,7.51){\fontsize{8.54}{10.24}\selectfont $0$}
  \put(108.31,7.51){\fontsize{8.54}{10.24}\selectfont $1$}
  \put(205.43,7.51){\fontsize{8.54}{10.24}\selectfont $2$}
  \put(9.67,53.00){\fontsize{8.54}{10.24}\selectfont $1$}
  \put(9.68,97.26){\fontsize{8.54}{10.24}\selectfont $2$}
  \put(311.58,7.51){\fontsize{8.54}{10.24}\selectfont $0$}
  \put(382.89,7.51){\fontsize{8.54}{10.24}\selectfont $1$}
  \put(449.69,7.51){\fontsize{8.54}{10.24}\selectfont $2$}
  \put(308.94,53.00){\fontsize{8.54}{10.24}\selectfont $1$}
  \put(309.35,97.26){\fontsize{8.54}{10.24}\selectfont $2$}
  \put(36.11,27.95){\fontsize{14.23}{17.07}\selectfont $M$}
  \end{picture}%
\else
  \setlength{\unitlength}{1bp}%
  \begin{picture}(489.32, 109.59)(0,0)
  \put(0,0){\includegraphics{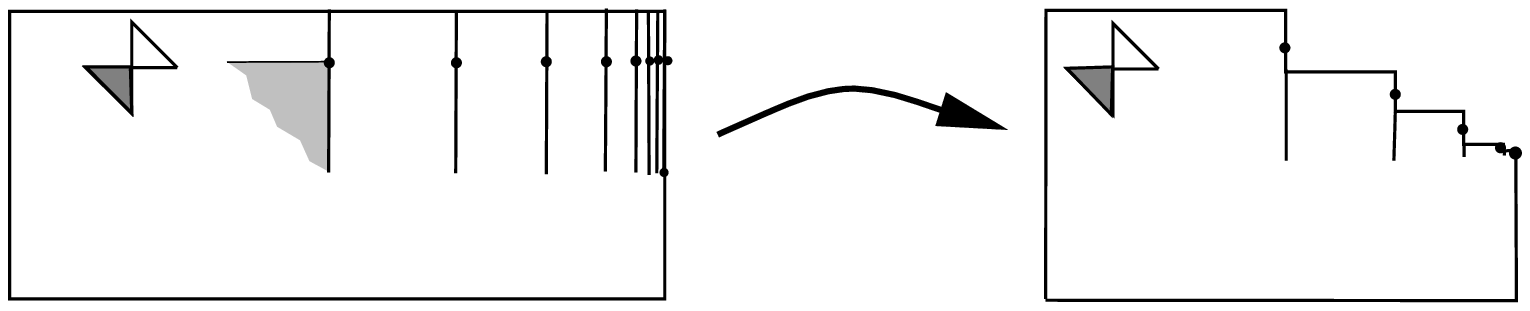}}
  \put(252.25,84.28){\fontsize{14.45}{17.35}\selectfont $\tilde{\i}$}
  \put(115.60,82.37){\fontsize{5.69}{6.83}\selectfont $p_1$}
  \put(153.60,82.37){\fontsize{5.69}{6.83}\selectfont $p_2$}
    \put(211.72,81.62){\fontsize{5.69}{6.83}\selectfont $q$}
\put(211.72,51.62){\fontsize{5.69}{6.83}\selectfont $p$}
  \put(112.85,44.67){\fontsize{8.54}{10.24}\selectfont $L_1$}
  \put(390.95,86.73){\fontsize{5.69}{6.83}\selectfont $\tilde{p}_1$}
  \put(422.95,73.73){\fontsize{5.69}{6.83}\selectfont $\tilde{p}_2$}
  \put(458.04,55.45){\fontsize{5.69}{6.83}\selectfont $\tilde{p}$}
  \put(90.33,66.45){\fontsize{8.54}{10.24}\selectfont $P_1$}
  \put(193.03,77.91){\fontsize{5.69}{6.83}\selectfont $p_n$}
  \put(441.16,62.38){\fontsize{5.69}{6.83}\selectfont $\tilde{p}_n$}
  \put(13.49,7.51){\fontsize{8.54}{10.24}\selectfont $0$}
  \put(108.31,7.51){\fontsize{8.54}{10.24}\selectfont $1$}
  \put(205.43,7.51){\fontsize{8.54}{10.24}\selectfont $2$}
  \put(9.67,53.00){\fontsize{8.54}{10.24}\selectfont $1$}
  \put(9.68,97.26){\fontsize{8.54}{10.24}\selectfont $2$}
  \put(311.58,7.51){\fontsize{8.54}{10.24}\selectfont $0$}
  \put(382.89,7.51){\fontsize{8.54}{10.24}\selectfont $1$}
  \put(449.69,7.51){\fontsize{8.54}{10.24}\selectfont $2$}
  \put(308.94,53.00){\fontsize{8.54}{10.24}\selectfont $1$}
  \put(309.35,97.26){\fontsize{8.54}{10.24}\selectfont $2$}
  \put(36.11,27.95){\fontsize{14.23}{17.07}\selectfont $M$}
  \end{picture}%
\fi \caption{\label{f6}}
\end{figure}

\begin{figure}
\centering \ifpdf
  \setlength{\unitlength}{1bp}%
  \begin{picture}(217.77, 206.45)(0,0)
  \put(0,0){\includegraphics{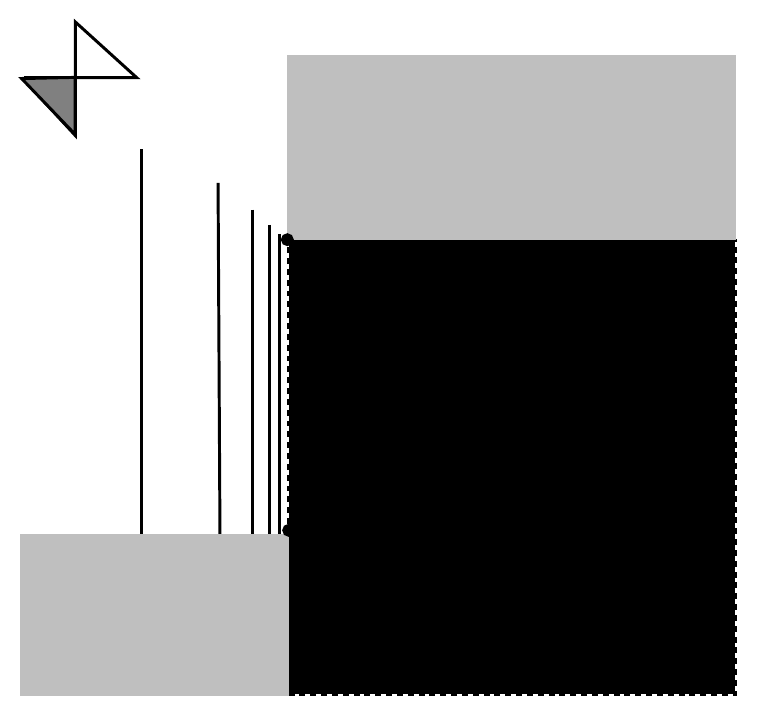}}
  \put(141.79,73.81){\fontsize{8.54}{10.24}\selectfont \textcolor[rgb]{1, 1, 1}{$H$}}
  \put(64.80,46.91){\fontsize{5.69}{6.83}\selectfont $(0,0)$}
  \put(83.50,141.07){\fontsize{5.69}{6.83}\selectfont $(0,1)$}
  \put(41.27,108.94){\fontsize{5.69}{6.83}\selectfont $L_n$}
  \put(8.24,10.28){\fontsize{14.23}{17.07}\selectfont $P$}
  \put(174.30,186.21){\fontsize{14.23}{17.07}\selectfont \makebox(0,0)[lt]{$F$\strut}}
  \end{picture}%
\else
  \setlength{\unitlength}{1bp}%
  \begin{picture}(217.77, 206.45)(0,0)
  \put(0,0){\includegraphics{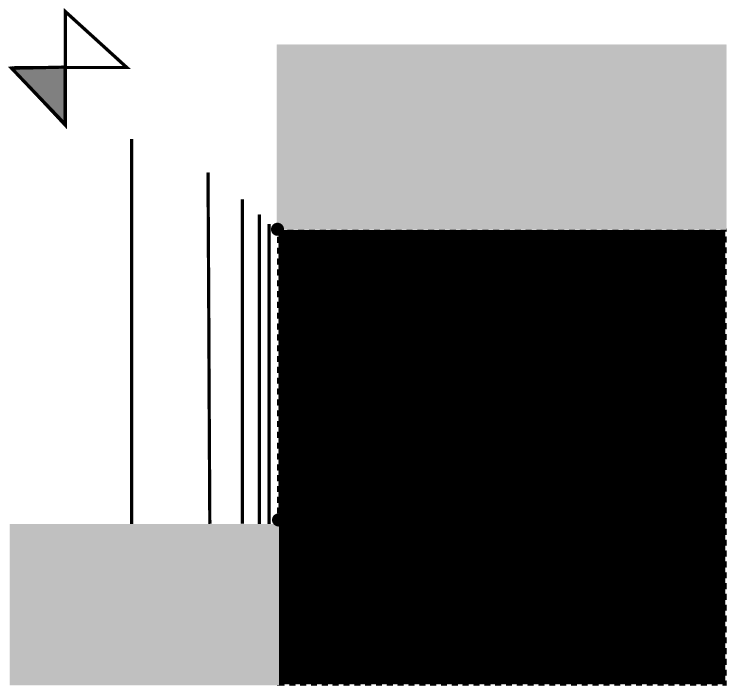}}
  \put(141.79,73.81){\fontsize{8.54}{10.24}\selectfont \textcolor[rgb]{1, 1, 1}{$H$}}
  \put(64.80,46.91){\fontsize{5.69}{6.83}\selectfont $(0,0)$}
  \put(83.50,141.07){\fontsize{5.69}{6.83}\selectfont $(0,1)$}
  \put(41.27,108.94){\fontsize{5.69}{6.83}\selectfont $L_n$}
  \put(8.24,10.28){\fontsize{14.23}{17.07}\selectfont $P$}
  \put(174.30,186.21){\fontsize{14.23}{17.07}\selectfont \makebox(0,0)[lt]{$F$\strut}}
  \end{picture}%
\fi \caption{\label{f3}}
\end{figure}

\begin{figure}
\centering
\ifpdf
  \setlength{\unitlength}{1bp}%
  \begin{picture}(458.77, 117.91)(20,0)
  \put(0,0){\includegraphics{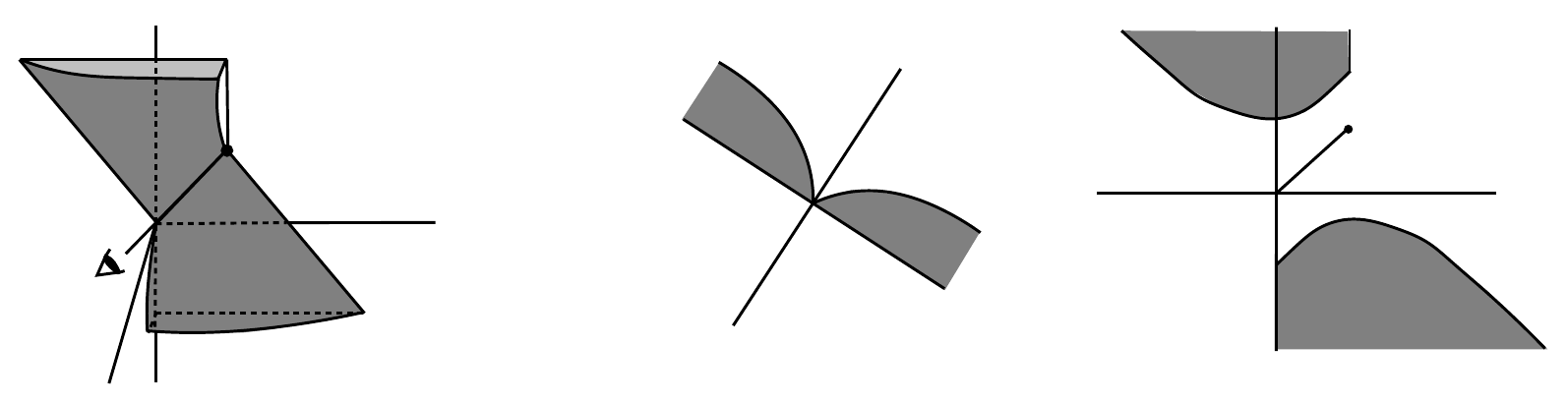}}
  \put(122.23,45.48){\fontsize{6.41}{7.70}\selectfont $y$}
  \put(21.49,10.10){\rotatebox{1.87}{\fontsize{6.41}{7.70}\selectfont \smash{\makebox[0pt][l]{$x$}}}}
  \put(47.60,107.23){\fontsize{6.41}{7.70}\selectfont $t$}
  \put(195.41,74.78){\rotatebox{358.95}{\fontsize{6.41}{7.70}\selectfont \smash{\makebox[0pt][l]{$v$}}}}
  \put(218.33,18.08){\rotatebox{0.16}{\fontsize{6.41}{7.70}\selectfont \smash{\makebox[0pt][l]{$x$}}}}
  \put(69.10,73.88){\fontsize{4.28}{5.13}\selectfont $(0,1,1)$}
  \put(321.65,55.30){\fontsize{6.41}{7.70}\selectfont $y$}
  \put(364.32,15.90){\fontsize{6.41}{7.70}\selectfont $z$}
  \put(59.77,62.18){\fontsize{6.41}{7.70}\selectfont $u$}
  \put(26.05,59.77){\fontsize{6.41}{7.70}\selectfont $v$}
  \end{picture}%
\else
  \setlength{\unitlength}{1bp}%
  \begin{picture}(458.77, 117.91)(0,0)
  \put(0,0){\includegraphics{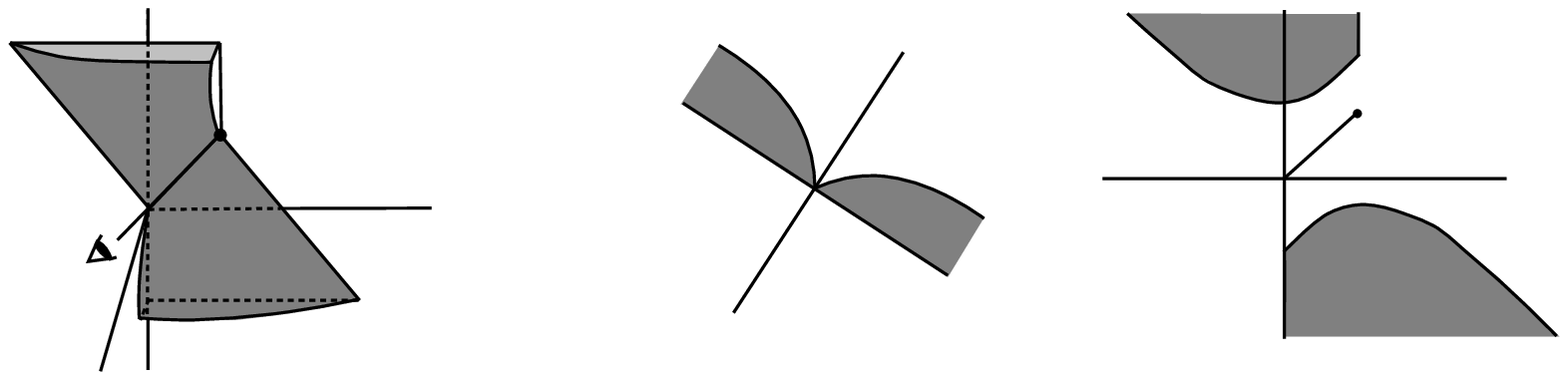}}
  \put(122.23,45.48){\fontsize{6.41}{7.70}\selectfont $y$}
  \put(21.49,10.10){\rotatebox{1.87}{\fontsize{6.41}{7.70}\selectfont \smash{\makebox[0pt][l]{$x$}}}}
  \put(47.60,107.23){\fontsize{6.41}{7.70}\selectfont $t$}
  \put(195.41,74.78){\rotatebox{358.95}{\fontsize{6.41}{7.70}\selectfont \smash{\makebox[0pt][l]{$v$}}}}
  \put(218.33,18.08){\rotatebox{0.16}{\fontsize{6.41}{7.70}\selectfont \smash{\makebox[0pt][l]{$x$}}}}
  \put(69.10,73.88){\fontsize{4.28}{5.13}\selectfont $(0,1,1)$}
  \put(321.65,55.30){\fontsize{6.41}{7.70}\selectfont $y$}
  \put(364.32,15.90){\fontsize{6.41}{7.70}\selectfont $z$}
  \put(59.77,62.18){\fontsize{6.41}{7.70}\selectfont $u$}
  \put(26.05,59.77){\fontsize{6.41}{7.70}\selectfont $v$}
  \end{picture}%
\fi \caption{\label{f4}}
\end{figure}

\newpage

\newpage
\begin{figure}
\centering \ifpdf
  \setlength{\unitlength}{1bp}%
  \begin{picture}(136.95, 171.11)(0,0)
  \put(0,0){\includegraphics{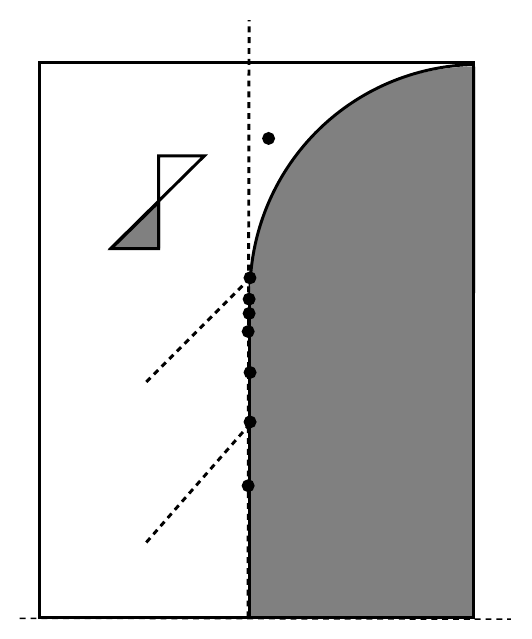}}
  \put(75.52,103.09){\rotatebox{0.00}{\fontsize{5.69}{6.83}\selectfont \smash{\makebox[0pt][l]{$(0,0)$}}}}
  \put(74.38,60.88){\fontsize{5.69}{6.83}\selectfont $v_n$}
  \put(73.38,170.88){\fontsize{8.69}{9.83}\selectfont $v$}
   \put(138.38,9.88){\fontsize{8.69}{9.83}\selectfont $u$}
  \put(45.25,21.27){\fontsize{5.69}{6.83}\selectfont $P_n$}
  \put(41.96,62.67){\fontsize{5.69}{6.83}\selectfont $P_0$}
 \put(88.98,159.29){\fontsize{5.69}{6.83}\selectfont $F_0$}
  \put(79.33,146.64){\fontsize{5.69}{6.83}\selectfont $q$}
  \end{picture}%
\else
  \setlength{\unitlength}{1bp}%
  \begin{picture}(136.95, 171.11)(0,0)
  \put(0,0){\includegraphics{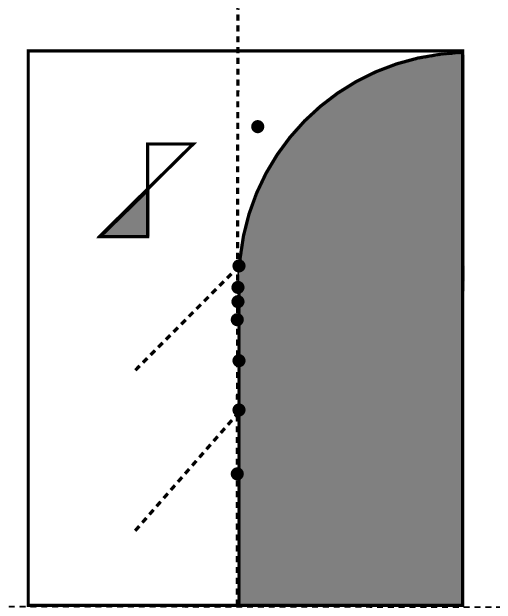}}
  \put(75.52,103.09){\rotatebox{0.00}{\fontsize{5.69}{6.83}\selectfont \smash{\makebox[0pt][l]{$(0,0)$}}}}
  \put(74.38,60.88){\fontsize{5.69}{6.83}\selectfont $v_n$}
  \put(73.38,170.88){\fontsize{8.69}{9.83}\selectfont $v$}
   \put(138.38,9.88){\fontsize{8.69}{9.83}\selectfont $u$}
  \put(45.25,21.27){\fontsize{5.69}{6.83}\selectfont $P_n$}
  \put(41.96,62.67){\fontsize{5.69}{6.83}\selectfont $P_0$}
  \put(88.98,159.29){\fontsize{5.69}{6.83}\selectfont $F_0$}
  \put(79.33,146.64){\fontsize{5.69}{6.83}\selectfont $q$}
  \end{picture}%
\fi \caption{\label{f7}}
\end{figure}


\section*{Acknowledgments}

Comments by Prof. S. Harris (St. Louis U.),  Prof. O. Palmas
(UNAM), Prof. D. Sol\'{i}s (UADY) and his student L. Ak\'e on
different parts of the article, as well as by Prof. L. Andersson
(A. Einstein Inst., Postdam) on Remark \ref{rlars}, are
acknowledged.

All the authors are partially supported by the Spanish Grants
MTM2010-18099 (MICINN) and P09-FQM-4496 (J. Andaluc\'{i}a), both
with FEDER funds. Also, the second-named author is supported by
Spanish MEC Grant AP2006-02237.


\begin{thebibliography}{99}

\bibitem{AF} V. Ala\~na, J.L. Flores,
The causal boundary of product spacetimes, {\em Gen. Rel.
Gravitation} {\bf 39} (2007), no. 10, 1697--1718.

\bibitem{AH} A. Ashtekar, R.O. Hansen, A unified treatment of null and spatial infinity in general
relativity. I. Universal structure, asymptotic symmetries, and
conserved quantities at spatial infinity, {\em J. Math. Phys.}
{\bf  19} (1978), no. 7, 1542--1566.

\bibitem{BEE} J.K. Beem, P.E. Ehrlich, K.L. Easley,
{\em Global Lorentzian geometry}, Monographs Textbooks Pure Appl.
Math. {\bf 202} (Dekker Inc., New York, 1996).


\bibitem{BMN} D. Berenstein, J. M. Maldacena and H. Nastase, Strings in flat
space and pp waves from N = 4 super Yang Mills, {\em J. High
Energy Phys.} {\bf 0204} (2002) 013.

\bibitem{BN} D. Berenstein and H. Nastase, {\em On lightcone string field theory from super Yang-Mills and
holography}. Available at arXiv:hep-th/0205048.


\bibitem{BeSa2} A.N. Bernal and M. S\'anchez, Globally hyperbolic spacetimes can
be defined as ``causal" instead of ``strongly causal'', {\it
Class. Quant. Grav.} {\bf  24} (2007) 745--750.


\bibitem{BS} R. Budic, R.K. Sachs, Causal boundaries for general relativistic
spacetimes, {\it J. Math. Phys.} {\bf 15} (1974) 1302--1309.


\bibitem{CFS} A.M. Candela, J.L. Flores, M. S\'anchez, On  general plane fronted waves. Geodesics, {\it Gen. Relat. Gravitation} {\bf 35} (2003) 631--649.




\bibitem{F} J.L. Flores, The Causal Boundary of spacetimes revisited, {\it Commun. Math. Phys.} {\bf 276} (2007) 611--643.

\bibitem{FH} JL Flores, S.G. Harris, Topology of the
causal boundary for standard static spacetimes, {\em Class. Quant.
Grav.} {\bf 24} (2007), no. 5, 1211--1260.

\bibitem{FHSst}  J.L. Flores, J. Herrera, M. S\'anchez, {\em Gromov, Cauchy and causal boundaries for Riemannian, Finslerian and
Lorentzian manifolds}. Available at arXiv:1011.1154.



\bibitem{FS} J.L. Flores, M. S\'anchez,
 The causal boundary of wave-type spacetimes,{\em J. High Energy Phys.} (2008), no. 3, 036, 43 pp

\bibitem{Fr} C. Frances, {\em The conformal boundary of ani-de Sitter
space-times}, in: AdS/CFT Correspondence: Einstein Metrics and
Their Conformal Boundaries, Ed. O. Biquard, Euro. Math. Soc.,
Zurich, (2005), 205-216.



\bibitem{GpScqg05}
A. Garc{\'\i}a-Parrado, J. M. M. Senovilla, Causal structures and
causal  boundaries, {\it Class. Quant. Grav.} {\bf 22} (2005)
R1--R84.


\bibitem{GeJMP}
R.P. Geroch, Local characterization of singularities in general
relativity, {\em J. Math. Phys.} {\bf 9} (1968) 450--465.

\bibitem{GKP}
R.P. Geroch, E.H. Kronheimer and R. Penrose, Ideal points in
spacetime, {\it Proc. Roy. Soc. Lond. A} {\bf 237} (1972) 545--67.

\bibitem{GLW}
R.P. Geroch,  C.B Liang,  R.M Wald,  Singular boundaries of
space-times, {\em J. Math. Phys.} {\bf 23} (1982), no. 3,
432--435.

\bibitem{goreham} A. Goreham,  Sequential convergence in
topological spaces, arxiv: math/0412558v1.

\bibitem{H1} S.G. Harris,
Universality of the future chronological boundary, {\em J. Math.
Phys.} {\bf 39} (1998), no. 10, 5427--5445.

\bibitem{H2} S.G. Harris, Topology of the future chronological boundary: universality for spacelike boundaries,
{\em Classical Quantum Gravity} {\bf 17} (2000), no. 3, 551--603.


\bibitem{HE} S.W. Hawking, G.F.R. Ellis, {\it The Large Scale
Structure of Space-Time}, Cambridge University, Cambridge, 1973.


\bibitem{KLLPRD} Z.Q. Kuang, J.Z. Li,   C.B. Liang, c-boundary of Taubs plane-symmetric static vacuum spacetime {\em Phys. Rev. D} {\bf 33}.
{Phys. Rev. D}, (1986) 1533-1537.

\bibitem{KLJMP} Z.Q. Kuang, C.B. Liang,
On the GKP and BS constructions of the $c$-boundary, {\em J. Math.
Phys.} {\bf 29} (1988), no. 2, 433--435.


\bibitem{KLPRD} Z.Q. Kuang, C.B. Liang,  On the R\'acz and Szabados
constructions of the $c$-boundary {\em Phys. Rev. D} (3) {\bf 46}
(1992), no. 10, 4253--4256.




\bibitem{MR1} D. Marolf, S. Ross, Plane Waves: To infinity and
beyond! {\it Class. Quant. Grav.} {\bf 19} (2002) 6289--6302.

\bibitem{MR} D. Marolf, S.R. Ross, A new recipe for causal
completions, {\it Class. Quant. Grav.} {\bf 20} (2003) 4085--4117.





\bibitem{MS} E. Minguzzi, M. S\'{a}nchez,  The causal hierarchy of
spacetimes, in {\em Recent developments in pseudo-Riemannian Geometry} (2008) 359--418. 
ESI Lect. in Math. Phys., European Mathematical Society Publishing
House. (Available at gr-qc/0609119).

\bibitem{MuSa} O. M\"uller, M. S\'anchez, Lorentzian manifolds isometrically embeddable in $L^N$, {\em Trans. Amer. Math.
Soc.}, {\bf 363} (2011), 5367--5379.

\bibitem{NO} K. Nomizu, H. Ozeki, The existence of complete Riemannian metrics, {\it Proc. Amer. Math. Soc.} {\bf 12} (1961) 889--891.

\bibitem{O} B. O'Neill, {\em Semi-Riemannian Geometry with applications to Relativity}, Academic Press, INC, 1983.


\bibitem{SS} S. Scott, P. Szekeres,  The abstract boundary---a new
approach to singularities of manifolds, {\em J. Geom. Phys.} {\bf
13} (1994), no. 3, 223--253.


\bibitem{Ra}
I. R\'acz, Causal boundary of space-times, {\it Phys. Rev. D} {\bf
36} (1987), 1673--1675


\bibitem{Ru} P. R\"ube, An example of a nontrivial causally simple space-time having interesting consequences for boundary constructions.
{\em J. Math. Phys.} {\bf 31} (1990), no. 4, 868--870.


\bibitem{S} M. S\'anchez, Causal boundaries and holography on wave
type spacetimes, {\em Nonlinear Anal.}, {\bf 71} (2009),
e1744-e1764.


\bibitem{Sc} B.G. Schmidt,
 A new definition of singular points in general relativity. {\em Gen.
Relat. Gravitation } {\bf 1} (1970/71), no. 3, 269--280.

\bibitem{Sc2} B.G. Schmidt,  Remarks about modifications of the $b$-boundary
definition, {\em Gen. Relat. Gravitation}{\bf  10} (1979), no. 12,
981--982.




\bibitem{Sz}
L.B. Szabados, Causal boundary for strongly causal spaces, {\it
Class. Quant. Grav.} {\bf 5} (1988) 121--34.

\bibitem{Sz2}
L.B. Szabados, Causal boundary for strongly causal spacetimes: II,
{\it Class. Quant. Grav.} {\bf 6} (1989) 77--91.

\bibitem{Wald} R.M. Wald, {\it General Relativity}. Chicago: The
University of Chicago Press (1984).

\end{thebibliography}
\end{document}